\input{aipcheck}

\documentclass[
    ,final            
  ]
  {aipproc}

\layoutstyle{6x9}

\usepackage{amssymb}

\begin{document}

\title{Spins, Stripes, and Superconductivity in Hole-Doped Cuprates}


\author{John M. Tranquada}{
  address={Condensed Matter Physics \&\ Materials Science Department\\
                     Brookhaven National Laboratory, Upton, NY 11973-5000, USA}
}


\begin{abstract}
One of the major themes in correlated electron physics over the last quarter century has been the problem of high-temperature superconductivity in hole-doped copper-oxide compounds.  Fundamental to this problem is the competition between antiferromagnetic spin correlations, a symptom of strong Coulomb interactions, and the kinetic energy of the doped carriers, which favors delocalization.  After discussing some of the early challenges in the field, I describe the experimental picture provided by a variety of spectroscopic and transport techniques.  Then I turn to the technique of neutron scattering, and discuss how it is used to determine spin correlations, especially in model systems of quantum magnetism.  Neutron scattering and complementary techniques have determined the extent to which antiferromagnetic spin correlations survive in the cuprate superconductors.  One experimental case involves the ordering of spin and charge stripes.  I first consider related measurements on model compounds, such as La$_{2-x}$Sr$_x$NiO$_{4+\delta}$, and then discuss the case of La$_{2-x}$Ba$_x$CuO$_4$.  In the latter system, recent transport studies have demonstrated that quasi-two-dimensional superconductivity coexists with the stripe order, but with frustrated phase order between the layers.  This has led to new concepts for the coexistence of spin order and superconductivity.  While the relevance of stripe correlations to high-temperature superconductivity remains a subject of controversy, there is no question that stripes are an intriguing example of electron matter that results from strong correlations.
\end{abstract}

\maketitle

\section{Opening Remarks}

Cuprate superconductivity is a fascinating field, with an enormous literature.  There have been a great many measurements done on a wide variety of cuprate families.  There are numerous theoretical perspectives on the nature of these materials and on the mechanism of superconductivity.

I approach this topic not as an unbiased outside observer, but as a participant who has been around from the beginning.   My purpose here is not to present a balanced review of all work and perspectives in the field, as there would be too much to cover.  Instead, I will present one story about hole-doped cuprates, based largely on experiment and especially on neutron scattering studies.  Since one of the main advantages of neutrons is their ability to follow dynamic antiferromagnetic spin correlations, it should come as no surprise that they will play a key role in the story.

I have tried to cite sufficient references (especially review articles) so that the curious reader can find further information and potentially a starting point in searching the broader literature.  I have not attempted to cite all relevant work, as that would have been impractical for a set of lectures.  In any case, I apologize in advance to any researcher in the field who feels slighted by my choice of references or topics.

\section{Introduction to the cuprates}

\subsection{Discovery and early questions}

\begin{figure}[t]
  \includegraphics[width=.3\textwidth]{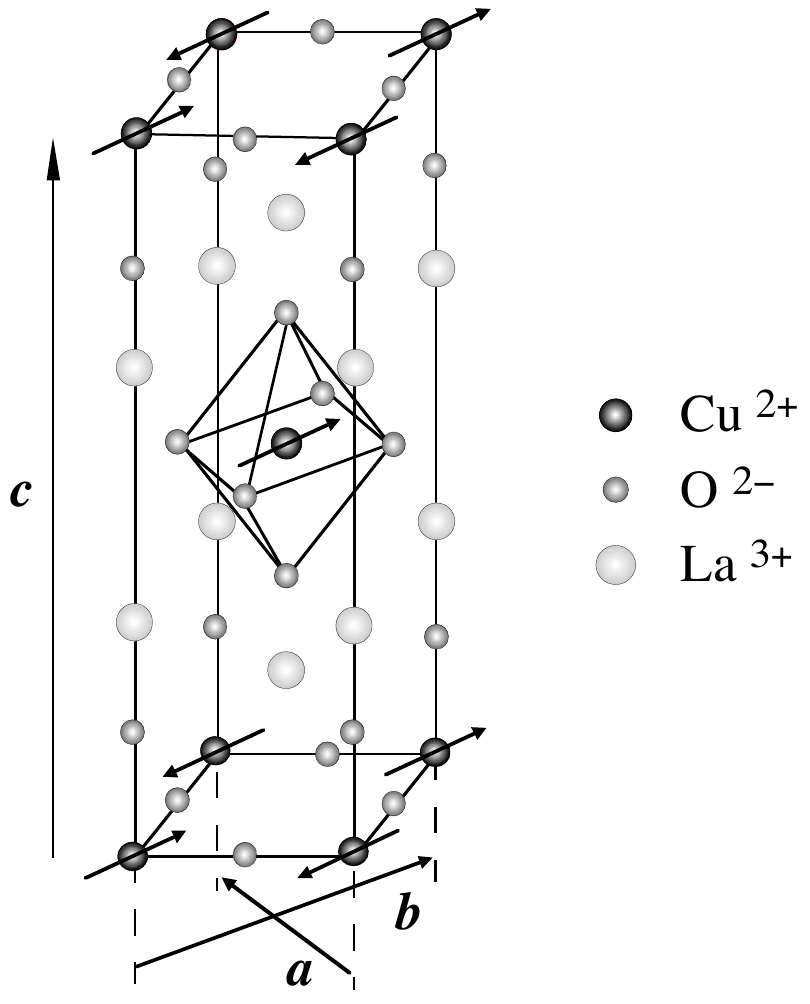}
  \caption{Atomic and magnetic structure of La$_2$CuO$_4$.  Arrows indicate the orientation of magnetic moments in the antiferromagnetic phase.   The lattice parameters shown correspond to the low-temperature-orthorhombic (LTO) crystal structure. }
  \label{fg:lco_struc}
\end{figure}

High-temperature superconductivity was originally discovered by Bednorz and M\"uller in 1986 \cite{bedn86}.    Motivated by theoretical proposals of bipolarons, they worked with a mixture of La-Ba-Cu-O.  Their hope was to dope Jahn-Teller-active sites into a cubic perovskite lattice; these polaronic centers might then pair and lead to superconductivity.  For example, in LaCuO$_3$ the Cu ions are nominally trivalent, with one hole in each of the two $3d$ orbitals with $e_g$ symmetry.\footnote{In a cubic crystal field, the five $3d$ orbitals are split into a pair of orbitals ($x^2-y^2$ and $3z^2-r^2$) with $e_g$ symmetry and a triplet of orbitals ($xy$, $yz$, and $zx$) with $t_{2g}$ symmetry.}  On the other hand, a Cu$^{2+}$ ion has just one hole that could go into either of the two $e_g$ orbitals.  In such a case, the energy is generally lowered by a structural distortion that breaks the degeneracy of the $e_g$ orbitals ({\it i.e.}, the Jahn-Teller effect). It turns out that the discovered superconducting phase is actually La$_{2-x}$Ba$_x$CuO$_4$ \cite{bedn88}, a layered cuprate with built-in broken symmetry (see Fig.~\ref{fg:lco_struc}). The CuO$_2$ layers are similar to those in the perovskite structure, but they are separated by La$_2$O$_2$ layers.  Each Cu has six O neighbors, but the octahedral arrangement is elongated along the $c$ axis, perpendicular to the planes.  As a result, the one hole on each Cu$^{2+}$ site occupies the $d_{x^2-y^2}$ orbital.

The experimental properties of the parent compound La$_2$CuO$_4$, and related La$_2$NiO$_4$, were not well understood in 1986.  For example, Singh {\it et al.} \cite{sing84} interpreted magnetic susceptibility and transport measurements as indicating charge-density-wave ordering below 200~K.  Following the discovery of superconductivity, a rather different picture was proposed by Anderson \cite{ande87}, who pointed out that La$_2$CuO$_4$ should be a Mott insulator with superexchange driving antiferromagnetic correlations among Cu moments.

\begin{figure}
  \includegraphics[width=0.3\textwidth]{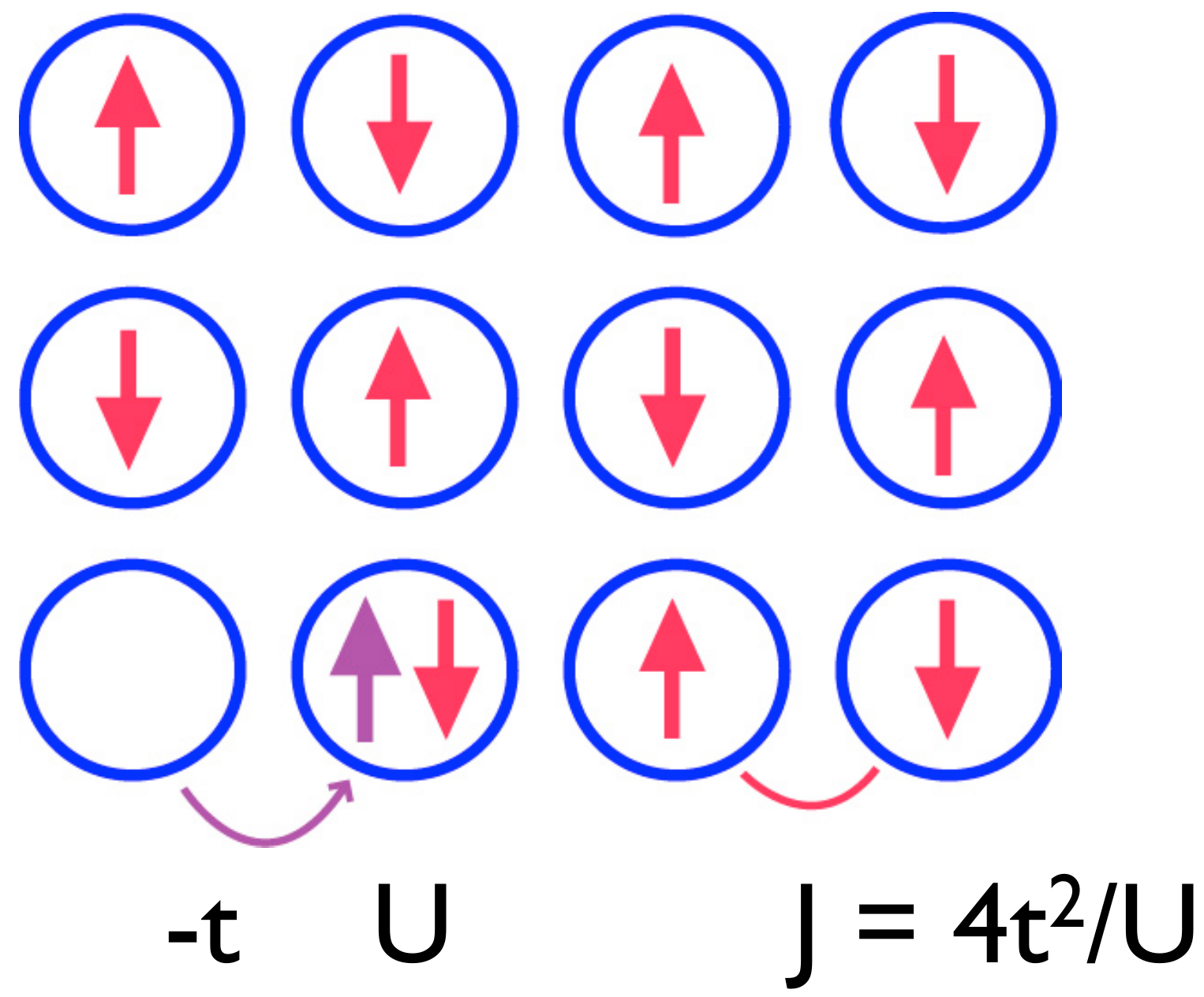}
  \caption{Cartoon of the Hubbard model and the superexchange mechanism.}
  \label{fg:hubbard}
 \end{figure}

Superexchange is understood in terms of the Hubbard model \cite{ande59}, with one orbital per Cu site; for now, we will ignore the oxygens.  As already mentioned, Cu$^{2+}$ in a tetragonal environment has one unpaired electron in the $3d_{x^2-y^2}$ orbital.   In conventional band theory, one focuses on the kinetic energy of the electrons.  An electron wants to delocalize and can reduce its energy by an amount $t$ by hopping from one site to the next; the band width corresponds to $8t$.  (Conventional band structure calculations predicted that La$_2$CuO$_4$ should essentially be metallic, with a single band crossing the Fermi level \cite{matt87,yu87}.)  In the Hubbard model, one also takes account of the Coulomb repulsion $U$ between two electrons on the same site.   If $U$ is greater than the band width, then the electrons will tend to be localized, one per site.  An electron can still lower its kinetic energy by  making a virtual hop to a nearest-neighbor site and back again, but it can only hop if its spin is antiparallel to that of an electron already on the neighboring site, due to the Pauli exclusion principle (see Fig.~\ref{fg:hubbard}). The effective Heisenberg exchange energy between neighboring antiparallel spins corresponds to the superexchange energy $J=4t^2/U$.  Thus, Anderson's analysis suggests that the parent compound La$_2$CuO$_4$ should fall into the class of antiferromagnetic insulators.\footnote{ Anderson actually proposed that quantum fluctuations would prevent antiferromagnetic order \cite{ande87}.  As we will discuss later, neutron diffraction measurements eventually demonstrated that La$_2$CuO$_4$ exhibits antiferromagnetic order, though with an ordered moment reduced by quantum fluctuations.} 

\begin{figure}
  \includegraphics[width=0.5\textwidth]{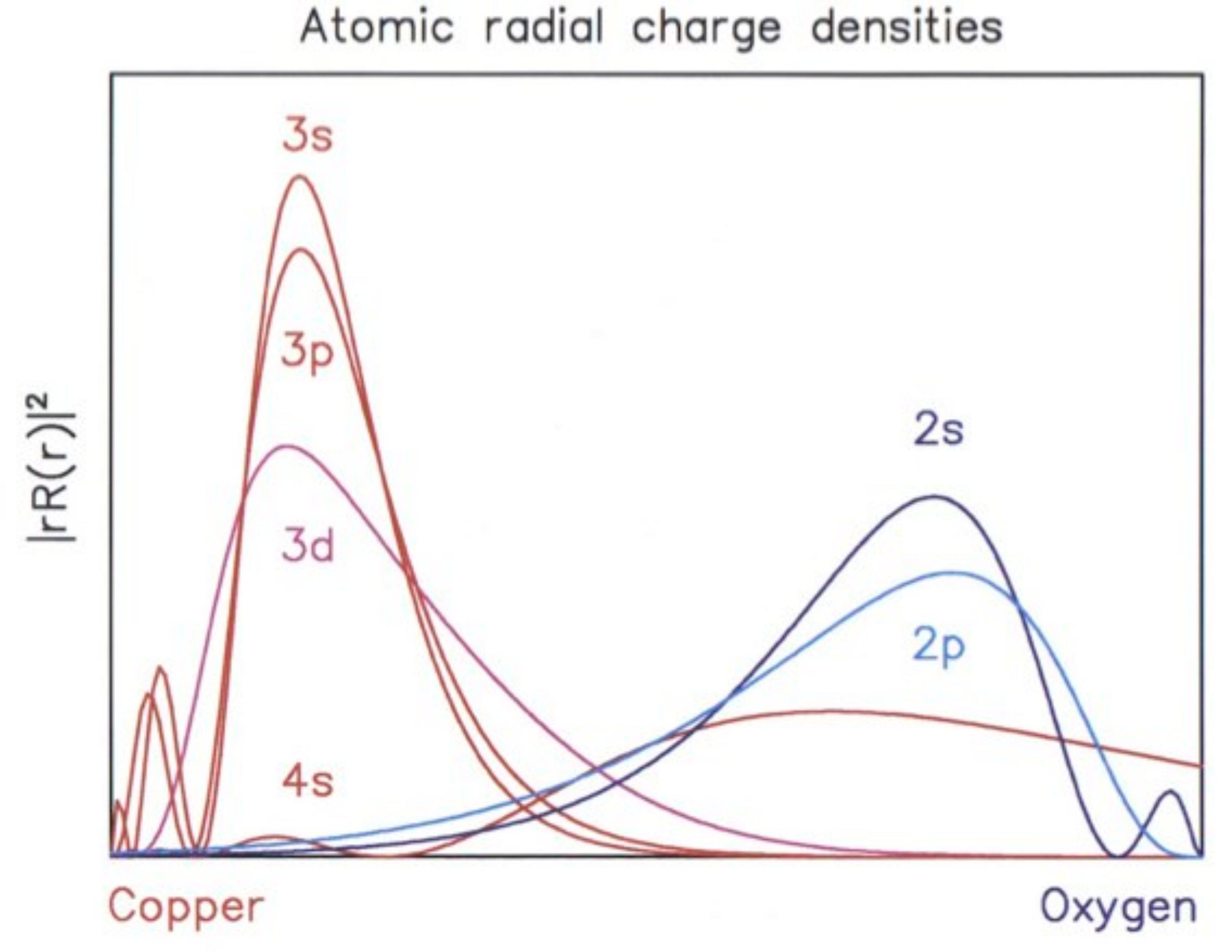}
  \caption{Radial charge densities for outer orbitals of Cu and O at a typical Cu-O bond spacing; obtained from parametrized tabulations calculated wave functions.}
  \label{fg:orbitals}
 \end{figure}

For a more complete picture, we also need to take account of the fact that there is an oxygen atom between each pair of coppers.  One can gain some appreciation for the situation by looking at the orbital radial charge densities at the typical Cu-O spacing, as illustrated in Fig.~\ref{fg:orbitals}.  The Cu $4s$ states strongly overlap with the O $2p$, so it is not surprising that those electrons can be treated as having been transferred to O; the oxygen ions have a valence of ${2-}$ when the $2p$ states are filled.  On the other hand, the peak density of Cu $3d$ states is inside of that  for the $3p$ and $3s$ states, so it is understandable that there are strong Coulomb interactions between Cu $3d$ electrons.  In addition, there is significant overlap between the Cu $3d$ and O $2p$ orbitals, which results in considerable hybridization.

Early on, there were many questions about the nature of the layered cuprates.  Is the parent material La$_2$CuO$_4$ closer to a conventional metal or to a Mott (or charge-transfer) insulator?  Is it antiferromagnetic?  Or would it avoid antiferromagnetic order because of quantum spin fluctuations?

\begin{figure}
  \includegraphics[width=0.5\textwidth]{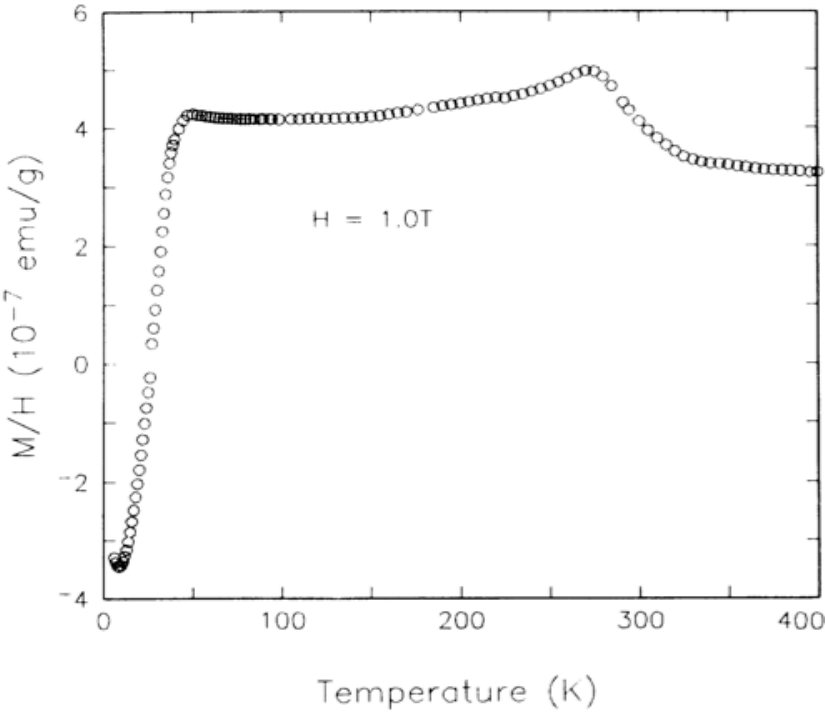}
  \caption{Plot of magnetic susceptibility vs.\ temperature indicating both antiferromagnetic ordering (peak at $\sim 270$~K) and superconductivity (diamagnetism below 30~K) in a sintered sample identified as La$_2$CuO$_4$ in the original study by Grant {\it et al.} \cite{gran87}, but now understood to contain (phase separated) excess oxygen.}
  \label{fg:grant}
 \end{figure}

Many of the initial experimental results were confusing.  For example, an early paper on nominally pure La$_2$CuO$_4$ found evidence for superconductivity \cite{gran87}; the magnetization data shown in Fig.~\ref{fg:grant} shows diamagnetism below 30~K.  It turns out that if one looks carefully at the data, there is also evidence of the antiferromagnetic transition at $\sim270$~K.\footnote{The evidence is a peak in the magnetization at the ordering temperature.  Such a peak is anomalous for a quasi-2D antiferromagnet, and is a consequence of a small ferrimagnetic canting of the spins due to Dzyaloshinskii-Moriya interactions \cite{kast98}.}  Is stoichiometric La$_2$CuO$_4$ actually superconducting? And does the superconductivity coexist with antiferromagnetic order?

This is an example of the importance of careful materials science in order to determine intrinsic properties of exciting new materials.  Early samples were polycrystalline, the nature of possible defects was not understood, and frequently there was a coexistence of multiple phases.  These challenges are common when a new class of materials attracts sudden attention.  With time, the situation in the cuprates has been clarified.

\begin{figure}
  \includegraphics[width=0.5\textwidth]{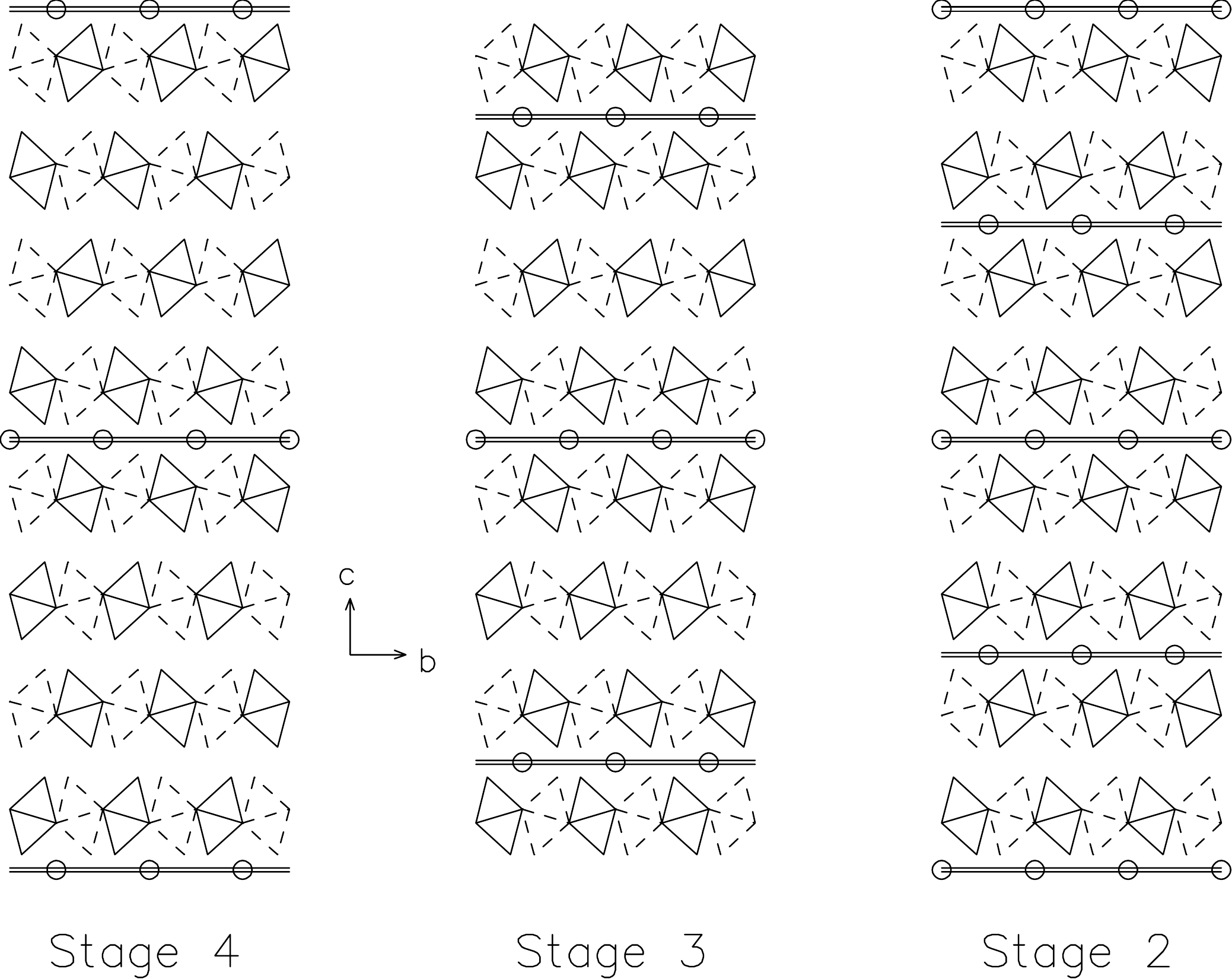}
  \caption{Schematic illustration of stage ordering in La$_2$NiO$_{4+\delta}$ involving layers of oxygen interstitials (indicated by circles) intercalated between layers of NiO$_6$ octahedra that have an ordered tilt pattern; the interstitial layers act as anti-phase domain walls for the tilt pattern.  For simplicity, La sites are not shown.  From \cite{tran94b}. }
  \label{fg:stage}
 \end{figure}

\begin{figure}
  \includegraphics[width=0.5\textwidth]{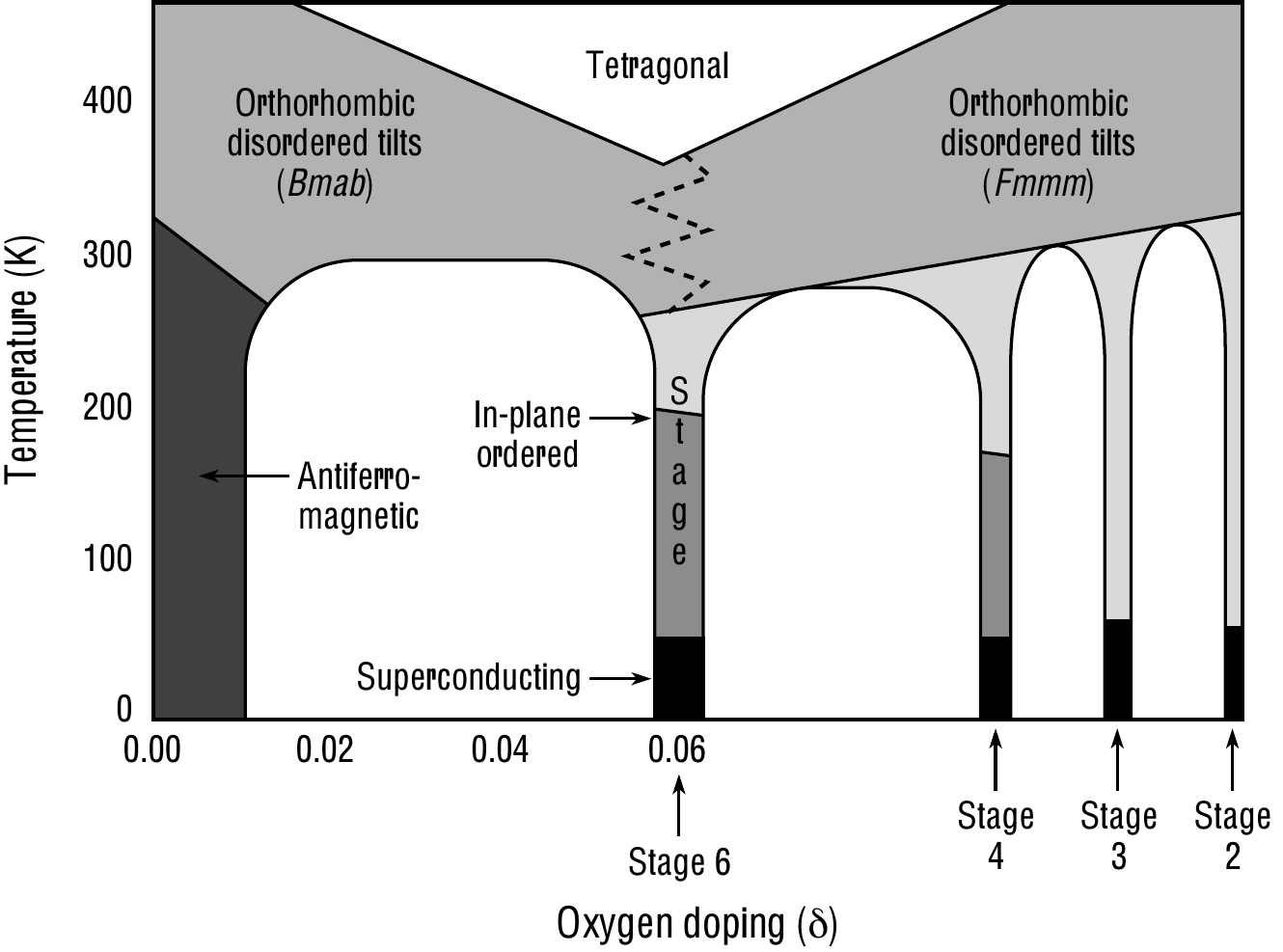}
  \caption{Phase diagram for La$_2$CuO$_{4+\delta}$ resulting from neutron diffraction studies by Wells {\it et al.}  \cite{well97}.  Reprinted with permission from AAAS. }
  \label{fg:lco_ph_diag}
 \end{figure}

In the case of La$_2$CuO$_4$, it turns out that excess oxygen atoms can enter the lattice as interstitials within the La$_2$O$_2$ layers, with a range of possible densities.\footnote{For a review of a considerable amount of experimental work on this problem, see \cite{tran98b}.}  Studies of the related material La$_2$NiO$_{4+\delta}$, by single-crystal neutron diffraction, demonstrated that the interstitial oxygens can order in staged structures analogous to the structures formed by intercalants in graphite \cite{tran94b}; the ordering is illustrated in Fig.~\ref{fg:stage}.  Similar ordered phases were then identified in La$_2$CuO$_{4+\delta}$ \cite{well96,well97}.  As shown in Fig.~\ref{fg:lco_ph_diag}, the phase diagram has several preferred interstitial concentrations; a sample with an intermediate concentration phase separates into domains of the two neighboring ordered states.  The magnetic insulator state survives only for very small oxygen excesses, then there is a miscibility gap separating the first superconducting phase.  Clearly, the sample studied by Grant {\it et al.} \cite{gran87} contained both of these phases.

\begin{figure}
  \includegraphics[width=0.5\textwidth]{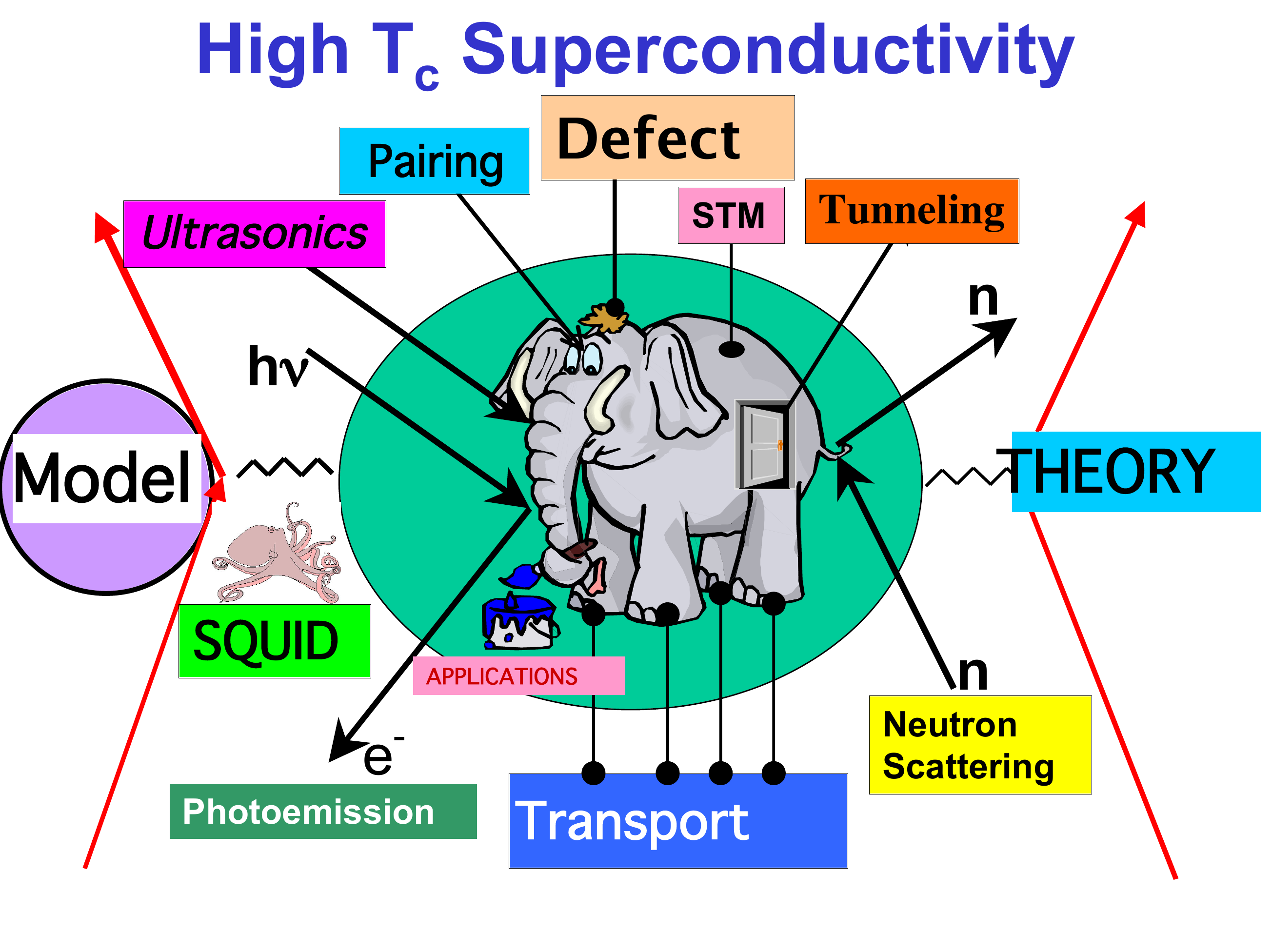}
  \caption{Cartoon illustrating the challenge of the blind physicists trying to identify the superconducting elephant using a range of techniques that each probe a different aspect.  From \cite{schu02}. }
  \label{fg:elephant}
 \end{figure}

Another challenge is represented through the allegory of the blind physicists and the superconducting elephant, as illustrated in Fig.~\ref{fg:elephant}.  Each experimentalist probes the mysterious system with a different technique.  Interpretation of the measurements typically relies on simplified standard models.  The collective result is an array of apparently conflicting views of cuprate superconductors.  It has taken many years to reconcile the disparate perspectives, and work towards a full understanding is still in progress.

An important lesson for theorists is that they should not accept everything that experimentalist report at face value.   One needs to ask whether the samples are well characterized.  What models have been used to interpret the data?  What assumptions have been made?  Rather than simply picking the experimental measurement one likes and focusing solely on it, greater progress can be made by considering how apparently conflicting observations might be reconciled.

On the other side, experimentalists should not be limited by theorists' predictions.  Initial expectations need to be tested with measurements on well-characterized samples.  Nevertheless, even wrong predictions can motivate important experimental discoveries.  The key is to approach surprises with a questioning attitude.

\subsection{Spectroscopic and transport characterizations of La$_{2-x}$Sr$_x$CuO$_4$}

\subsubsection{Optical spectroscopy}

Let us examine some of the experimental characterizations of cuprate superconductors.  The results for different cuprate families tend to be similar, so we will focus on La$_{2-x}$Sr$_x$CuO$_4$ and closely related compounds.  We will begin with various electronic spectroscopies and transport measurements. 

\begin{figure}
  \includegraphics[width=0.4\textwidth]{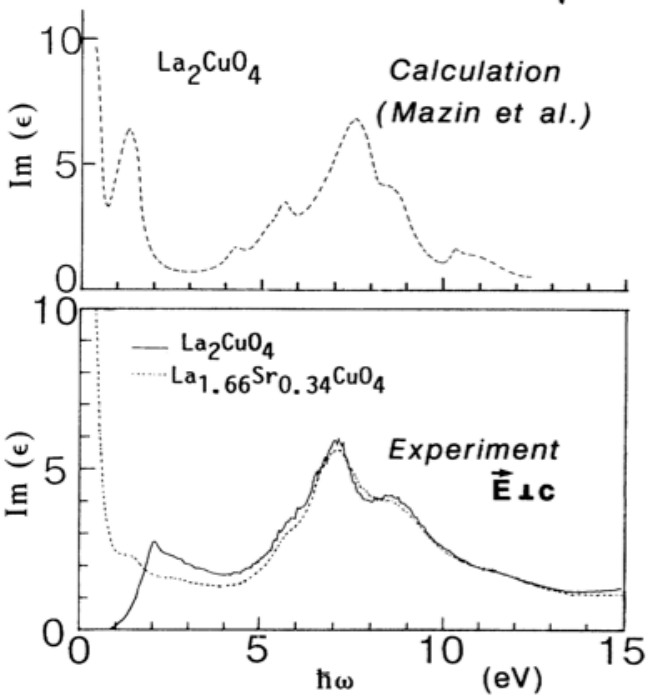}
  \caption{Plots of $\textrm{Im}(\epsilon)$, the imaginary part of the dielectric function, which equals the real part of the optical conductivity $\sigma(\omega)$ divided by $\omega$, for polarization parallel to the CuO$_2$ planes in La$_2$CuO$_4$ and La$_{1.66}$Sr$_{0.34}$CuO$_4$ from experiment (bottom) and from a calculation (top) based on density-functional theory for La$_2$CuO$_4$ \cite{mazi88}, as reported by Uchida {\it et al.} \cite{uchi91}. }
  \label{fg:lco_optical}
 \end{figure}

An early study of optical conductivity\footnote{Optical studies typically involve a measurement of reflectivity from a well-oriented surface of a single crystal.  The optical conductivity is obtained from a Kramers-Kronig transformation of the reflectivity measured over a large frequency range.  It probes the joint densities of filled and empty states for transitions involving negligible momentum transfer (${\bf q}\approx 0$).  For a review, see Basov and Timusk \cite{baso05}.} on a series of single crystals at room temperature was performed by Uchida {\it et al.} \cite{uchi91}.  Figure~\ref{fg:lco_optical} shows a comparison of the experimental spectra for undoped La$_2$CuO$_4$ and highly-overdoped La$_{1.66}$Sr$_{0.34}$CuO$_{4}$ (bottom) with a calculation based on density functional theory \cite{mazi88}.  From the experiment, one can see that La$_2$CuO$_4$ is an insulator with a gap of $\sim2$~eV.  The calculation does fairly well at describing the high-energy spectrum, but it incorrectly predicts metallic conductivity at low frequency.

A system that is insulating due to Coulomb repulsion among electrons within a single band is commonly labelled a Mott insulator.   This categorization is often applied to the cuprate parent compounds; however, they are more accurately described as charge-transfer insulators, following the analysis of Zaanen, Sawatzky, and Allen \cite{zaan85}.   If $d_i^n$ indicates $n$ electrons in the $3d$ states of site $i$, then the Coulomb energy $U$ associated with a Mott excitation can be denoted as $d_i^n d_j^n \leftrightarrow d_i^{n-1} d_j^{n+1}$.  In addition, one needs to consider a charge transfer between a metal atom and a ligand site $L$ (where the ligands are oxygen in our case) with a ligand hole denoted by $\underline{L}$.  The charge-transfer excitation, with energy $\Delta$, can be written $d_i^n \rightarrow d_i^{n+1}\underline{L}$.   When $\Delta < U$, the optical gap corresponds to $\Delta$ rather than $U$, and one has a charge-transfer insulator.  This is the case that applies to cuprates.


We have already mentioned that a cubic crystal field causes a splitting of the $3d$ orbitals into two groups: $e_g$ symmetry ($x^2-y^2$, $3z^2-r^2$) and $t_{2g}$ ($xy$, $xz$, $yz$), with $e_g$ at higher energy.  With a tetragonal elongation of the CuO$_6$ octahedra, the ${3z^2-r^2}$ orbital is lowered in energy relative to the ${x^2-y^2}$, so that the one hole sits in the latter orbital.  It has recently become possible to determine the energy splittings between Cu $3d$ orbitals with resonant inelastic x-ray scattering at the Cu $L_3$ edge \cite{more11}.  Recent {\it ab initio} calculations are in good agreement with the measurements \cite{hozo11}.    For La$_2$CuO$_4$, measurements show that the $d$-$d$ excitation energies from the ${x^2-y^2}$ state are 1.7~eV to ${3z^2-r^2}$, 1.8~eV to $xy$, and 2.1~eV to $xz/yz$.

\begin{figure}
  \includegraphics[width=0.3\textwidth]{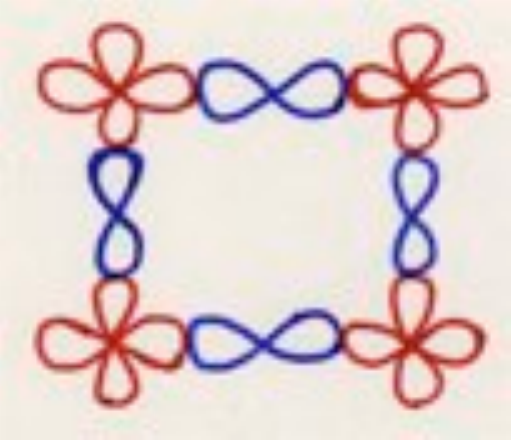}
  \caption{Diagram illustrating the arrangement of the Cu $3d_{x^2-y^2}$ and O $2p_\sigma$ orbitals that are the basis of the states close the Fermi level in CuO$_2$ planes.}
  \label{fg:hybrid}
 \end{figure}

The states near the chemical potential are formed from hybridization between Cu $3d_{x^2-y^2}$ and the O $2p$ orbitals that point towards them (the $p_\sigma$ orbitals), as illustrated in Fig.~\ref{fg:hybrid}.  From a tight-binding perspective, these states contribute to 3 bands, with only one crossing the Fermi level.  Experimentally, one has to dope holes into the CuO$_2$ planes of La$_2$CuO$_4$ in order to get electronic states at the Fermi level.

Ionic valence counting provides a useful approach to the charge distribution in cuprates.  For example, in the case of La$_2$CuO$_4$ one has (La$^{3+}$)$_2$(Cu$^{2+}$)(O$^{2-}$)$_4$; adding up the ionic charges in a formula unit, one finds that the system is neutral.  Counting up the charges in the CuO$_2$ planes, one notes that each CuO$_2$ unit has a charge of $2-$, indicating that the planes pull charge away from the La$_2$O$_2$ layers.  Substituting for La$^{3+}$ by Sr$^{2+}$ or Ba$^{2+}$ reduces the amount of charge in the system, so that electrons are removed from (holes are added to) the CuO$_2$ planes.  For La$_{2-x}$Sr$_x$CuO$_4$, the density of holes per Cu, $p$, is equal to $x$.

\begin{figure}
  \includegraphics[width=0.7\textwidth]{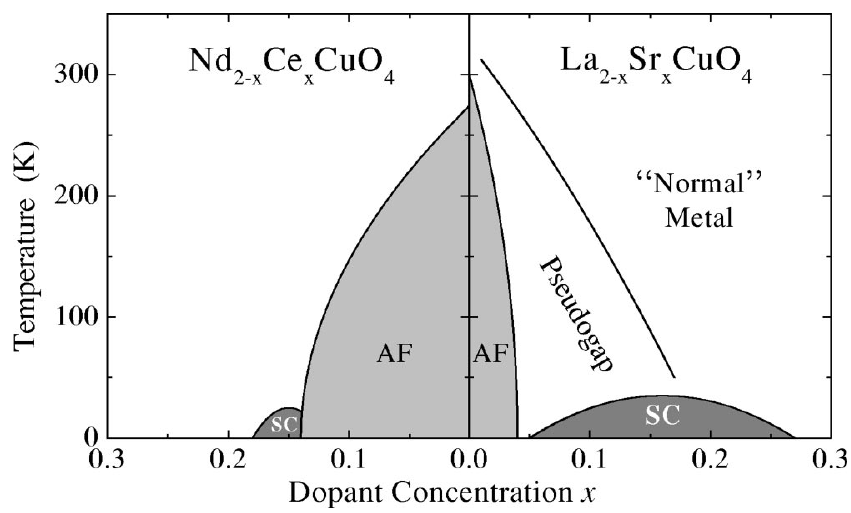}
  \caption{Phase diagram for hole-doping, as in La$_{2-x}$Sr$_x$CuO$_4$, and electron-doping, as in Nd$_{2-x}$Ce$_x$CuO$_4$, from Dascelli {\it et al.} \cite{dama03}.  AF = antiferromagnetic order, SC = superconducting order; the pseudogap phase is a regime with a depressed density of states at the Fermi level compared with the predictions of conventional band theory.}
  \label{fg:ph_diag}
 \end{figure}

It is common to plot the phase diagram of cuprates as a function of temperature and carrier concentration, as shown in Fig.~\ref{fg:ph_diag}.  Introducing holes into the parent compound, one finds experimentally that the antiferromagnetic order is destroyed by a small density, followed by the development of superconductivity, with the maximum superconducting transition temperature, $T_c$, occurring for $x\approx 0.16$.

\begin{figure}
  \includegraphics[width=0.5\textwidth]{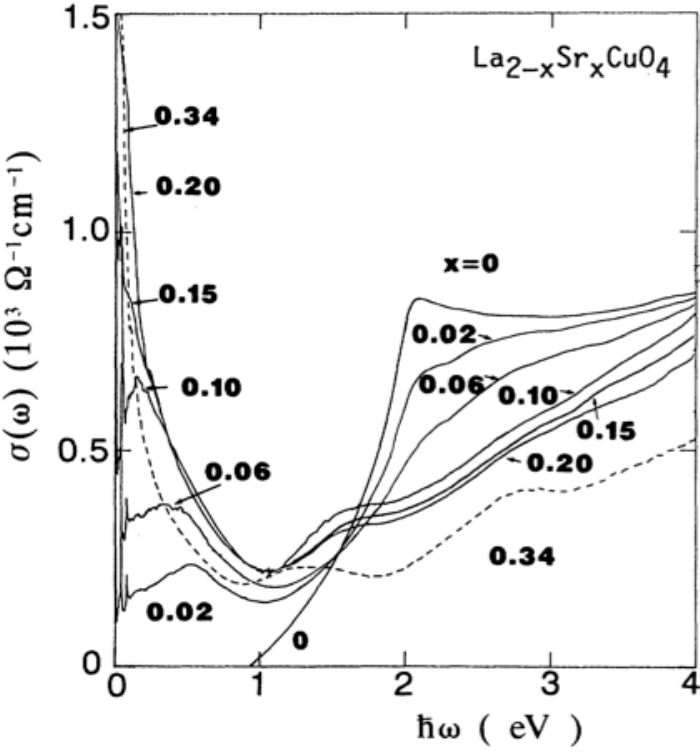}
  \caption{In-plane optical conductivity vs.\ energy for La$_{2-x}$Sr$_x$CuO$_4$ for a range of dopings, from Uchida {\it et al.} \cite{uchi91}. }
  \label{fg:lsco_optical}
 \end{figure}

Let us return to the in-plane optical conductivity, which has been measured in La$_{2-x}$Sr$_x$CuO$_4$ (LSCO) as a function of doping by Uchida {\it et al.} \cite{uchi91}, as shown in Fig.~\ref{fg:lsco_optical}.  Initial doping appears to introduce states within the charge transfer gap.  Eventually, a Drude peak develops; that is, a peak at zero frequency corresponding to metallic conductivity.  

\begin{figure}
  \includegraphics[width=0.7\textwidth]{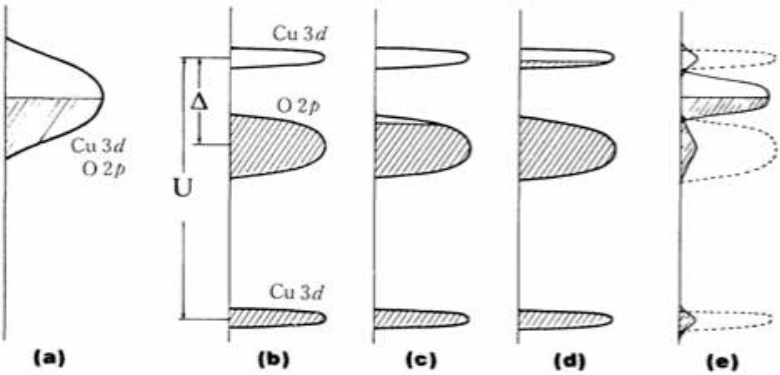}
  \caption{Effective densities of states underlying the optical excitations for several different models, from Uchida {\it et al.} \cite{uchi91}. }
  \label{fg:charge_transfer}
 \end{figure}

Useful schematic diagrams for interpreting the observations are presented in Fig.~\ref{fg:charge_transfer}.  The conventional band-theory prediction is represented in (a), from which one would expect to La$_2$CuO$_4$ to be metallic with a carrier density of one per Cu site.  A more realistic picture is shown in (b), indicating the splitting by energy $U$ between the upper and lower Hubbard bands, with the O $2p$ bands in-between, resulting in a charge transfer gap of $\Delta\approx 2$~eV.   With doping, a rigid band picture would predict that the Fermi level would move down into the O $2p$ band, for hole doping as in (c), or up into the upper Hubbard band, for electron doping as in (d).  Instead, experiment suggests that weight is transferred into new mid-gap states, as indicated in (e).

\begin{figure}
  \includegraphics[width=0.5\textwidth]{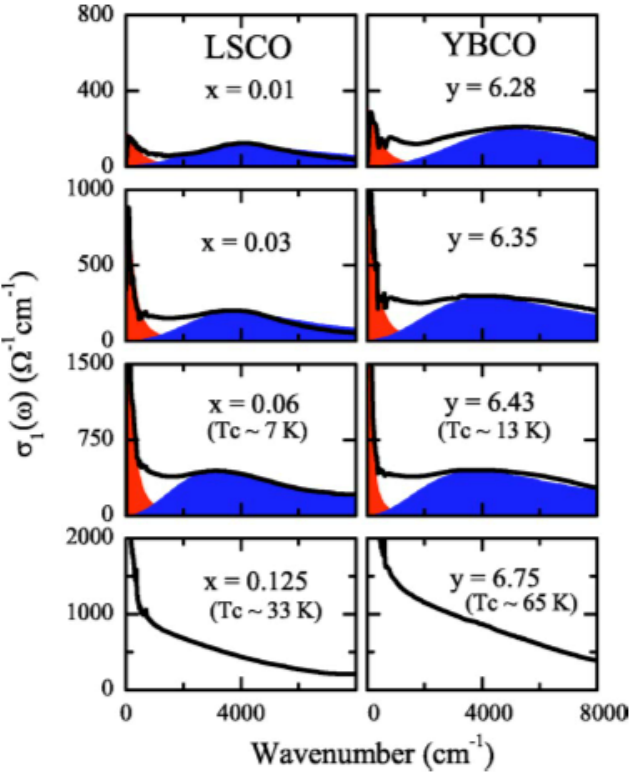}
  \caption{In-plane optical conductivity measured at $T\approx T_c$ in LSCO (left) and YBa$_2$Cu$_3$O$_{6+x}$ (YBCO).  Phonon peaks have been removed, so that only the electronic conductivity is presented.  (Note that in YBCO, $x$ is different from the hole concentration $p$.) For the samples at lower doping, the shading indicates a distinction between a Drude peak, corresponding to mobile carriers, and a mid-infrared peak \cite{lee05}.}
  \label{fg:lsco_ybco}
 \end{figure}

\begin{figure}
  \includegraphics[width=0.7\textwidth]{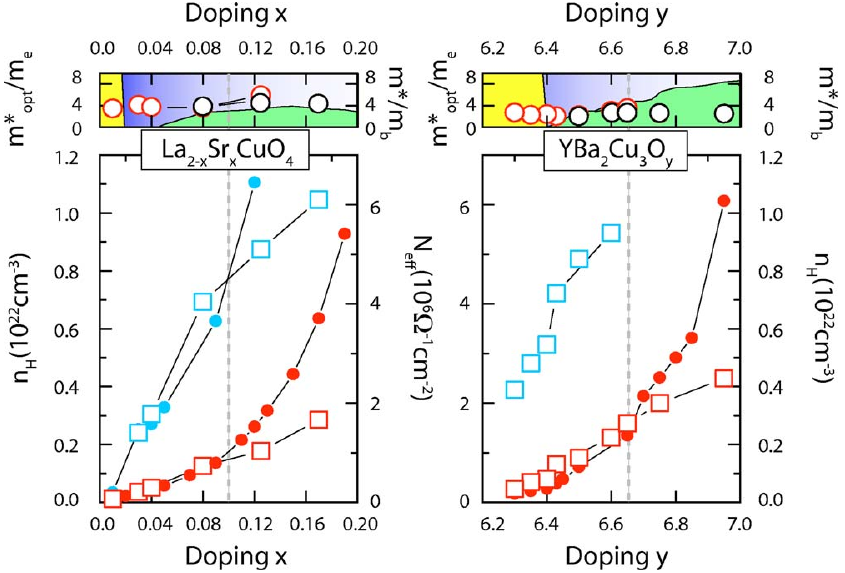}
  \caption{Lower panels: plots of effective carrier density vs. doping in LSCO (left) and YBCO (right).  The open squares are obtained from integrating the optical conductivity to $650\textrm{ cm}^{-1}\approx 80$~meV (lower points) and to $10,000\textrm{ cm}^{-1}\approx1.25$~eV (upper points).   The filled circles show the density of carriers determined from the Hall coefficient $R_{\rm H}$ measured at 250 K (lower points) and at 720 K (upper points). Upper panels: estimates of the effective mass. From Padilla {\it et al.} \cite{padi05}.}
  \label{fg:eff_carrier}
 \end{figure}

Measurements of the in-plane optical conductivity at low temperature allow one to distinguish a Drude peak from ``mid-IR'' states, as illustrated for underdoped LSCO and YBCO in Fig.~\ref{fg:lsco_ybco} \cite{lee05}.  The mid-IR weight is large compared to the Drude weight.  To quantify this effect, the effective carrier density within a spectral band from zero up to frequency $\omega$ is given by 
\begin{equation}
  N_{\rm eff}(\omega) = \int_0^\omega d\omega'\, \sigma_1(\omega')
\end{equation}
Evaluating $N_{\rm eff}$ in the Drude peak, one obtains the lower set of open squares in the lower panels of Fig.~\ref{fg:eff_carrier}.  These results show a trend very similar to the carrier density $n_{\rm H}$ determined from the Hall coefficient, $R_{\rm H}$, measured below room temperature (lower filled circles).  In contrast, integrating the optical conductivity through the mid-IR range yields the upper sets of open squares, which match the $n_{\rm H}$ obtained from $R_{\rm H}$ at high temperature (upper filled circles).

The lesson from optical conductivity is this: for modest doping and temperature, only the dopant-induced holes are mobile ({\it i.e.}, $p\approx x$ in LSCO).  Note that this is very different from what one would expect from conventional band theory, which would predict a density of $1+x$ holes.

\subsubsection{X-ray absorption spectroscopy}

\begin{figure}
  \includegraphics[width=0.5\textwidth]{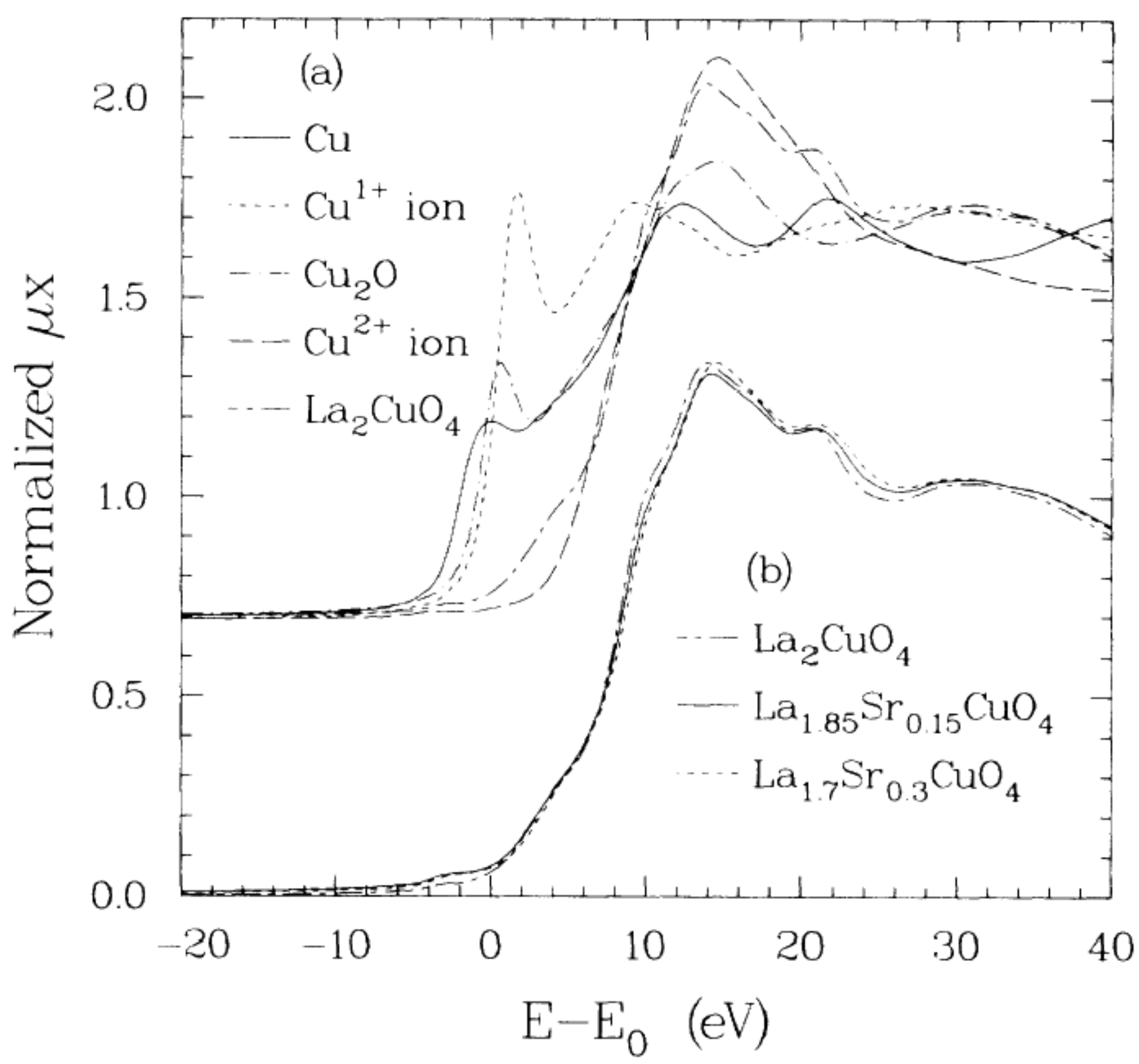}
  \caption{(a) Cu $K$ edge spectra for reference materials with various nominal valences, demonstrating the sensitivity to valence change.  (b)  Similar spectra for LSCO, from undoped to overdoped, indicating a lack of any significant change.  From \cite{tran87a}.}
  \label{fg:cuk}
 \end{figure}

\begin{figure}
  \includegraphics[width=0.8\textwidth]{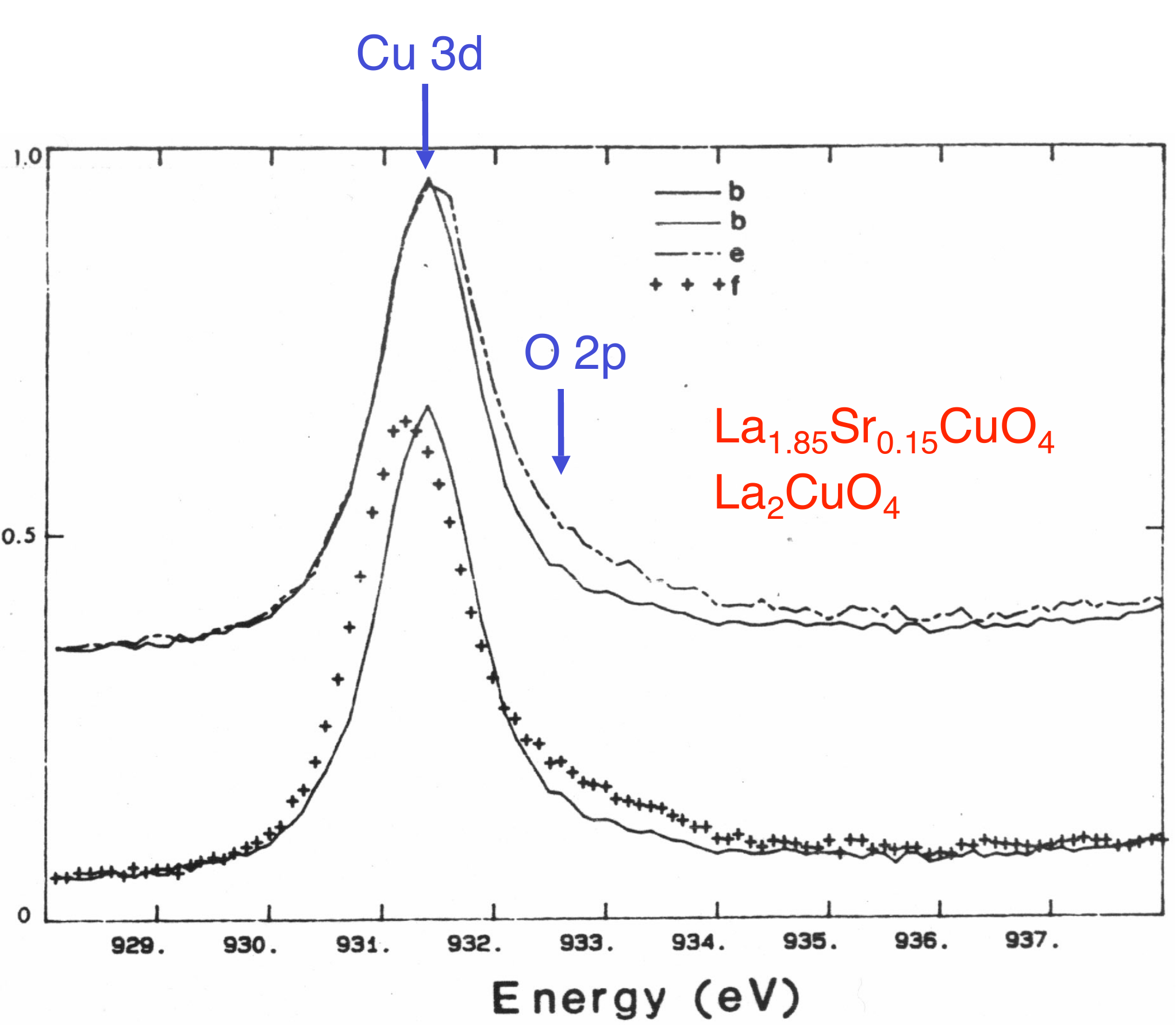}
  \caption{Cu $L_3$ x-ray absorption spectra, reprinted from Bianconi {\it et al.} \cite{bian88} with permission from Elsevier. For the upper two curves, the solid line corresponds to La$_2$CuO$4$ and the dashed line to La$_{1.85}$Sr$_{0.15}$CuO$_4$.  Note that there is no change in the leading edge or main peak, which are sensitive to empty $3d$ states.  Doping enhances the absorption on the high-energy side, corresponding to O $2p$ holes.  The empty $3d$ states are pulled down in energy relative to these by the potential of the x-ray-induced core hole.}
  \label{fg:cuL}
 \end{figure}

To determine the orbital character of the holes, one can use x-ray absorption spectroscopy.  For example, measurements at the Cu $K$ edge involve transitions from the $1s$ core level to states with $p$ symmetry with respect to the absorbing Cu site.  Measurements on LSCO showed an absence of significant change with doping \cite{tran87a}, as indicated in Fig.~\ref{fg:cuk}.  Measurements at the Cu $L_3$ edge detect transitions from a $2p$ core level to $d$ symmetry final states.  Bianconi {\it et al.} \cite{bian88} found that doping does not change the density of unfilled $3d$ states, which contribute to the strong low-energy peak, but that it does introduce holes at slightly higher energy, corresponding to hybridized O $2p$ states, as apparent in Fig.~\ref{fg:cuL}. 

\begin{figure}
  \includegraphics[width=0.5\textwidth]{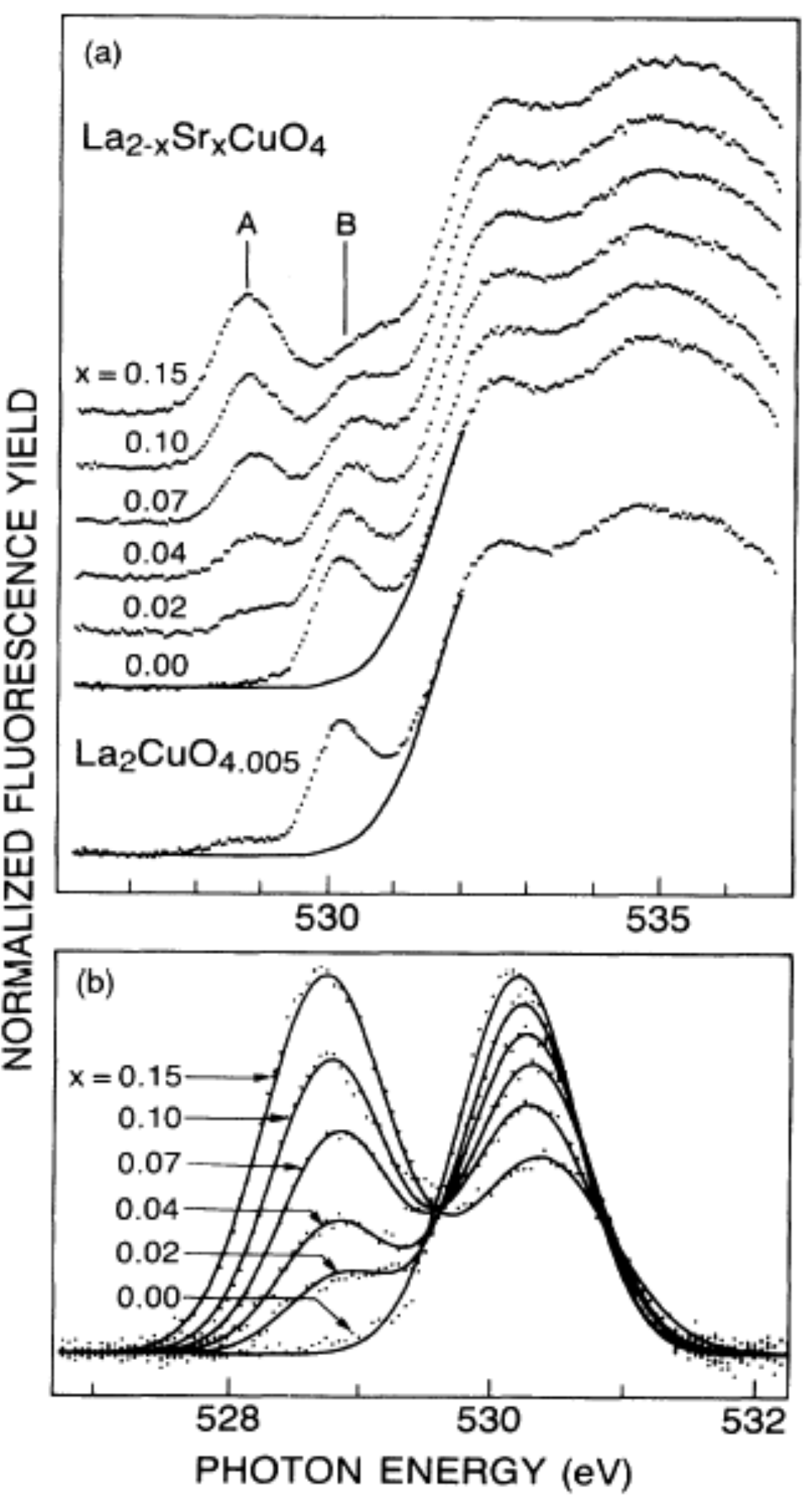}
  \caption{(a) Oxygen $K$ edge x-ray absorption spectra measured by fluorescence yield on ceramic samples of LSCO for various dopings.  (b) Expanded view of the pre-edge peaks, after subtraction of the solid curve indicated in (a).  From Chen {\it et al.} \cite{chen91}.}
  \label{fg:ok}
 \end{figure}

The significance of the O $2p$ character was firmly established by measurements of O $K$-edge spectra in LSCO by Chen {\it et al.} \cite{chen91}, as shown in Fig.~\ref{fg:ok}.  Here the excitation is from the O $1s$ level to states with $p$-symmetry with respective to the absorbing atom.  Two peaks are observed in the pre-edge region, with A labeling the O $2p$ peak and B the upper Hubbard band.  In the undoped state, peak A is absent while B is strong.  With doping peak A grows while B decreases. The growth of A indicates the O $2p$ character of the doped holes.  Later measurements on single crystals, taking advantage of the polarization sensitivity of the absorption process, have demonstrated that the holes are dominantly within the CuO$_2$ planes, with little weight on apical oxygens \cite{pell93}.

The lesson from x-ray absorption spectroscopy is that the dopant-induced holes have strong O $2p$ character.  Combining this with the optical spectroscopy and Hall-effect results, we can conclude that the Cu $3d$ holes of the parent insulator remain localized at low temperature, even after substantial doping.

\subsubsection{Angle-resolved photoemission spectroscopy (ARPES)}

The dispersion of electronic states near the Fermi level can be measured by angle-resolved photoemission.  In this technique, one typically shines ultraviolet photons on a flat crystal surface and measures emitted electrons as a function of angle, analyzing the electron current as a function of kinetic energy.   Light sources include the He lamp (photon energy 20 eV), synchrotrons (tunable from UV to soft x-ray to hard x-ray), and lasers (6-7 eV).\footnote{For a review, see Damascelli, Shen, and Hussain \cite{dama03}.}

\begin{figure}
  {\includegraphics[width=0.3\textwidth]{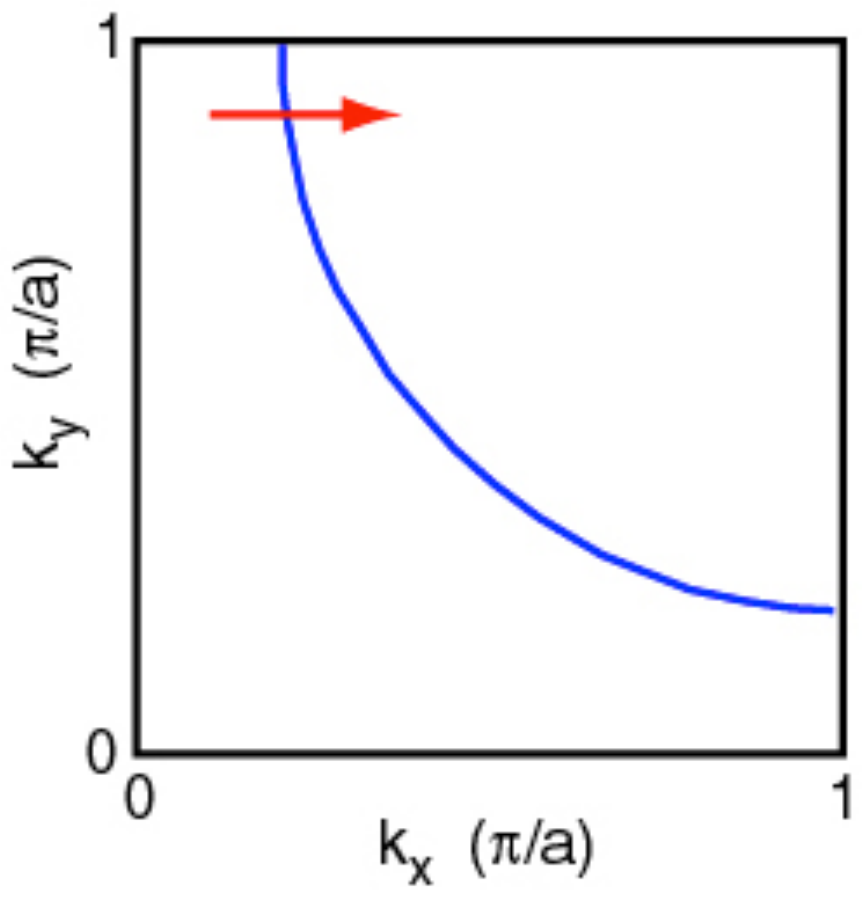}\hskip10pt
  \includegraphics[width=0.3\textwidth]{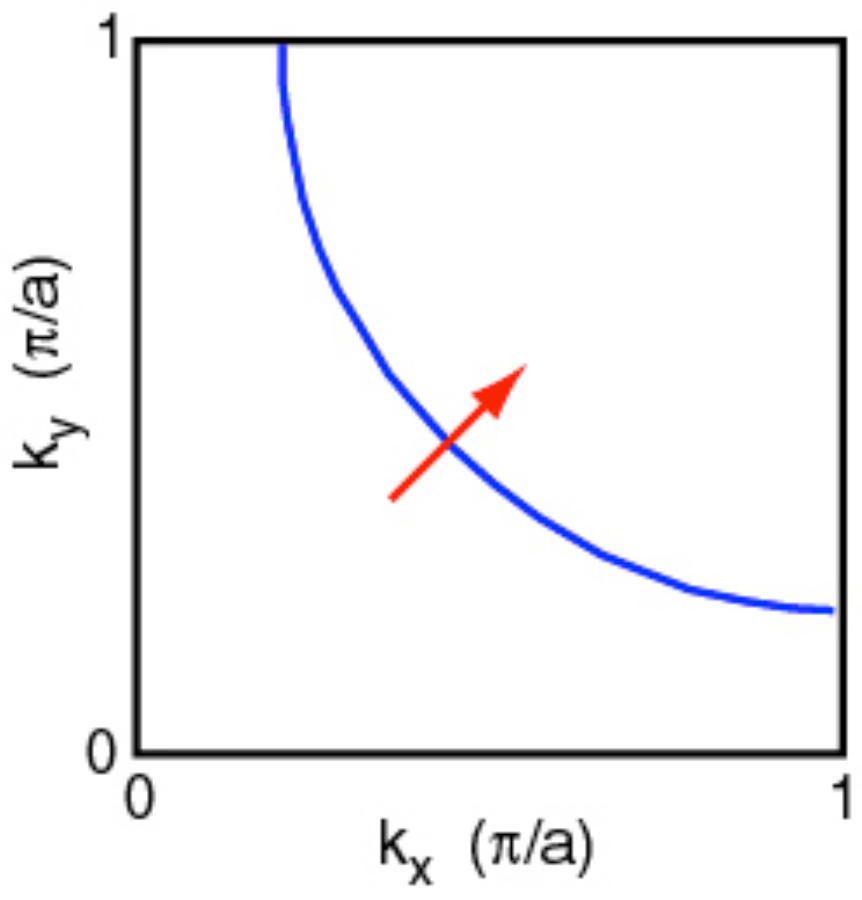}}
  \caption{Schematic diagrams of the nominal Fermi surface for LSCO in one quadrant of reciprocal space.  Arrows indicate the nodal direction (right) and antinodal direction (left).}
  \label{fg:fs_sk}
 \end{figure}

\begin{figure}
 { \includegraphics[width=0.5\textwidth]{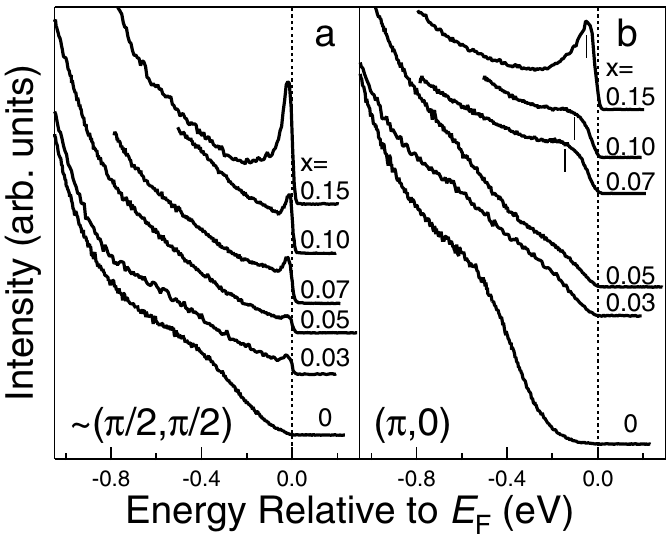}\hskip20pt
  \includegraphics[width=0.35\textwidth]{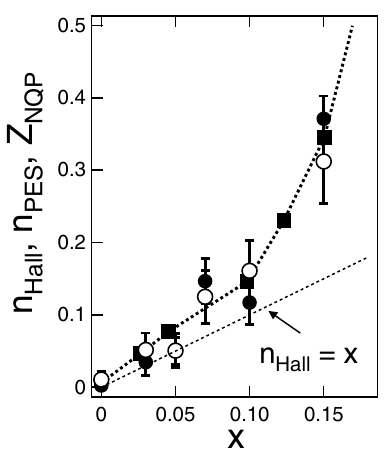}}
  \caption{ARPES results from Yoshida {\it et al.} \cite{yosh03}.  Left panel: Spectral function in LSCO for $x=0$ (bottom) to 0.15 (top) at the nodal point (a) and antinodal region (b).  Right panel: doping dependence of the nodal quasiparticle spectral weight $Z_{\rm NQP}$ (open circles), spectral weight integrated at $E_{\rm F}$ over the (second) Brillouin zone $n_{\rm PES}$ (filled circles), and hole concentration from the Hall coefficient $n_{\rm Hall}$ (filled squares).}
  \label{fg:n_an}
 \end{figure}

Figure~\ref{fg:fs_sk} shows a schematic of the nominal Fermi surface for LSCO as predicted by band structure calculations.  It is useful to distinguish between different regions of the Fermi surface.  The language used is based on the $d$-symmetry pair wave function observed in cuprates.  The gap goes to zero ({\it i.e.}, has a node) along the diagonal direction, indicated by the arrow in the right-hand panel of Fig.~\ref{fg:fs_sk}; the gap maximum occurs in the ``anti-nodal'' region, as indicated on the left.  Even in the normal state, there are differences in the spectral function measured in these two regions.  The left-hand panel in Fig.~\ref{fg:n_an} shows the spectral function along the nodal (a) and antinodal (b) directions for LSCO with $x$ from zero to 0.15 \cite{yosh03}.  Along the nodal direction, a sharp peak is seen near the Fermi level even for $x=0.03$, whereas in the antinodal direction a sharp peak at the Fermi level does not appear until one approaches optimal doping ($x=0.15$).  The antinodal spectra tend to show a diffuse, gapped behavior, providing one definition of the pseudogap.  In the right-hand panel, the open circles indicate the doping dependence of the nodal spectral weight, which, in the underdoped regime, is comparable to the doping dependence of the carrier density obtained from the Hall effect.

\begin{figure}
  {\includegraphics[width=0.4\textwidth]{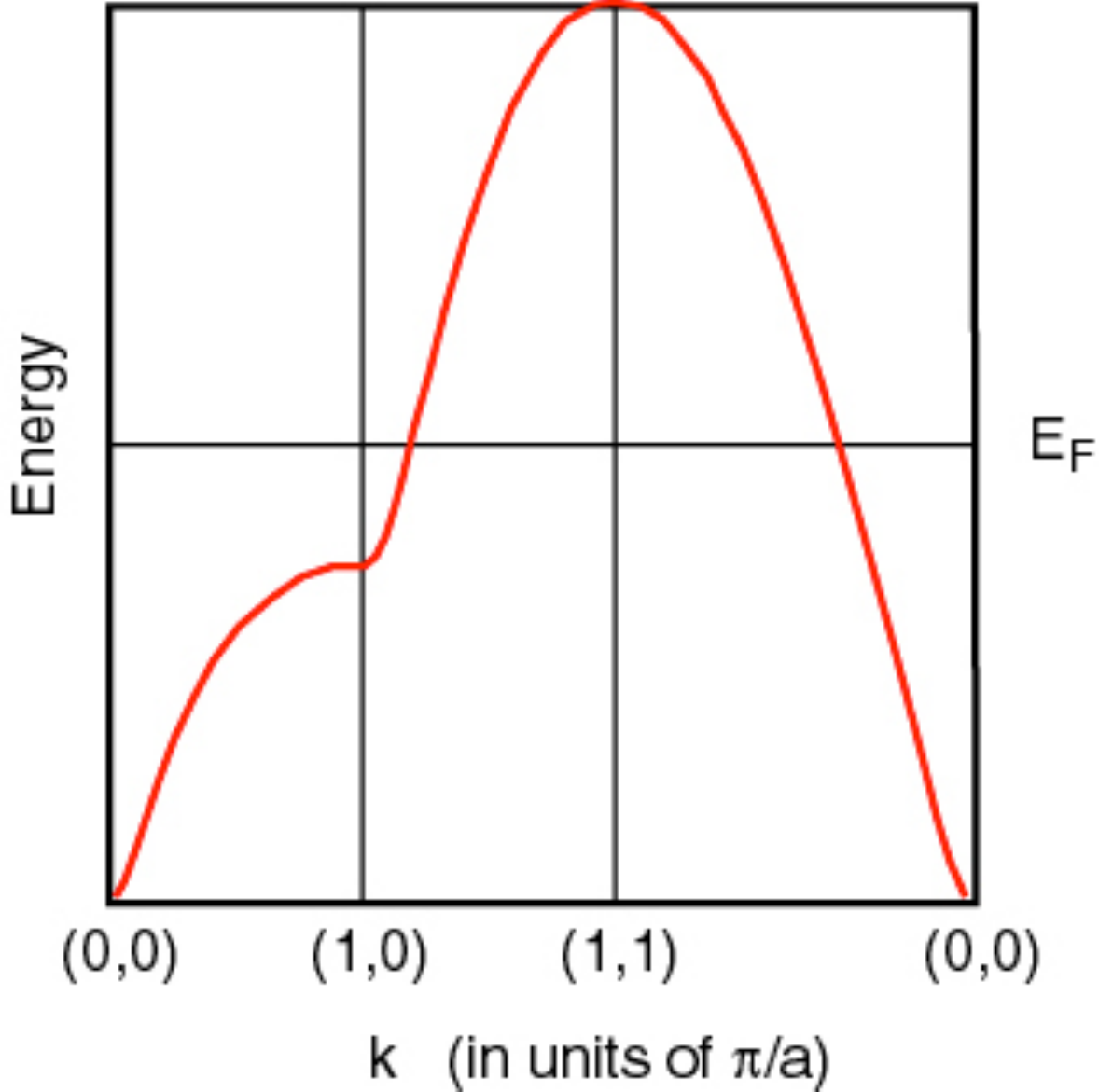} \hskip20pt
  \includegraphics[width=0.5\textwidth]{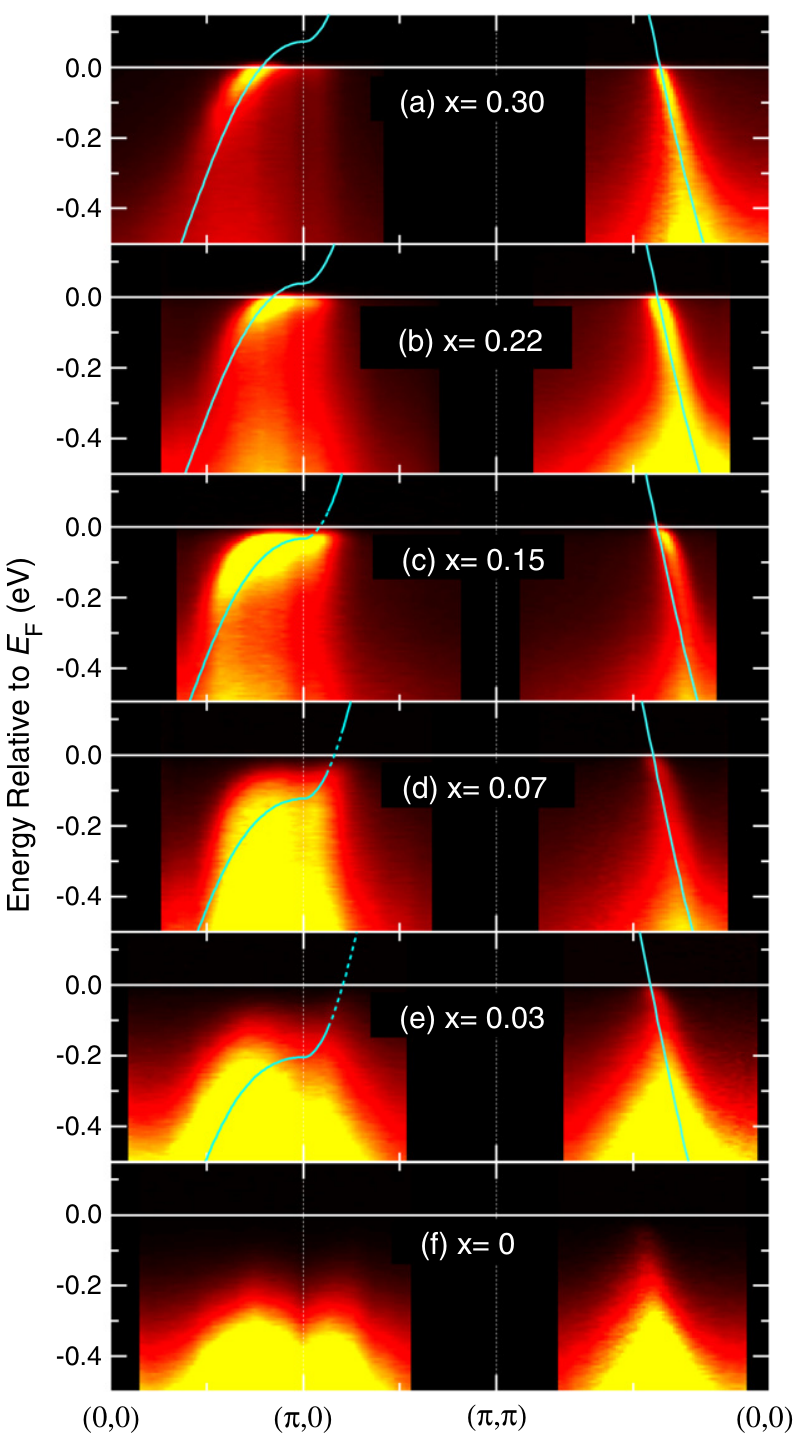}}
  \caption{(Left) Nominal dispersion of the electronic band that crosses the Fermi level for a typical single-layer cuprate, plotted along high-symmetry directions of reciprocal space. The high-symmetry points are labelled in units of $\pi/a$, where $a\approx 3.8$~\AA\ is the lattice parameter. (Right) Spectral weight as a function of $E-E_{\rm F}$ and {\bf k}, along high-symmetry directions measured by ARPES on LSCO for various dopings.  The high-symmetry wave vectors are labelled in units that assume $a=1$.  The thin lines through the data correspond to tight-binding calculations.  From Yoshida {\it et al.} \cite{yosh07}, \copyright\ IOP Publishing.  Reproduced with permission.  All rights reserved.}
  \label{fg:qp_disp}
 \end{figure}

One can also measure the variation of electronic energy as a function of wave vector, {\bf k}.  The left panel of Figure~\ref{fg:qp_disp} shows the nominal dispersion that one might obtain from a tight-binding calculation for a single-layer cuprate.  ARPES measurements of the filled states are shown on the right-hand side of Fig.~\ref{fg:qp_disp} for LSCO with several dopant concentrations \cite{yosh07}.  In the undoped system, the spectral weight is diffuse and far below the Fermi energy, $E_{\rm F}$.  With doping, a sharp, roughly-linear dispersion develops along the nodal direction [(0,0) to $(\pi,\pi)$].  In the antinodal region, near $(\pi,0)$, a region of fairly flat dispersion just below $E_{\rm F}$ becomes apparent at $x=0.15$, but this feature shifts through $E_{\rm F}$ for $x=0.22$ and above.

\begin{figure}
  \includegraphics[width=0.6\textwidth]{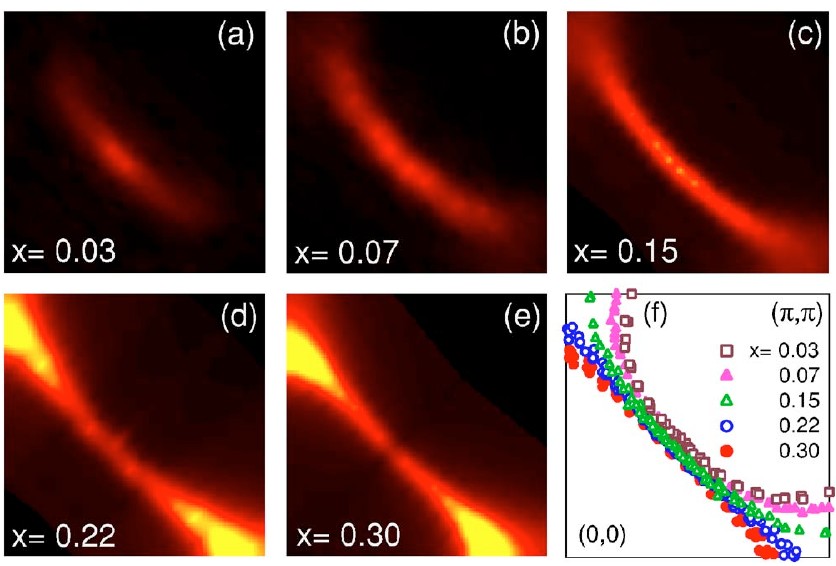}
  \caption{(a)-(e) Spectral weight along the Fermi surface (in the first quadrant) measured by ARPES on LSCO crystals for various dopings, from Yoshida {\it et al.}  \cite{yosh06}. (f) Location of the Fermi surface for various dopings determined from the measurements.}
  \label{fg:fsvsx}
 \end{figure}

Figure~\ref{fg:fsvsx} shows ARPES measurements of spectral weight along a quadrant of the Fermi surface for a range of dopings in LSCO \cite{yosh06}.  At low doping, the weight is only apparent an a small arc near the nodal point.  With doping, the nodal arc expands, until it covers most of the nominal Fermi surface near optimal doping.  When the flat, antinodal dispersion moves through $E_{\rm F}$ near $x=0.22$ as seen on the right side of Fig.~\ref{fg:qp_disp}, the shape of the Fermi surface changes.  For lower doping, the nominal Fermi surface is centered around $(\pi,\pi)$, while for $x=0.22$ and above the Fermi surface is closed around (0,0).  This corresponds to a change from a hole-like to an electron-like Fermi surface.

While we are discussing ARPES, we should also consider the fact that ARPES studies have measured the $d$-wave gap that develops about $E_{\rm F}$ in the superconducting state.  Note that ARPES directly measures the angle-dependent magnitude of the gap, but does not detect the phase.  For underdoped samples, the gap closes along the nodal arc at $T_c$, but an antinodal pseudogap remains up to a characteristic temperature $T^*$ (which decreases with doping)  \cite{kani06,lee07,yang08}.  

A significant lesson from the ARPES studies is that, for underdoped cuprates, the mobile carrier density is associated with states along the nodal arc.

\subsubsection{Transport and magnetic susceptibility measurements}

\begin{figure}
  {\includegraphics[width=0.45\textwidth]{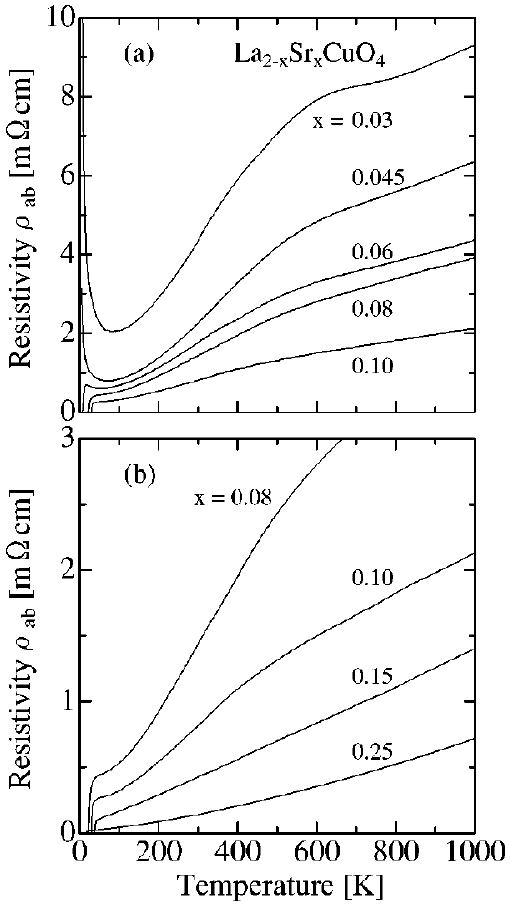}\hskip20pt
  \includegraphics[width=0.45\textwidth]{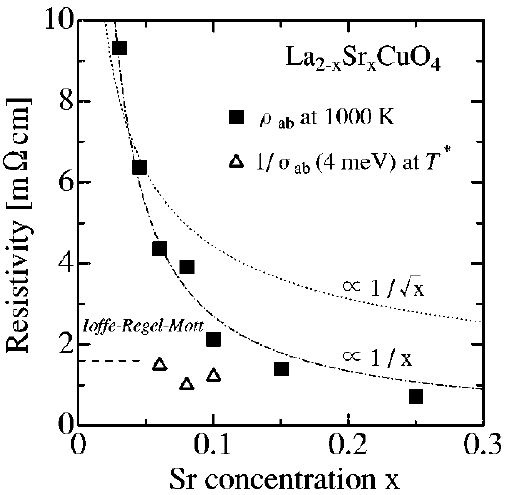}}
  \caption{(Left) In-plane resistivity measured on single crystals of LSCO for various dopings by Takenaka {\it et al.} \cite{take03}.  (Right) In-plane resistivity at 1000~K for LSCO vs.\ doping.  The dashed line indicates the Ioffe-Regel limit. }
  \label{fg:rho_take}
 \end{figure}

The in-plane resistivity, $\rho_{\rm ab}$, of underdoped cuprates is unusual for its large magnitude.   In a Fermi liquid, the resistivity increases with temperature due to a growing scattering rate, typically due to scattering from phonons.  The Fermi wave length is always greater than the mean free path, so that the resistivity remains below the Ioffe-Regel limit.  When the electron-phonon scattering becomes large, the resistivity tends to saturate.  The situation is different in cuprates.  Examples of $\rho_{\rm ab}$ for LSCO at a range of dopings from Takenaka {\it et al.} \cite{take03} are shown in the left-hand panels of Fig.~\ref{fg:rho_take}.  The resistivity becomes quite large at high temperature and does not saturate.   The magnitude of $\rho_{\rm ab}$ at 1000 K exceeds the Ioffe-Regel limit for a considerable range of doping, as shown in the right-hand panel of Fig.~\ref{fg:rho_take}.  For this reason, underdoped cuprates have been called bad metals \cite{emer95b}.

Optical conductivity measurements show that there is no Drude peak in underdoped cuprates at high temperature \cite{take03,lee05}.  One could argue that there is no violation of the Ioffe-Regel limit because there are no quasiparticles at high temperature.  The important point, however, is that the underdoped cuprates are incoherent metals at high temperature.  Coherence develops only at lower temperatures, coinciding with the so-called pseudogap state.

\begin{figure}
  {\includegraphics[width=0.5\textwidth]{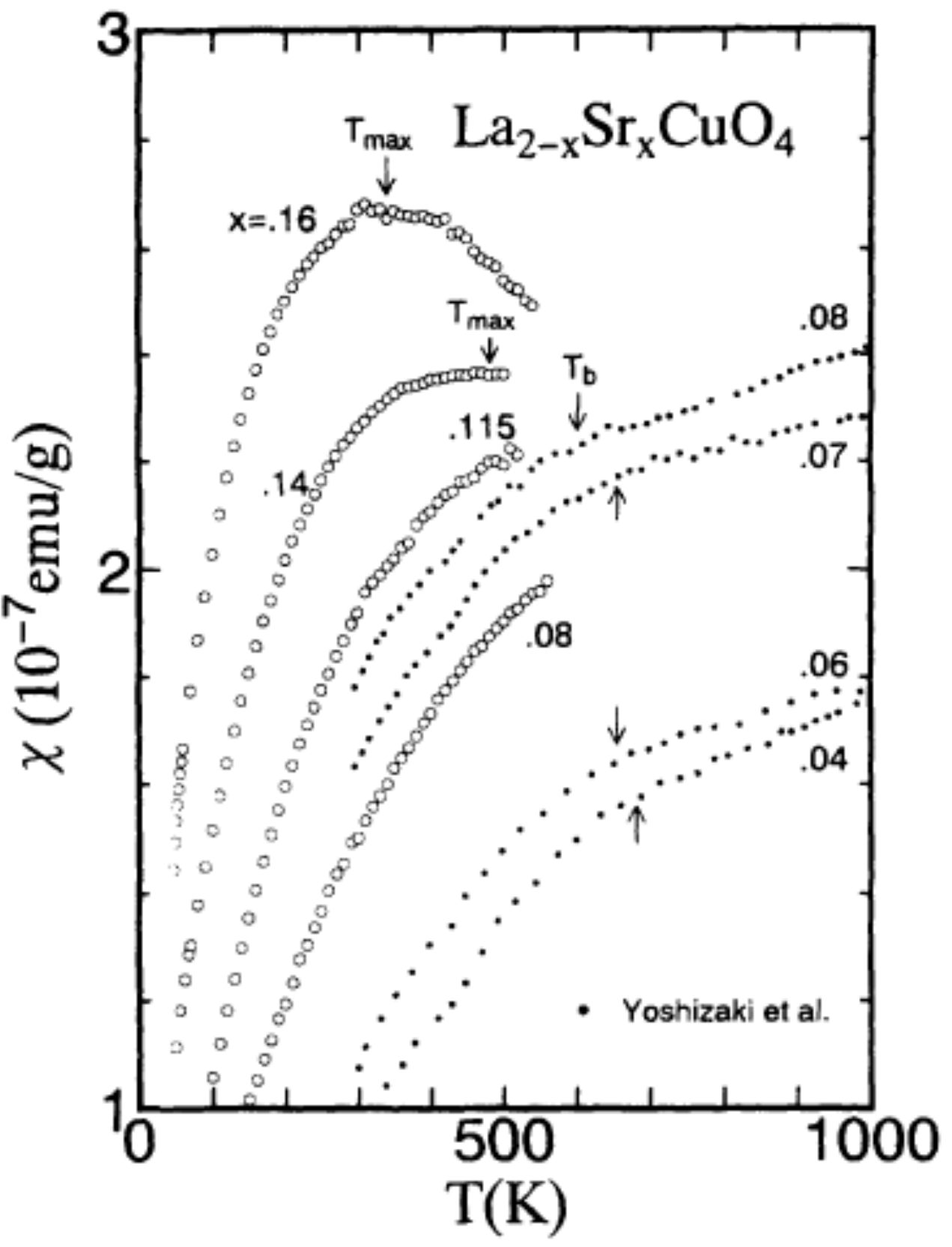}\hskip20pt
  \includegraphics[width=0.4\textwidth]{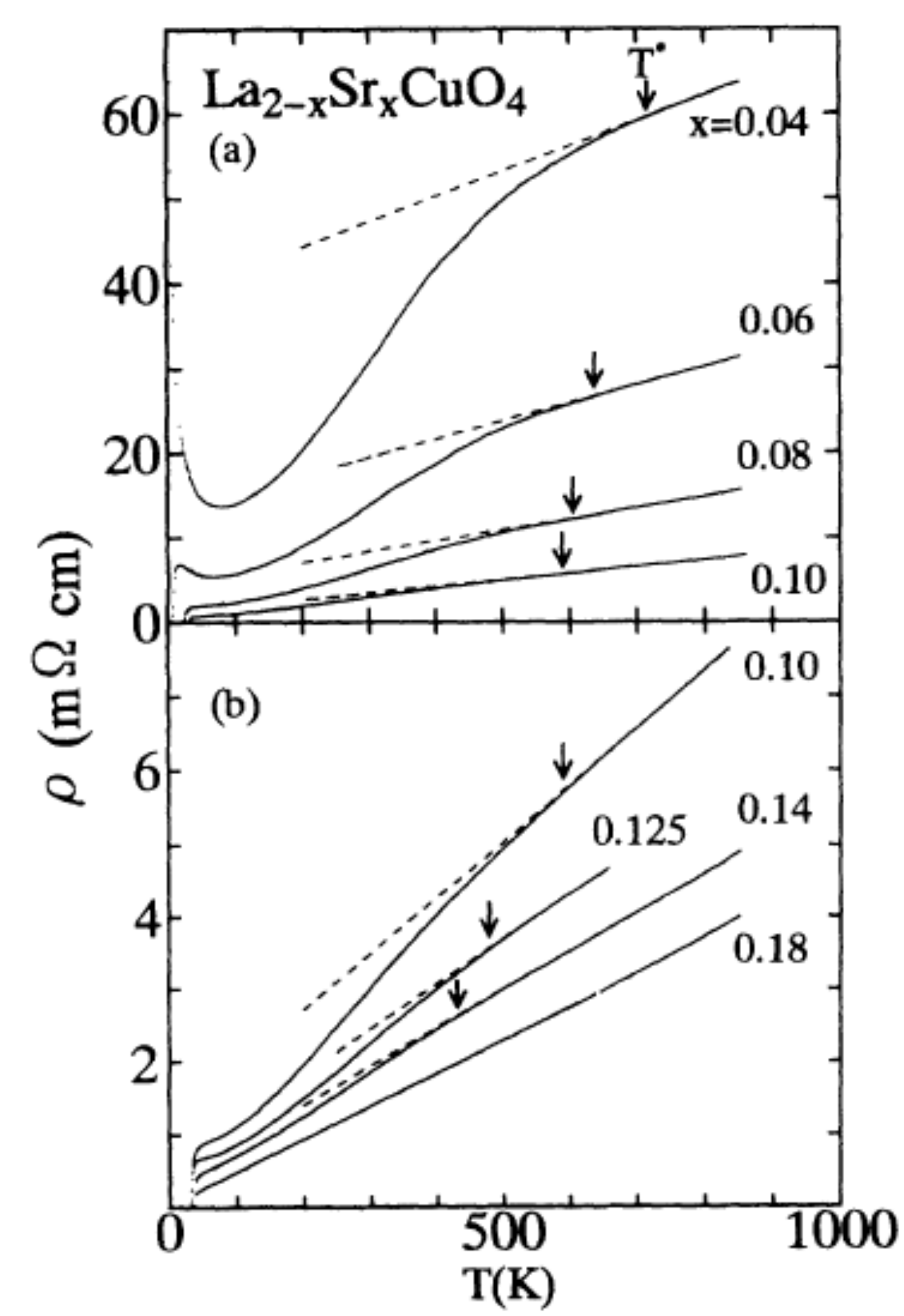}}
  \caption{(Left) Bulk susceptibility in LSCO for a range of doping levels; arrows point to the features identified with the onset of the pseudogap phase at $T^*$.  (Right) Resistivity measurements for LSCO, with $T^*$ indicated by arrows.  From Nakano {\it et al.} \cite{naka94}. }
  \label{fg:chi_max}
 \end{figure}

Various properties have been associated with the onset of the pseudogap state with cooling.  (The onset temperature is typically labelled $T^*$.)  For example, there is the maximum in the bulk susceptibility and the deviation of $\rho_{\rm ab}$ from linear temperature dependence at high temperature.  Examples of such measurements from Nakano {\it et al.} \cite{naka94} are shown in Fig.~\ref{fg:chi_max}.

\begin{figure}
  \includegraphics[width=0.6\textwidth]{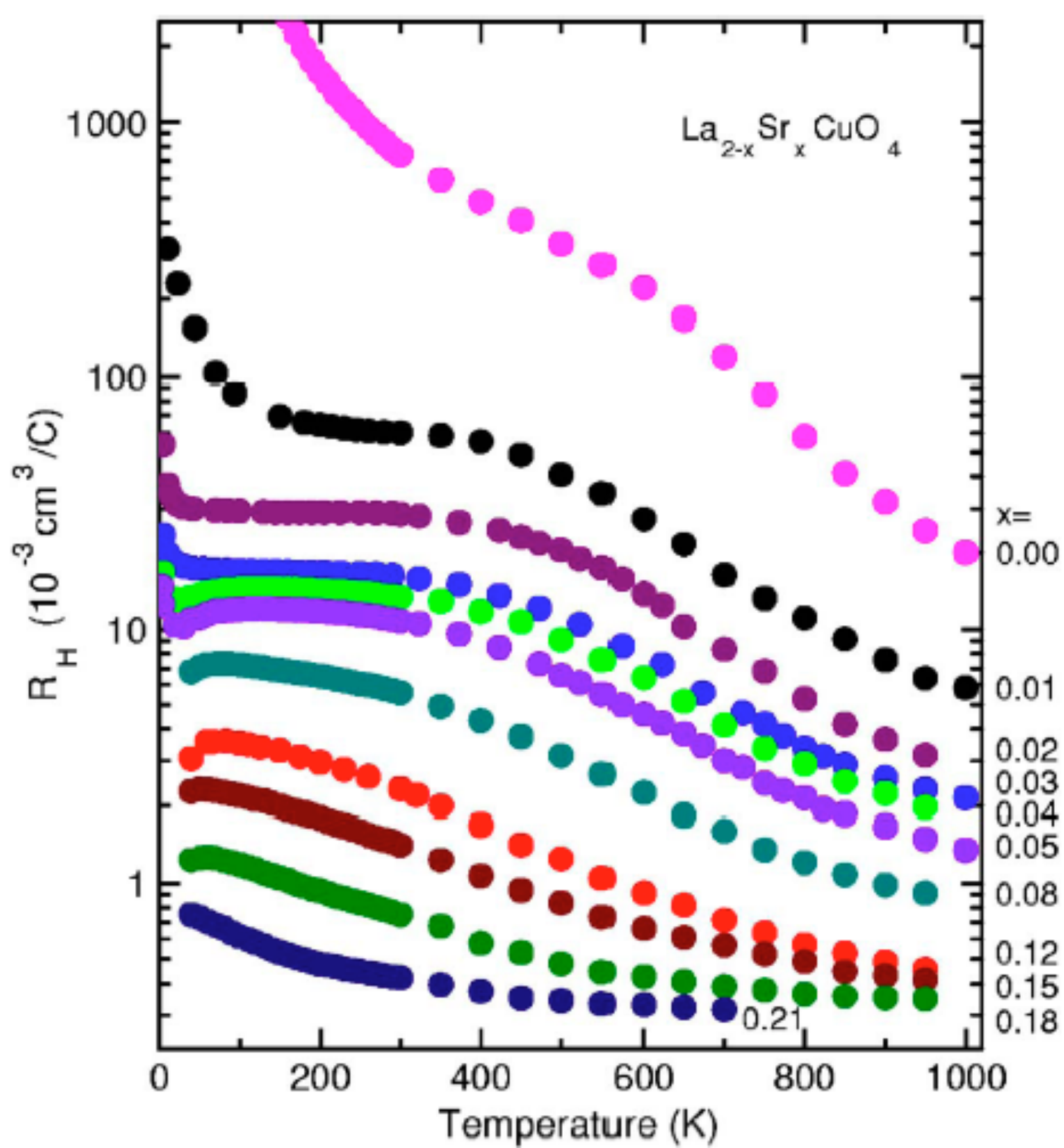}
  \caption{Temperature dependence of $R_{\rm H}$ measured in LSCO for a range of doping by Ando {\it et al.} \cite{ando04}.}
  \label{fg:lsco_hall}
 \end{figure}

\begin{figure}
  {\includegraphics[width=0.41\textwidth]{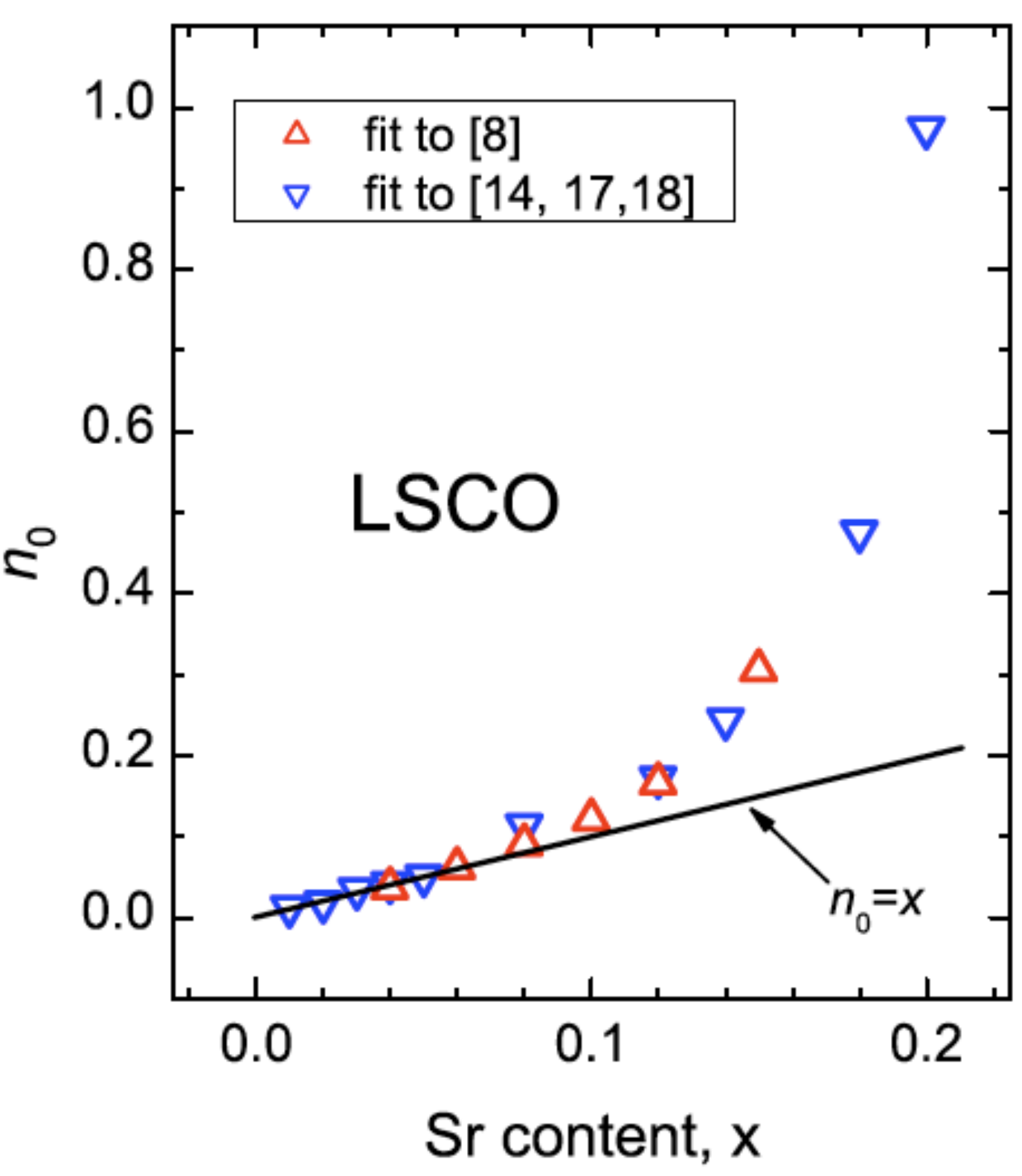}\hskip20pt
  \includegraphics[width=0.49\textwidth]{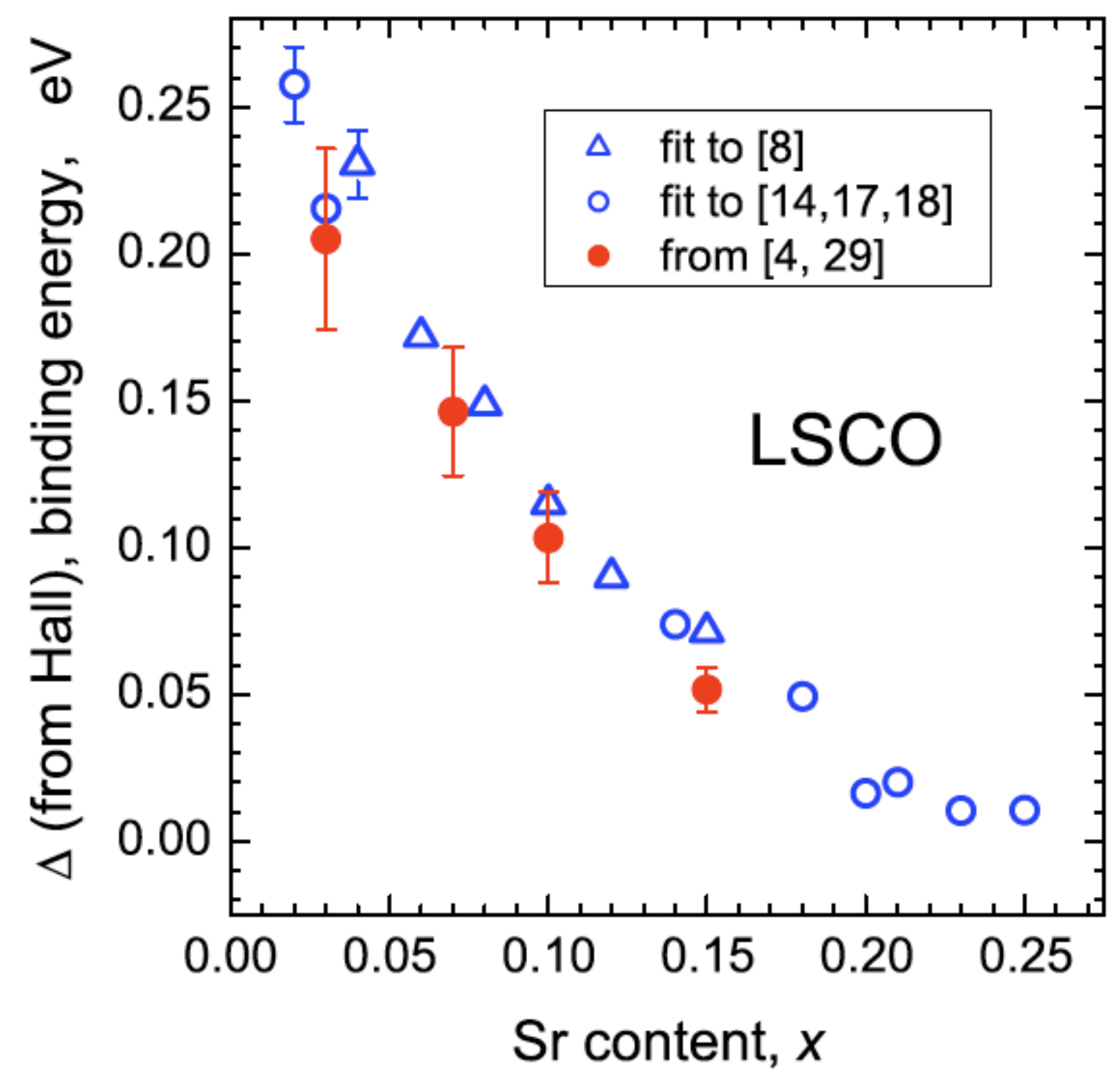}}
  \caption{Parameters obtained by Gorkov {\it et al.} \cite{gork06} from fits of Eq.~(\ref{eq:hall}) to the temperature dependence of $R_{\rm H}$.  (Left) Temperature-independent carrier density $n_0(x)$.  (Right) Triangles: gap $\Delta$ from fits to the temperature dependence of $R_{\rm H}$; Filled circles: antinodal gap from ARPES measurements.}
  \label{fg:gorkov1}
 \end{figure}

There is also interesting $T$ dependence in the Hall coefficient \cite{ando04}, as shown in Fig.~\ref{fg:lsco_hall}.  One can estimate the density of holes, $n_{\rm H}$, from measurements of the Hall coefficient with the formula $R_{\rm H}=1/n_{\rm H}ec$.  The data in Fig.~\ref{fg:lsco_hall} imply that $n_{\rm H}$ increases with temperature.  Gor'kov and Teitel'baum \cite{gork06} have fit temperature-dependent Hall effect data with the formula
\begin{equation}
  n_{\rm H}(x,T) = n_0(x) + n_1(x)e^{-\Delta(x)/T}.
  \label{eq:hall}
\end{equation}
As shown on the left-hand side of Fig.~\ref{fg:gorkov1},  the temperature-independent component, $n_0$, is proportional to $x$ for low doping.  The gap $\Delta$ for the thermally-excited carriers is shown on the right-hand side of Fig.~\ref{fg:gorkov1}; it is quantitatively quite similar to the antinodal pseudogap from ARPES measurements.  

So far we have established that at high temperature, underdoped cuprates have an incoherent metal phase in which there is an enhanced effective carrier density but no quasiparticles.   On cooling below a characteristic temperature $T^*$, the effective carrier density drops to a value comparable to the doped hole density, the resistivity drops substantially, and the susceptibility starts to decrease.  The optical conductivity studies previously discussed show that a Drude peak develops and narrows in frequency with continued cooling.

Let us consider the significance of the thermal evolution of the magnetic susceptibility.  In a Fermi liquid, each electron makes a temperature-independent contribution known as the Pauli susceptibility.  A decrease in the susceptibility would signify a reduction in the density of free carriers.  Indeed, this interpretation was applied to the initial discovery of the pseudogap, which was identified in terms of the decrease in spin susceptibility \cite{allo89}.  There is, however, an alternative interpretation.  In the parent compound La$_2$CuO$_4$, there are no free carriers, but there are local moments associated with the Cu$^{2+}$ ions.   When the superexchange energy $J$ becomes comparable to $kT$, neighboring spins will develop antiferromagnetic correlations, which grow with cooling.   Correspondingly, the spin susceptibility will decrease.   In the hole-doped system, some Cu moments may still survive.  (How these moments can survive within the metallic phase is one of the key issues that will be addressed in the following sections.)  The correlations among the spins now depend not only on the temperature, but also on the rate at which carriers move through the system, flipping spins.  Thus, the carriers will impact $\chi(T)$.

\begin{figure}
  {\includegraphics[width=0.45\textwidth]{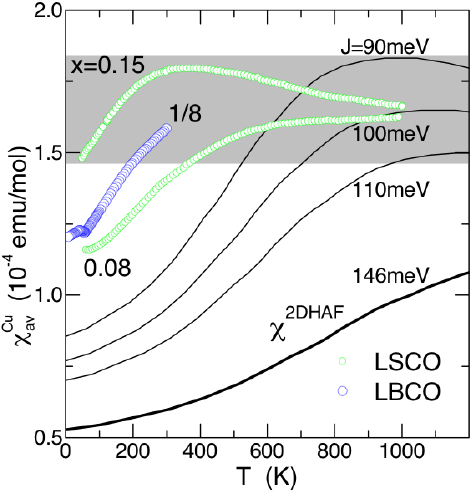}\hskip30pt
  \includegraphics[width=0.45\textwidth]{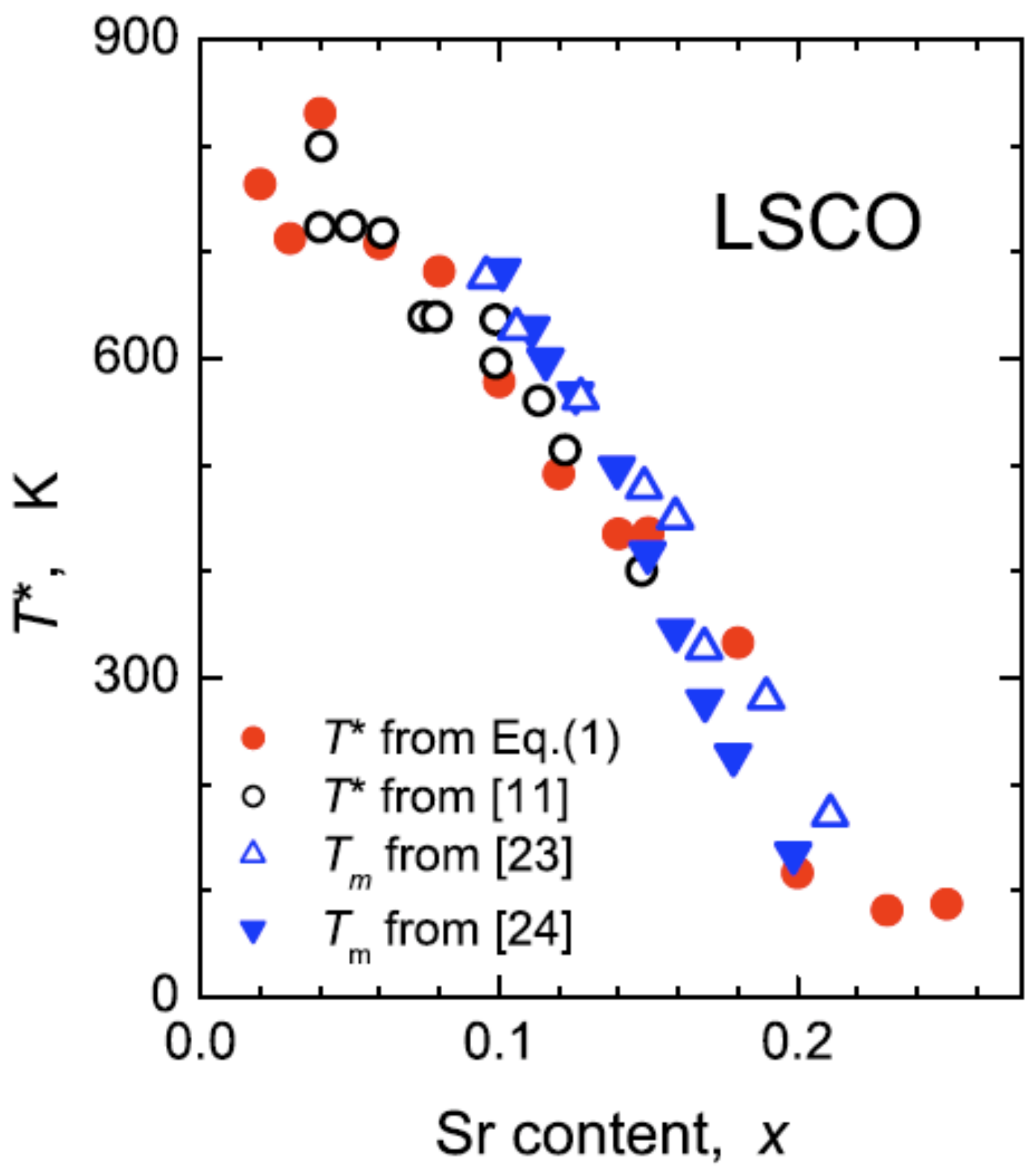}}
  \caption{(Left) Isotropic spin susceptibility for a single crystal of LBCO with $x=1/8$ compared with results from polycrystalline samples of LSCO with $x=0.08$ and 0.15.  Solid curves are model calculations for a 2D Heisenberg antiferromagnet with different values of $J$.  From H\"ucker {\it et al.} \cite{huck08}. (Right) Doping dependence of $T^*$ in LSCO from analysis of $R_{\rm H}(T)$, $\rho_{\rm ab}(T)$, and $\chi(T)$, from Gorkov {\it et al.} \cite{gork06}.}
  \label{fg:lsbco_chi}
 \end{figure}

H\"ucker {\it et al.} \cite{huck08} have measured and analyzed the anisotropic magnetic susceptibility for single crystals of La$_{2-x}$Ba$_x$CuO$_4$ (LBCO).  The anisotropy is consistent with the response one would expect from local magnetic moments.  From the analysis, it is possible to extract the isotropic spin susceptibility.  An example for $x=1/8$ is shown in the left panel of Fig.~\ref{fg:lsbco_chi}, together with results for LSCO with $x=0.08$ and $0.15$.  The magnitude of $\chi$ is inversely proportional to the effective spin coupling $J$, while the peak occurs at the temperature where local antiferromagnetic spin correlations begin to develop.  (The solid lines in the figure indicate the spin susceptibility calculated for a two-dimensional (2D) Heisenberg antiferromagnet with several different values of effective $J$, as labelled.)

The right-hand panel of Fig.~\ref{fg:lsbco_chi} shows values of $T^*$ obtained by Gorkov {\it et al.} \cite{gork06} from the temperature dependence of $R_{\rm H}$ (filled circles) and compared with values from the crossover temperature in the resistivity (open circles) and from the maximum in $\chi(T)$.   Below $T^*$, one has the development of coherent states associated with the nodal arc identified by ARPES.  At the same time, antiferromagnetic spin correlations develop among local Cu moments.  To understand the nature of these magnetic correlations, we will have to consider what has been learned from neutron scattering studies.

\subsubsection{Summary}

For underdoped cuprates, the following observations apply to the pseudogap phase from which the superconductivity develops:
\begin{itemize}
\item Only dopant-induced holes contribute to transport.
\item These holes have strong O $2p$ character.
\item They are associated with the nodal arc at the Fermi level.
\item They develop coherence on cooling.
\end{itemize}

\section{Neutron scattering}

Neutron scattering is a powerful technique for investigating atomic and magnetic structures, as well as lattice and spin dynamics.   Thermal neutrons, with a typical energy of 30 meV, have a wave length of  order 1.65~\AA, which is well matched to common interatomic spacings; it follows that thermal neutrons are quite useful for Bragg diffraction studies of crystal structure.  Unlike x-rays or electrons, which are scattered by the electronic charge density of atoms, neutrons scatter from atomic nuclei via the strong force.  The nuclear scattering cross section is sensitive to isotope as well as element, and there is no simple formula to characterize it; however, the typical magnitude of the cross section is roughly independent of atomic number.   This is beneficial when measuring the structure of a compound such as La$_2$CuO$_4$, where there is a large spread in the atomic number of the constituent elements, from $Z=16$ for O to $Z=57$ for La.   For x-ray scattering, the weight per atom in the diffraction pattern is proportional to $Z^2$, providing 13 times less sensitivity to O compared to La, whereas neutrons have roughly uniform sensitivity to all of these elements.

The neutron also has a spin of ${\scriptstyle\frac12}$, which means that it can scatter from atomic magnetic moments via the dipole-dipole interaction.  When large ordered magnetic moments are present in a sample, magnetic diffraction can be of the same strength as nuclear diffraction.  This is in contrast to x-ray scattering, where the magnetic cross section (without taking advantage of anomalous scattering near an absorption edge) is reduced relative to charge scattering by a factor of $\alpha^2=(1/137)^2$.

The energy of thermal neutrons is comparable to typical phonon and spin-wave energies in solids, so that only modest energy resolution is needed to in order to characterize phonon and spin-wave dispersions.  On the other hand, the overall scale of the scattering cross section is quite weak, so that one generally needs rather large samples in order to efficiently measure excitation spectra.

To create neutron beams, one has to extract neutrons from atomic nuclei.  This can be done either through the fission process in a nuclear reactor, where each uranium fission produces a couple of neutrons, or by knocking neutrons out of heavy-metal nuclei (typically tungsten or mercury) with a high energy proton beam, producing 10 or 20 neutrons per proton collision, as in a spallation source.   

The rest of this section will focus primarily on applications of neutron scattering to studies of cuprates.  More details on the theory and practice of neutron scattering are available in references \cite{zali13,shir02,squi12}.

\subsection{Neutron scattering cross section}

\begin{figure}[t]
  \includegraphics[width=0.7\textwidth]{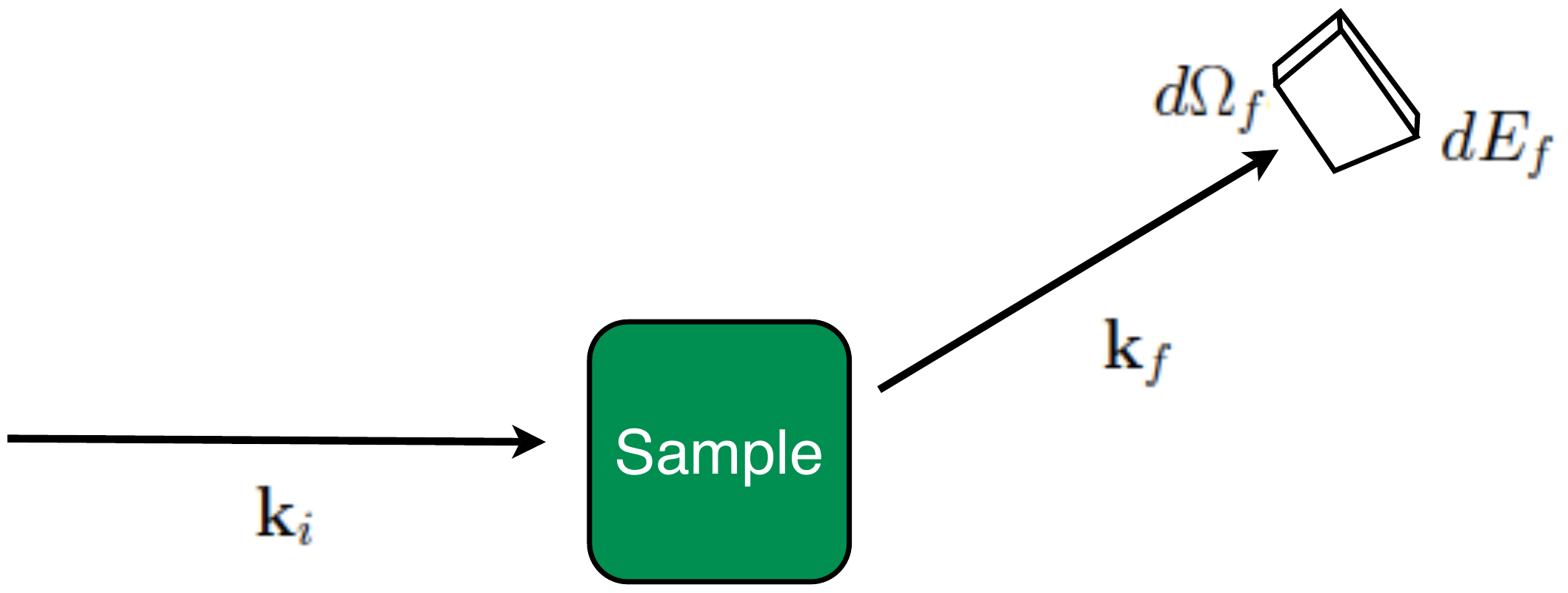}
  \caption{Diagram of scattering process.}
  \label{fg:neut1}
 \end{figure}

\begin{figure}[t]
  \includegraphics[width=0.3\textwidth]{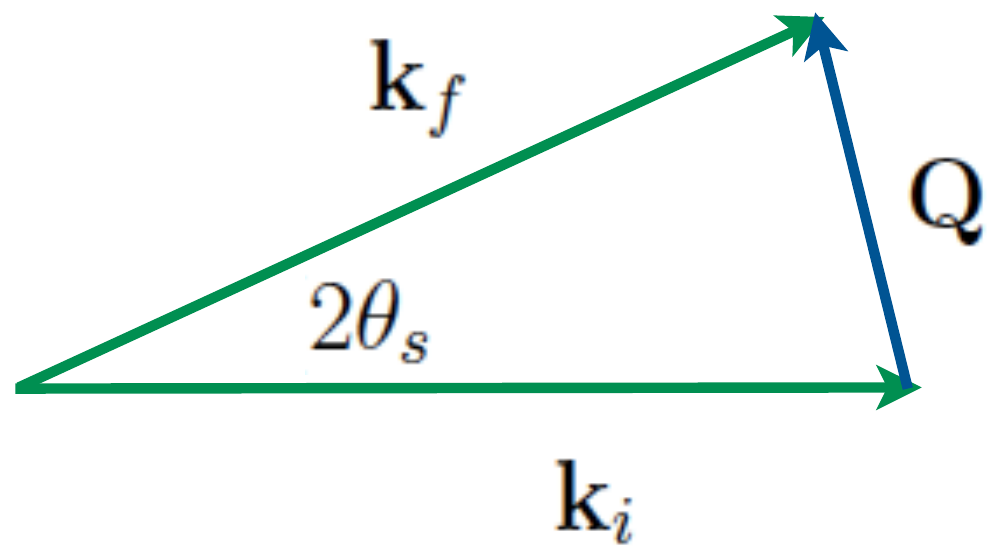}
  \caption{Scattering triangle.}
  \label{fg:neut2}
 \end{figure}

In a neutron scattering experiment, as illustrated in Fig.~\ref{fg:neut1}, the neutrons incident on the sample are characterized by an average momentum $\hbar{\bf k}_i$, where the magnitude of the wave vector is inversely proportional to the neutron wave length, $k=2\pi/\lambda$; the neutron energy is given by 
\begin{equation}
  E = {\hbar^2 k^2 \over 2m_{\rm n}},
\end{equation}
where $m_{\rm n}$ is the mass of the neutron.   One detects scattered neutrons with wave vector ${\bf k}_f$.  The probability of scattering into a differential solid angle $d\Omega_f$ and energy $dE_f$ corresponds to the double-differential cross section $d^2\sigma/d\Omega_f dE_f$.  To describe this function, it is helpful to introduce the neutron momentum transfer
\begin{equation}
 \hbar{\bf Q} = \hbar{\bf k}_f - \hbar{\bf k}_i,
\end{equation}
as illustrated in Fig.~\ref{fg:neut2}.  In terms of the scattering angle $2\theta_s$, a useful formula is
\begin{equation}
  |{\bf Q}| = k_i^2 + k_f^2 - 2k_ik_f\cos(2\theta_s).
\end{equation}
The energy transfer to the sample is
\begin{eqnarray}
  \hbar\omega & = & E_i - E_f, \\
   & = & {\hbar^2\over 2m_{\rm n}}(k_i^2-k_f^2).
\end{eqnarray}
  
It turns out that the differential cross section can be written in the relatively simple form 
\begin{equation}
  \frac{d^2\sigma}{d\Omega_fdE_f} =  N \frac{k_f}{k_i}{\sigma\over4\pi} {\cal S}({\bf Q},\omega),
  \label{eq:Sdef}
\end{equation}
where $N$ is the number of atoms in the sample, $\sigma$ is the cross section for a particular scattering process, and ${\cal S}({\bf Q},\omega)$ is the dynamical structure factor.  One can use this formula both for nuclear and magnetic scattering, but with different definitions of $\sigma$ and ${\cal S}$.

\subsection{Nuclear scattering and diffraction}

For nuclear scattering from a sample containing a single element (and single isotope), one has $\sigma = 4\pi b^2$, where $b$ is the nuclear scattering length, and
\begin{equation}
 {\cal S}({\bf Q},\omega) = \frac{1}{2\pi\hbar N} \sum_{ll'}\int_{-\infty}^{\infty} dt\,
 \langle e^{-i{\bf Q}\cdot{\bf r}_{l'}(0)} e^{i{\bf Q}\cdot{\bf r}_l(t)}\rangle\, e^{-i\omega t},
\end{equation}
where the angle brackets indicate a thermal and configurational average and $l$ indexes atomic positions.  Integrating over all time, one obtains the elastic cross section
\begin{equation}
   {\cal S}_{\rm el}({\bf Q},\omega) = \delta(\hbar\omega) \frac1N
   \left\langle\sum_{ll'} e^{i\bf{Q}\cdot({\bf r}_l-{\bf r}_{l'})}\right\rangle.
\end{equation}
For a monatomic lattice, one has
\begin{equation}
  {\cal S}_{\rm el}({\bf Q},\omega) = \delta(\hbar\omega) {(2\pi)^3\over v_0}
    \sum_{\bf G} \delta({\bf Q}-{\bf G}),
\end{equation}
where {\bf G} is a reciprocal lattice vector and $v_0$ is the unit cell volume.  This can be generalized to account for disorder due to phonons (by including the Debye-Waller factor) and for a unit cell with multiple atoms (by introducing a structure factor given by a sum over atoms in the unit cell, taking account of the appropriate scattering length and Debye-Waller factor for each site, weighted by a phase factor that depends on position).

\subsection{Inelastic scattering}

The dynamic structure factor must satisfy the detailed balance relation:
\begin{equation}
  {\cal S}(-{\bf Q},-\omega) = e^{-\hbar\omega/k_{\rm B}T} {\cal S}({\bf Q},\omega).
\end{equation}
Using the fluctuation-dissipation theorem, one can relate it to the imaginary part of a generalized susceptibility,
\begin{equation}
  {\cal S}({\bf Q},\omega) = {\chi''({\bf Q},\omega)\over 1-e^{-\hbar\omega/k_B T}}.
  \label{eq:chi}
\end{equation}
To be analytic, one must have
\begin{equation}
  \chi''({\bf Q},-\omega) = -\chi''({\bf Q},\omega).
\end{equation}
For a crystal lattice, $\chi''({\bf Q},\omega)$ describes the phonons, which generally have a weak dependence on temperature, whereas the thermal occupancy of a phonon state can vary greatly.

\subsection{Phonons}

Phonons have a dispersion $\omega_{{\bf q}s}$, where $s$ labels the phonon modes and {\bf q} is defined relative to a reciprocal lattice vector:
\begin{equation}
  {\bf Q} = {\bf G} + {\bf q}.
\end{equation}
For a monatomic lattice, one has
\begin{eqnarray}
  \chi''({\bf Q},\omega) & = & \frac12{(2\pi)^3\over v_0}
  \sum_{{\bf G},{\bf q}}\delta({\bf Q}-{\bf q}-{\bf G})
 \sum_s{1\over\omega_{{\bf q}s}} {({\bf Q}\cdot\mbox{\boldmath $\xi$}_s)^2\over m}
  \nonumber\\
  & & 
  \quad \times \left[\delta\left(\omega-\omega_{{\bf q}s}\right)
  -\delta\left(\omega+\omega_{{\bf q}s}\right)\right],
  \label{eq:phonon}
\end{eqnarray}
where $\mbox{\boldmath $\xi$}_s$ is the eigenvector for mode $s$.  Again, it is possible to generalize this to a multi-atom unit cell.  One can see that the intensity varies inversely with the mode frequency and depends on the angle between {\bf Q} and the eigenvector.   To measure longitudinal phonons, one needs ${\bf Q} || {\bf q}$, while transverse phonon measurements require ${\bf Q} \perp {\bf q}$.

One can certainly generalize Eq.~(\ref{eq:phonon}) for a crystal with a multi-atom unit cell.  Each phonon mode then has a structure factor in which phase factors depending on position within the unit cell are weighted by the phonon eigenvector, inverse mass, and nuclear scattering length.  The dispersion must be the same in every Brillouin zone, but the intensity (obtained from the square of the structure factor) can have a complicated dependence on {\bf Q}.

\begin{figure}[t]
  \includegraphics[width=0.8\textwidth]{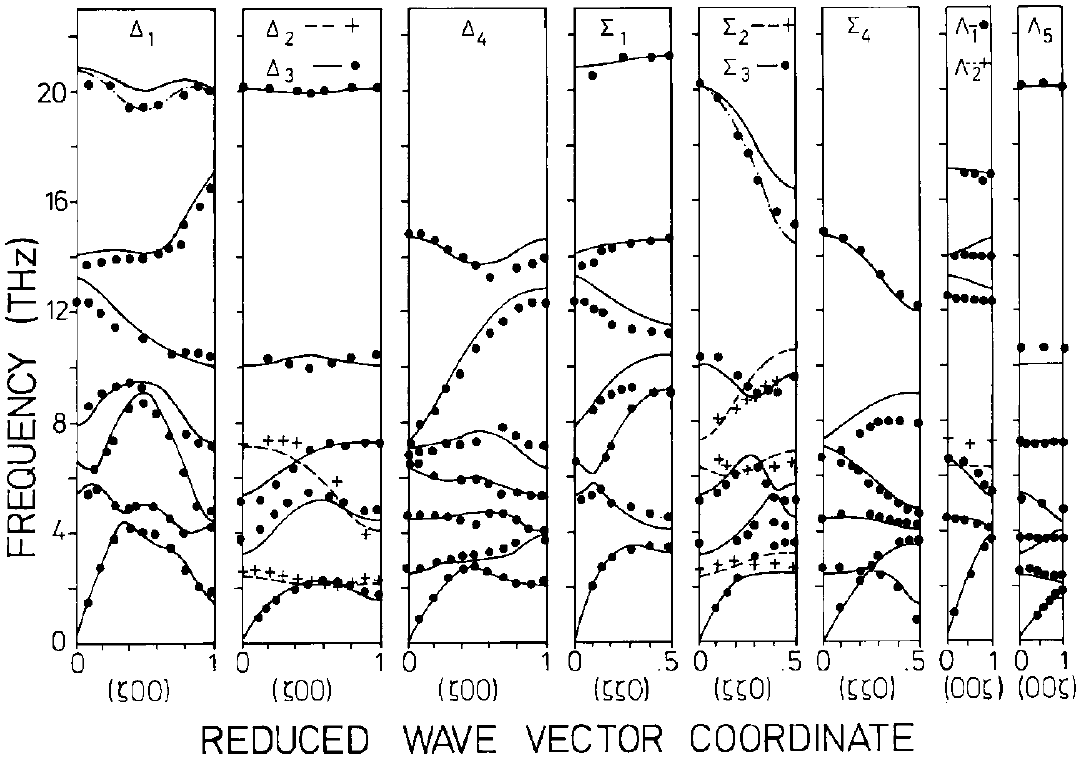}
  \caption{Phonon dispersions along high-symmetry directions for La$_2$CuO$_4$ from Chaplot {\it et al.} \cite{chap95}.  Circles, crosses: phonon energies determined by inelastic neutron scattering, with mode symmetries indicated at the top of each panel.  Solid lines: dispersions calculated from an interatomic force-constant model (based on the ionic shell model) with fitted parameters.  Dot-dashed line: Same model, but with inclusion of a quadrupolar interaction for Cu-O bonds.  Note that 1 THz = 4.14 meV.}
  \label{fg:phonon}
 \end{figure}

Figure~\ref{fg:phonon} shows an example of phonon dispersions in La$_2$CuO$_4$ measured by inelastic neutron scattering, compared with dispersions calculated from an interatomic potential model with fitted parameters.  For a system with $n$ atoms in the unit cell, there are a total of $3n$ phonon modes.  The longitudinal Cu-O bonding stretching mode is the highest-energy branch with $\Delta_1$ symmetry, up at $\sim20$~THz $\approx 83$~meV.
 
\subsection{Magnetic scattering}

For the case of magnetic scattering, Eq.~(\ref{eq:Sdef}) still applies, but now
\begin{equation}
  {\sigma\over 4\pi} =   \left({\gamma r_0\over2}\right) gf({\bf Q}),
\end{equation}
where
\begin{equation}
  {\gamma r_0\over 2} = 0.2695 \times 10^{-12}~\mbox{\rm cm},
\end{equation}
with $r_0=e^2/m_{\rm e}c^2$ being the classical electron radius and $\gamma=1.913$ is the neutron's gyromagnetic ratio; $g$ is the Land\'e $g$ factor (with a typical value $g\sim2$); and the magnetic form factor $f({\bf Q})$ is the Fourier transform of the the normalized unpaired spin density $\rho_s({\bf r})$ on an atom,
\begin{equation}
  f({\bf Q}) = \int \rho_s({\bf r}) e^{i{\bf Q}\cdot{\bf r}}d{\bf r},
\end{equation}
with
\begin{equation}
  f(0) \equiv 1.
\end{equation}
The magnetic spins in a sample have a vector character.  For the dipole-dipole interaction, it turns out that only the components of atomic spins that are perpendicular to {\bf Q} contribute to the cross section.  To take this into account, we must generalize the dynamic structure factor as follows:
\begin{equation}
  {\cal S}({\bf Q},\omega) \rightarrow \sum_{\alpha,\beta}(\delta_{\alpha,\beta}-Q_\alpha Q_\beta/Q^2){\cal S}^{\alpha,\beta}({\bf Q},\omega),
\end{equation}
with
\begin{equation}
 {\cal S}^{\alpha\beta}({\bf Q},\omega) = {1\over 2\pi\hbar}\int_{-\infty}^\infty dt\,
  e^{-i\omega t} \sum_l e^{i{\bf Q}\cdot{\bf r}_l} \langle
  S_0^\alpha(0)S_l^\beta(t)\rangle.
\end{equation}

Integrating ${\cal S}({\bf Q},\omega)$ over all frequencies, one obtains an instantaneous correlation function,
\begin{equation}
  {\cal S}^{\alpha\beta}({\bf Q},t=0) = \int_{-\infty}^\infty d\omega\,\,  {\cal S}^{\alpha\beta}({\bf Q},\omega).
\end{equation}
If one also integrates over a Brillouin zone (BZ) in reciprocal space, a simple sum rule is obtained:
\begin{equation}
   \int_{-\infty}^\infty d\omega \int_{\rm BZ}d{\bf Q}\,\,
  {\cal S}^{\alpha\beta}({\bf Q},\omega) = {(2\pi)^3\over3\hbar v_0}S(S+1)\delta_{\alpha\beta}.
\end{equation}


\subsection{Magnetic diffraction}

In the case of ferromagnetic order, the magnetic and nuclear scattering both occur at the same reciprocal lattice vectors.  To describe the intensity of Bragg reflections, one must write down a structure factor that includes both the nuclear and magnetic contributions.  The sign of the magnetic part depends on the direction of the neutron spin, so that the magnitude of the structure factor will depend on the neutron spin polarization.   If one uses a spin-polarized incident beam, then it is possible to analyze the nuclear and magnetic contributions to a particular Bragg reflection by performing measurements with two opposite orientations of the sample magnetization (controlled with an applied magnetic field).

Here we are more interested in the case of antiferromagnetic order, and we will limit the discussion to the case of equivalent collinear spins.  Let $\hat{\bf Q}= {\bf Q}/Q$; then, for an average spin ${\bf S}$ on one sub lattice, the vector that contributes to diffraction is defined by
\begin{equation}
  {\bf S}_\bot = {\bf S} - \hat{\bf Q}(\hat{\bf Q}\cdot {\bf S}).
\end{equation}
For an antiferromagnet, we must have at least two magnetic sites per unit cell, described by basis vectors ${\bf d}_j$.  When the magnetic unit cell is larger than the nuclear unit cell, the magnetic Bragg peaks will form a superlattice in reciprocal space, described by ${\bf G}_m$.  The cross section for antiferromagnetic diffraction can be written
\begin{equation}
   \left.{d\sigma\over d\Omega_f}\right|^{\rm el}_{\rm coh} = 
  N_m{(2\pi)^3\over v_m}\sum_{{\bf G}_m}\delta({\bf Q}-{\bf G}_m) 
  |{\bf F}_M({\bf G}_m)|^2,
\end{equation}
where
\begin{equation}
  {\bf F}_M = {\bf S}_\bot \sum_j s_j \,e^{i{\bf Q}\cdot {\bf d}_j},
\end{equation}
with $s_j=\pm1$, according to the pattern of the spin order.
If the magnetic structure has lower symmetry than the nuclear structure, then multiple magnetic domains will generally be present in a sample, and one must allow for an averaging over domains when evaluating the magnetic structure factor.  For collinear spins, this means one must average over equivalent orientations of ${\bf S}_\bot$.  If we label the angle between {\bf S} and {\bf Q} as $\eta$, then the quantity that appears in the average differential cross section is
\begin{equation}
  \langle |{\bf S}_\bot|^2\rangle = S^2(1-\langle\cos^2\eta\rangle).
\end{equation}

\subsection{Antiferromagnetic order in La$_2$CuO$_4$}

\begin{figure}[t]
  \includegraphics[width=0.5\textwidth]{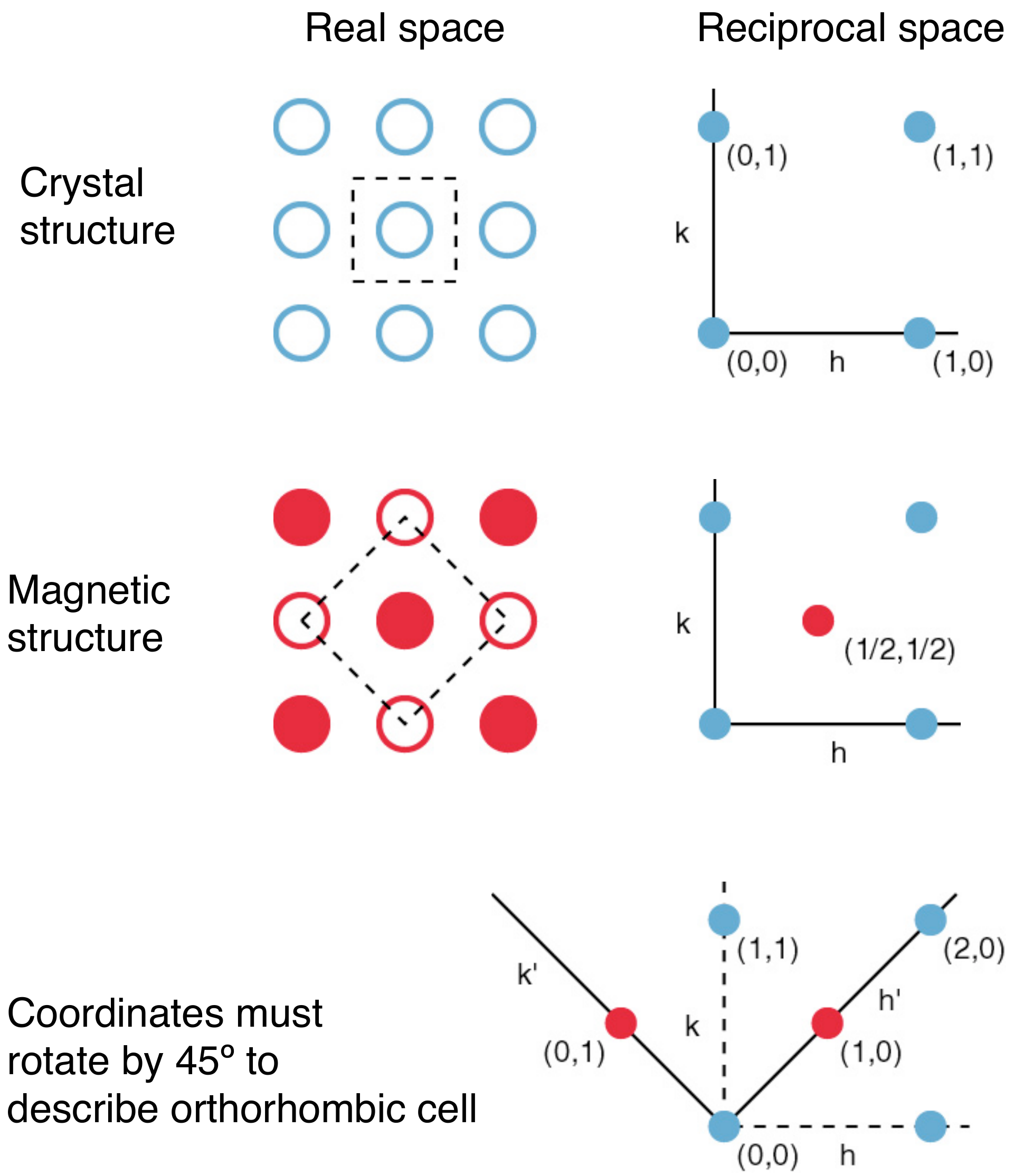}
  \caption{Illustration of the relationship between real-space structure and reciprocal lattice vectors for the copper atoms in a CuO$_2$ plane of La$_2$CuO$_4$.  In the panels indicating real space structure, the dotted lines denote the unit cell.  For the magnetic structure, filled and open circles denote up and down spins.}
  \label{fg:AF_cell}
 \end{figure}

Let us consider the case of antiferromagnetic (AF) ordering in the CuO$_2$ planes of La$_2$CuO$_4$.  As illustrated in Fig.~\ref{fg:AF_cell}, the magnetic ordering doubles the unit cell, resulting in new magnetic superlattice peaks at $({\scriptstyle\frac12},{\scriptstyle\frac12})$ in reciprocal space.   There happen to be two CuO$_2$ layers per unit cell in the three-dimensional (3D) crystal structure, with one layer centered with respect to the other.  The 3D AF ordering wave vector must take account of the relative phasing of the spins in the two layers.

Antiferromagnetism is not the only possible cause of superlattice peaks.  It so happens that structural distortions can change the size of the nuclear unit cell, as well.  In the case of La$_2$CuO$_4$, illustrated in Fig.~\ref{fg:lco_struc}, the CuO$_6$ octahedra can rotate slightly about a (110) axis of the tetragonal structure.  The octahedral tilts cause an in-plane doubling of the nuclear unit cell, together with an orthorhombic distortion with lattice vectors along the diagonals of the Cu plaquettes, as indicated in Fig.~\ref{fg:lco_struc}.  In the orthorhombic structure, the in-plane magnetic wave vectors are indexed as (1,0) and (0,1), as indicated in Fig.~\ref{fg:AF_cell}, and these are no longer equivalent.  The 3D magnetic structure, with spins indicated by the arrows in Fig.~\ref{fg:lco_struc}, is characterized by the wave vector (100) \cite{vakn87}.

The octahedral tilts also lead to superlattice peaks, which are of the type (012) (using the $Bmab$ space group).  In powder diffraction, the structural superlattice peaks are weak, but significantly more prominent than the magnetic peaks, which are smaller than the background \cite{yang87}.  An early diffraction study mis-identified the structural peaks as antiferromagnetic peaks \cite{yama87b}.  The use of polarized neutron diffraction eventually confirmed the proper identification \cite{mits87}.

%
 
So far, we have ignored the effect of fluctuations on Bragg intensities, but we cannot get away with that in the case of AF order in the cuprates.  Given that one has just $S={\scriptstyle\frac12}$ per Cu atom, the magnetic peak intensities should be rather weak.  It turns out that there are also rather large zero-point fluctuations of the spin directions that reduce the average spin, $\langle S\rangle$, yet further. Theory predicts for the $S={\scriptstyle\frac12}$ Heisenberg AF that $\langle S\rangle = 0.30$ \cite{mano91}, and neutron diffraction results on a variety of cuprates are roughly consistent with this \cite{tran07}. The total neutron scattering weight should be proportional to $\langle S^2\rangle = S(S+1) = \frac34$, whereas the elastic scattering is proportional to just $\langle S\rangle^2 = 0.09$.  It follows that 88\%\ of the sum-rule weight should be in inelastic scattering, which we will consider after a small digression.

\subsection{Digression: $\mu^+$ Spin Rotation}

Another effective probe of magnetic order is muon spin rotation ($\mu$SR) spectroscopy.  Muons have a lifetime of just 2.2 $\mu$s, but they can be prepared readily at a proton accelerator.   One generally uses positive muons, one at a time.  A muon injected into a sample rapidly diffuses to the most electronegative interstitial site in the lattice.  If there is a local magnetic field due to ordered moments, the spin of the muon will precess about the field.  When the muon decays, the resulting positron is emitted preferentially along the direction of the muon's spin.  By putting a pair of detectors on opposite sides of the sample, one can detect when the muon spin is pointing along a particular axis.  Binning the detected muons as a function of time from the initial implantation, one can directly measure the muon precession frequency.

\begin{figure}[t]
  {\includegraphics[width=0.4\textwidth]{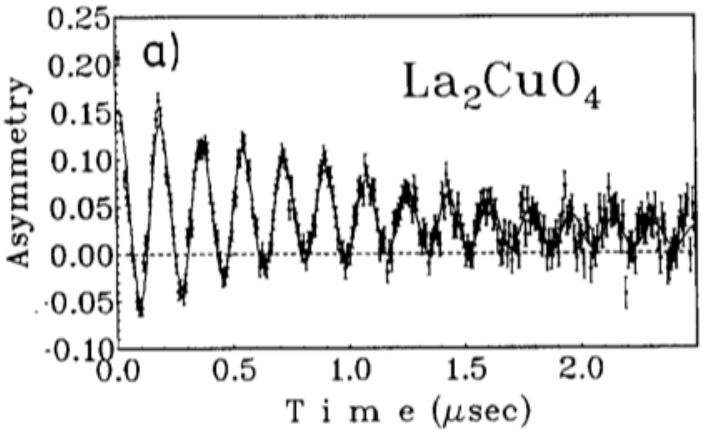} \hskip30pt
  \includegraphics[width=0.4\textwidth]{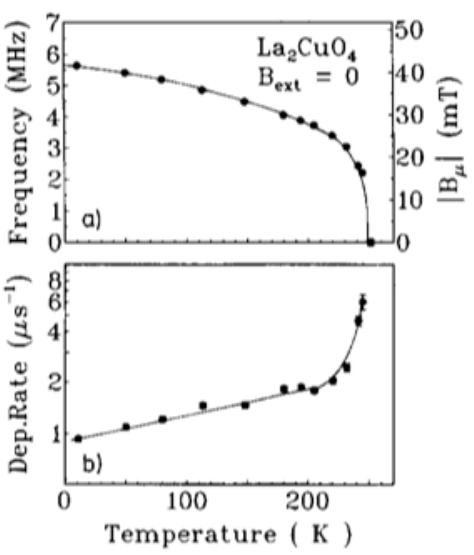}}
  \caption{(Left) $\mu$SR spectrum measured on a polycrystalline sample of  La$_2$CuO$_4$ at 11~K.  (Right) (a) Precession frequency measured as a function of temperature, showing the gradual decay of magnetic order, and a loss of order at $\sim250$~K.  (b) The depolarization rate of the precessing muons.  Reprinted from Budnick {\it et al.} \cite{budn87} with permission from Elsevier.}
  \label{fg:lco_musr1}
 \end{figure}

The left side of Fig.~\ref{fg:lco_musr1} shows the $\mu$SR spectrum measured at low temperature in La$_2$CuO$_4$ \cite{budn87}.  The precession frequency of $\sim6$~MHz is proportional to the local magnetic field at the muon site.  The fact that there is a well defined local field indicates that there must be magnetic order, though the measurement, by itself, does not distinguish the type of order ({\it i.e.}, antiferromagnetic vs.\ ferromagnetic).   

If one knows that the muon's position within the lattice, then it is possible to relate the average magnetic moment per Cu to the dipolar magnetic field at the muon site.  While there are challenges to using the $\mu$SR precession frequency as an absolute measure of local moments, it serves very well as a relative measure.  For example, one can measure the change in magnetic order as a function of temperature, as shown in the upper right of Fig.~\ref{fg:lco_musr1}.

\subsection{Antiferromagnetic spin waves}

Spin waves are purely transverse fluctuations relative to the magnetically-ordered state.  (Longitudinal fluctuations correspond to modulations of the order parameter.)  If we define the direction of the ordered spins to be $z$, then we have
\begin{equation}
  \sum_{\alpha,\beta}(\delta_{\alpha,\beta}-\hat{Q}_\alpha\hat{Q}_\beta)
  {\cal S}^{\alpha\beta}({\bf Q},\omega) = {\textstyle \frac12}(1+\hat{Q}_z^2){\cal S}_{\rm sw}({\bf Q},\omega).
\end{equation}
Antiferromagnetic spin waves near zone center have a linear dispersion
\begin{equation}
   \hbar\omega_{\bf q} = \hbar cq,
\end{equation}
with spin-wave velocity
\begin{equation}
  c = zJSa/\hbar,
\end{equation}
where $z$ is the number of nearest neighbors ($=4$ for a square lattice).  Using Eq.~(\ref{eq:chi}), we can write ${\cal S}_{\rm sw}$ in terms of $\chi_{\rm sw}''$, with 
\begin{eqnarray}
  \chi_{\rm sw}''({\bf Q},\omega) & = & S \sum_{{\bf G}_m,{\bf q}}
 {\hbar\omega_0\over\hbar\omega_{\bf q}}
  \left[\delta({\bf Q}-{\bf q}-{\bf G}_m)
  \delta(\omega-\omega_{\bf q}) \right. \nonumber\\
  & & \qquad\qquad + \left.
  \delta({\bf Q}+{\bf q}-{\bf G}_m)\delta(\omega+\omega_{\bf q})\right],
\end{eqnarray}
and
\begin{equation}
  \hbar\omega_0 = 2zJS.
\end{equation}

\begin{figure}[t]
  \includegraphics[width=0.5\textwidth]{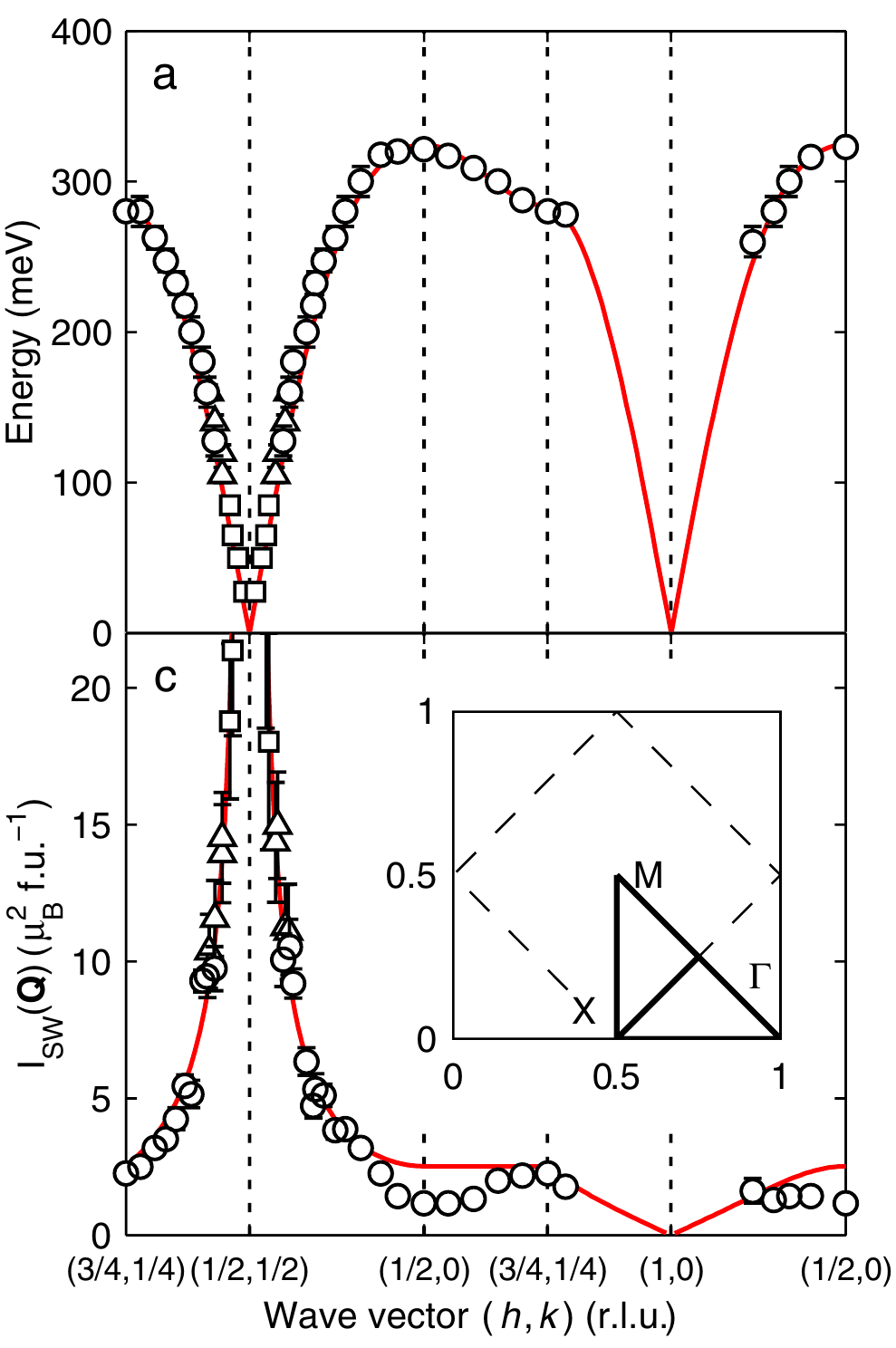}
  \caption{(a) Low-temperature spin wave dispersion and (c) intensity along high-symmetry directions in La$_2$CuO$_4$ measured with inelastic neutron scattering by Headings {\it et al.} \cite{head10}.  Circles: experimental results, lines: fits with spin wave theory. }
  \label{fg:lco_sw}
 \end{figure}

\begin{figure}[t]
  \includegraphics[width=0.3\textwidth]{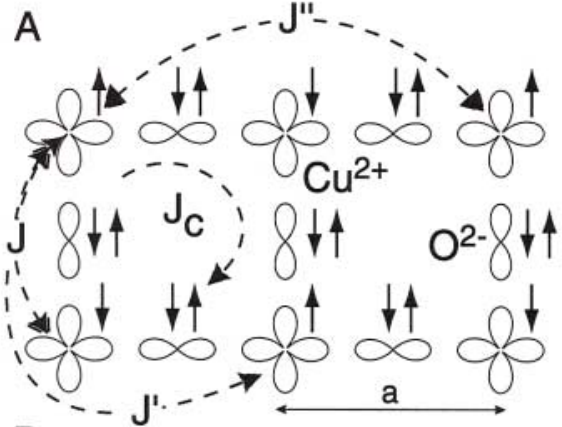}
  \caption{Sketch indicating possible exchange terms in the CuO$_2$ plane, from Coldea {\it et al.} \cite{cold01}.}
  \label{fg:lco_ex}
\end{figure}

Figure~\ref{fg:lco_sw} shows neutron scattering measurements of spin-wave energies and intensities along high-symmetry directions in reciprocal space for La$_2$CuO$_4$ \cite{head10}.  The dispersion along the zone boundary, between $({\scriptstyle \frac12},0)$ and $({\scriptstyle \frac34},{\scriptstyle \frac14})$, indicates that the spin Hamiltonian must contain more than nearest-neighbor interactions.  Figure~\ref{fg:lco_ex} shows longer range exchange parameters $J'$ and $J''$, as well as 4-spin-cyclic exchange $J_c$.  A fit to the measured dispersion [line in Fig.~\ref{fg:lco_sw}(a)] yields $J=143\pm2$~meV, $J'=J''=2.9\pm0.2$~meV, and $J_c=58\pm4$~meV.  The spin-wave model also describes the intensities remarkably well, except perhaps near $({\scriptstyle \frac12},0)$, where there is some broadening of the peaks.

It is common to associate strong antiferromagnetic superexchange with strong coupling ({\it i.e.}, $U\rightarrow\infty$); however, this is not correct.  Consider the case of the single-band Hubbard model at half-filling, for which $J=4t^2/U$.  In the limit of infinite $U$, one obtains $J=0$.  Instead, large $J$ requires intermediate coupling.  For example, suppose $U$ is equal to the bandwidth, $8t$; it follows that $J=t/2$.  Then to obtain $J=143$~meV, as in La$_2$CuO$_4$, one needs $t\approx 0.3$~eV.  These values are fairly close to those extracted from constrained density-functional calculations \cite{hybe90}.   Of course, the Heisenberg exchange model gives a good description of the half-filled Hubbard model only in the limit of large $U$.  In the case of intermediate coupling, one expects higher order interactions to become relevant.  Indeed, the relative magnitude of the $J_c$, the 4-spin cyclic exchange, found in fits to the spin wave dispersion is consistent with theoretical estimates \cite{cold01}.

\begin{figure}[t]
  \includegraphics[width=0.8\textwidth]{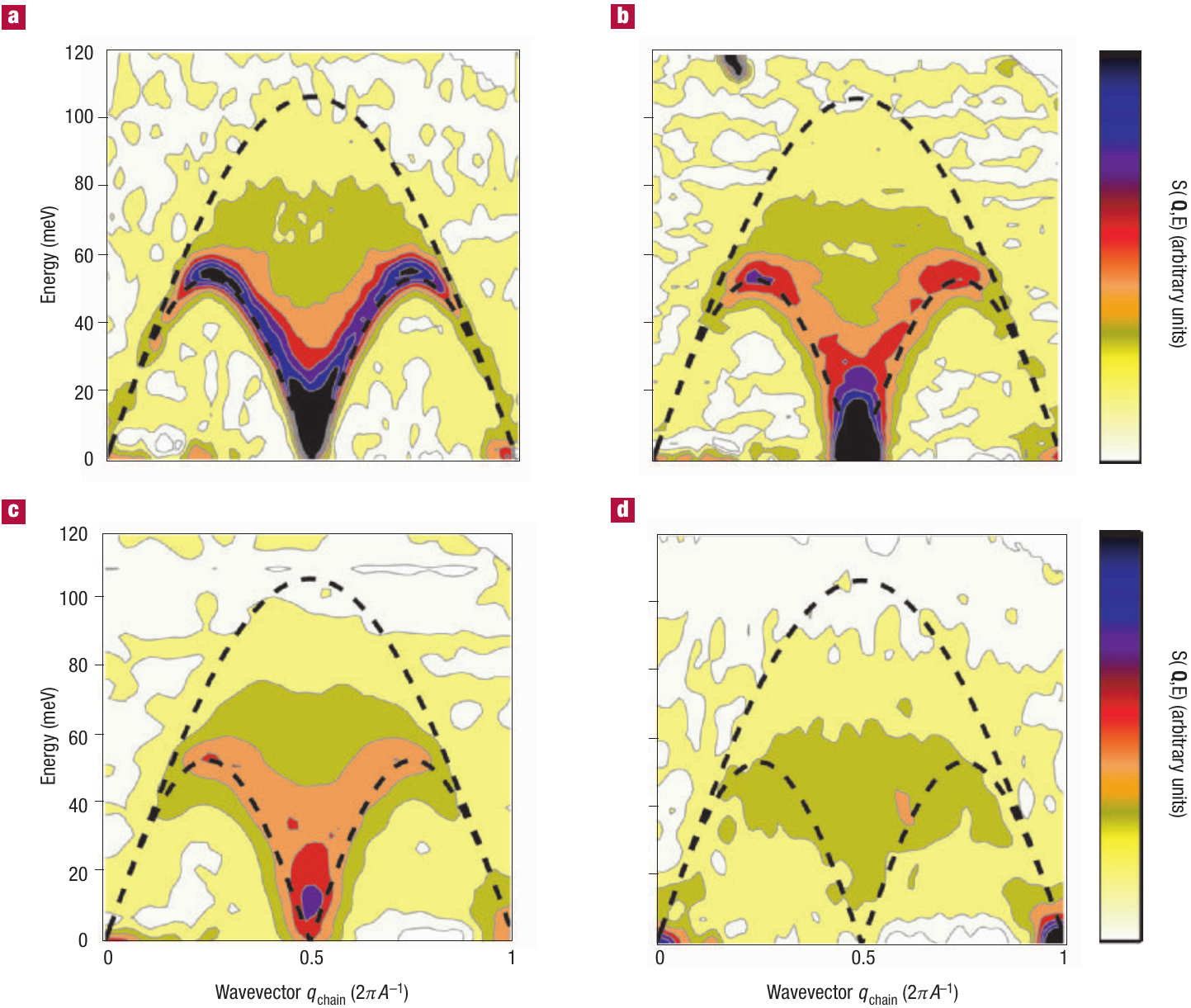}
  \caption{Inelastic neutron scattering spectra of KCuF$_3$ measured at temperatures of (a) 6~K, (b) 50~K, (c) 150~K, (d) 300~K, from Lake {\it et al.}  \cite{lake05}. Reprinted by permission from Macmillan Publishers Ltd.  Scattered intensity is plotted as a function of excitation energy and wave vector along the spin chains.  There is a continuum of two-spinon excitations between the two sets of dashed lines.  The lower dashed line has the form of an antiferromagnetic spin-wave dispersion.}
  \label{fg:kcuf3}
 \end{figure}
 
\begin{figure}[t]
  \includegraphics[width=0.5\textwidth]{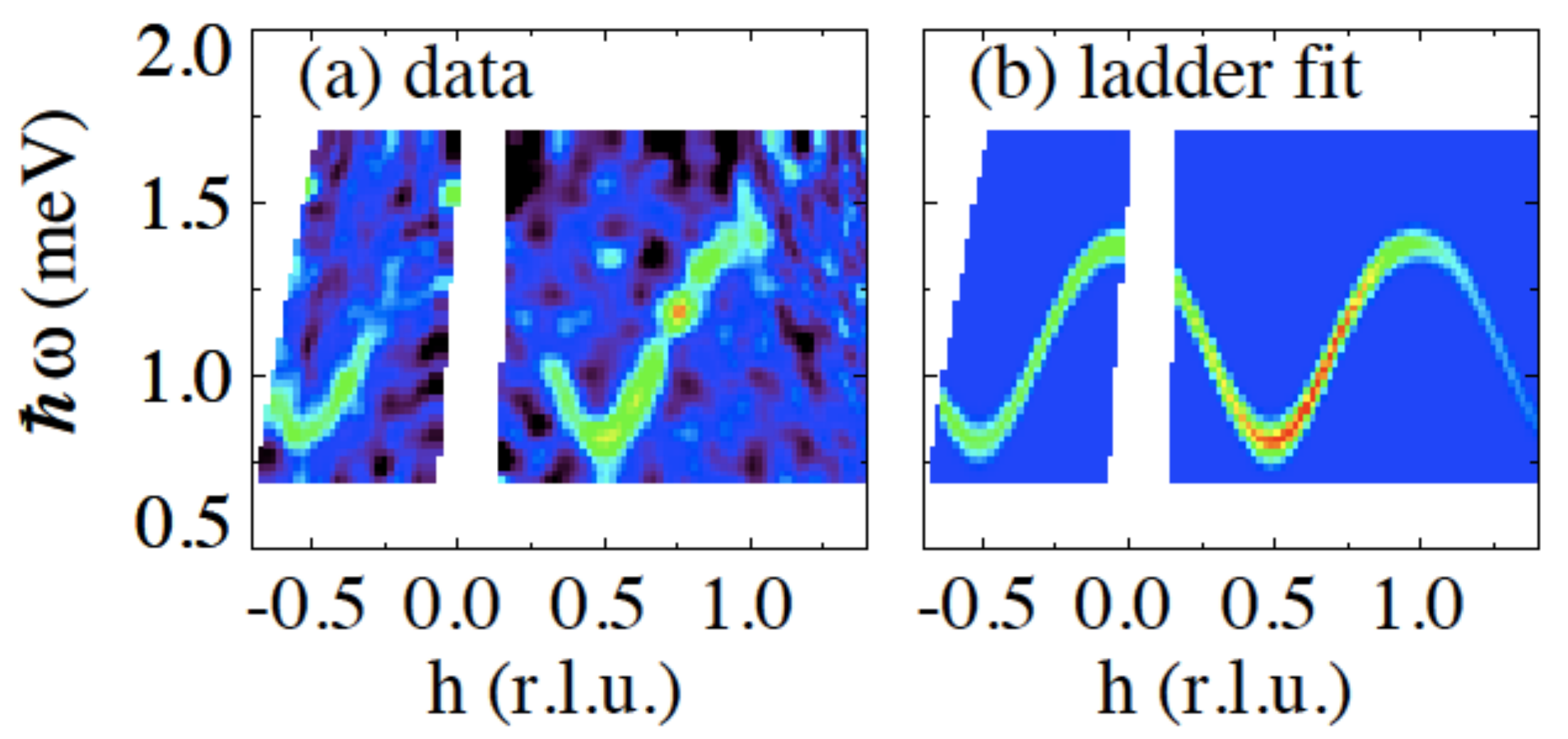}
  \caption{Neutron-scattering intensity for spin excitations in the spin-${\scriptstyle \frac12}$ 2-leg ladder compound (C$_5$D$_{12}$N)$_2$CuBr$_4$ from Savici {\it et al.} \cite{savi09}.}
  \label{fg:ladder}
 \end{figure}

\subsection{Impact of dimensionality}

An intriguing variety of spin-excitation spectra occur in $S={\scriptstyle \frac12}$ systems with low dimensionality.  A classic example is given by KCuF$_3$, in which orbital ordering of the half-filled Cu $3d$ orbital leads to almost-decoupled antiferromagnetic chains of Cu atoms.  The fundamental excitation of such a chain is a spinon, a $\Delta S={\scriptstyle \frac12}$ excitation corresponding to a half-twist of the spin chain.  Since neutrons can only create excitations with $\Delta S =1$, the observed spectrum consists of a two-spinon continuum, as illustrated in Fig.~\ref{fg:kcuf3}.  The lower bound of the continuum, where the intensity is strongest, corresponds to the spectrum one would expect for conventional spin waves, but with a renormalized spin-wave velocity. 

Such quantum spin chains are an example of a quantum critical system, where there is no static spin order but the spin excitations are gapless.  For a real system such as KCuF$_3$, weak couplings between the chains lead to confinement of the spinons and antiferromagnetic order at low temperature \cite{zali05b}.

If one takes a pair of spin-${\scriptstyle \frac12}$ chains and couples them together to form a so-called spin ladder, the spectrum is remarkably transformed.   Now the spins tend to form singlet pairs, and the lowest excitation involves the creation of a triplet.  With increasing energy, the triplet can disperse along the length of the ladder.  An experimental example of such a spectrum is shown in Fig.~\ref{fg:ladder}.   Because of the large spin gap, a substantial finite coupling between spin ladders in an array is required to obtain an ordered ground state \cite{twor99}.
 
\begin{figure}[t]
  \includegraphics[width=0.5\textwidth]{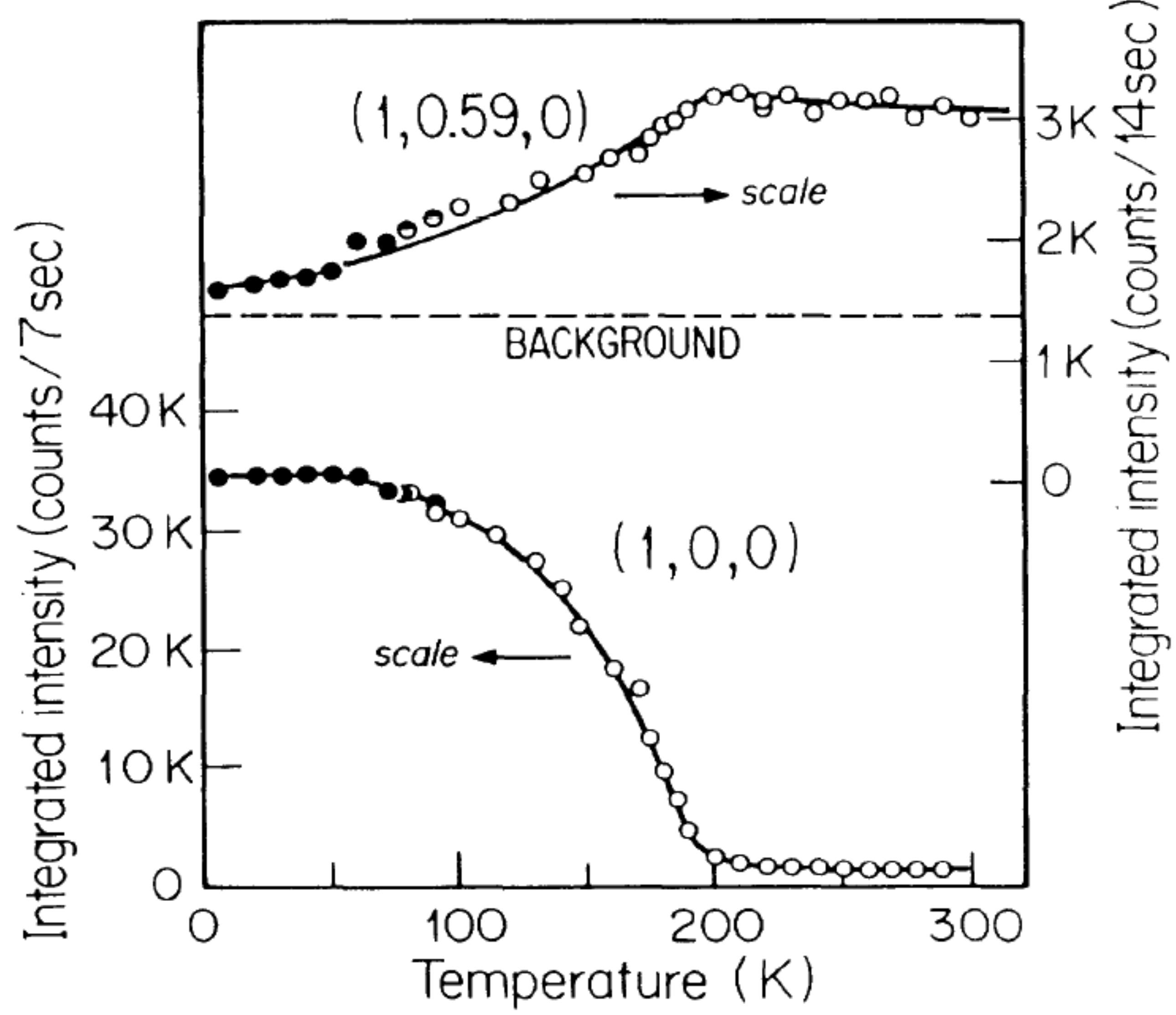}
  \caption{Bottom: integrated intensity of the (100) AF Bragg peak vs.\ temperature in a single crystal of La$_2$CuO$_4$ with $T_{\rm N}\approx 200$~K; top: integration of inelastic scattering from excitations along ${\bf Q}= (1,0,0.59\pm l)$, in coordinates of $(2\pi/a,2\pi/b,2\pi/c)$, from Shirane {\it et al.} \cite{shir87}.  This experiment was performed on one of the first large single crystals of La$_2$CuO$_4$, and the low value of $T_{\rm N}$ is likely due to unintentional impurities on the Cu sites due to contamination from the crucible in which the crystal was grown.}
  \label{fg:shir}
 \end{figure}

As already mentioned, Anderson predicted that the $S={\scriptstyle \frac12}$ 2D Heisenberg AF might have a disordered ground state \cite{ande87}.  Experiments eventually demonstrated that the AF ordering temperature of La$_2$CuO$_4$, $T_{\rm N} =325$~K \cite{keim92a}, is much lower than what one might anticipate from $J/k\sim1500$~K.  With such a large $J$, one would expect to find strong 2D spin correlations surviving to temperatures far above $T_{\rm N}$.

For a system with only 2D correlations, the scattering is essentially independent of the momentum transfer perpendicular to the layers.  Shirane {\it et al.} \cite{shir87} used an experimental trick to integrate spin-fluctuation scattered from the CuO$_2$ planes in La$_2$CuO$_4$ as a function of temperature; their results are shown in Fig.~\ref{fg:shir}, in comparison with the thermal evolution of the intensity of an AF Bragg peak.  The intensity of the 2D spin fluctuations grows slowly with cooling, and then decreases rapidly for $T<T_{\rm N}$, as low-energy weight is transferred to the Bragg peaks.
 
\begin{figure}[t]
  \includegraphics[width=0.5\textwidth]{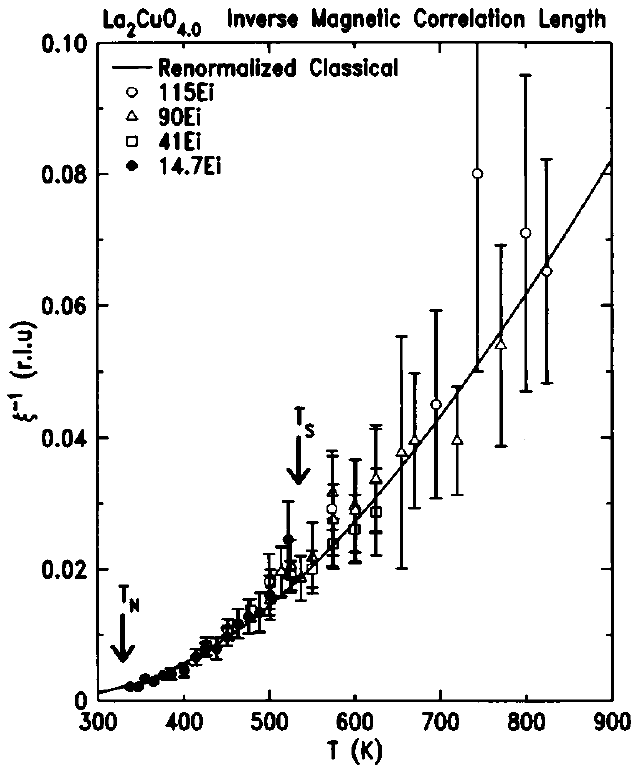}
  \caption{Experimental measurement of the inverse of the 2D spin-spin correlation length, $\xi^{-1}$, in La$_2$CuO$_4$ as a function of temperature. Reprinted from Birgeneau {\it et al.} \cite{birg95} with permission from Elsevier.  The line through the data corresponds to a fit to a theoretical form similar to Eq.~(\ref{eq:xi}) with $J=135$~meV.  $T_s$ labels the structural transition from tetragonal to orthorhombic due to ordering of octahedral tilts.}
  \label{fg:birg}
 \end{figure}

An isolated 2D Heisenberg AF should not order at finite temperature, according to the Mermin-Wagner theorem \cite{merm66}.  Nevertheless, a small coupling between layers can drive 3D ordering as soon as the 2D spin-spin correlation length $\xi$ in the planes becomes substantial.  As a result, there is no obvious critical scattering at the transition.

The theoretical form for the temperature dependence of $\xi$ depends on the nature of the ground state of the 2D system.  In the case where order should occur at $T=0$, the prediction is \cite{chak89,hase91}
\begin{equation}
  {\xi\over a} \approx A e^{BJ/kT},
  \label{eq:xi}
\end{equation}
where $a\approx5.4$~\AA\ is the lattice constant, $A\approx 0.5$, and $B\approx 1.1$.  Figure~\ref{fg:birg} shows a comparison of experimental measurements of $\xi^{-1}(T)$ with a fit to the theoretical prediction \cite{kast98,birg95}.   The agreement is clearly consistent with an extrapolated divergence of $\xi(T)$ at $T=0$.

There is now both experimental and theoretical agreement that CuO$_2$ planes should order at $T=0$ \cite{mano91}.  Nevertheless, we have noted that quantum fluctuations are large and have observable effects.  Chakravarty {\it et al.} \cite{chak88} evaluated the possibility that an increase in fluctuations might lead to a quantum critical point (QCP), at which the ground state would become disordered.  The concept of QCPs has become a recurring theme in analyses of the phase diagrams of high temperature superconductors.

 
\section{Evolution of spin correlations in cuprates with doping}

Looking back at the phase diagram in Fig.~\ref{fg:ph_diag}, one can see that hole doping causes a very rapid destruction of AF order.  In this section, we will investigate the nature of that destruction, and we will begin to consider the character of the correlations that survive.

\subsection{Magnetic dilution}

Each hole doped into a CuO$_2$ plane has its own spin.  Zhang and Rice \cite{zhan88} proposed that each hole forms a bound singlet state with a Cu$^{2+}$ ion, effectively creating a nonmagnetic site.  This bound hole impacts the magnetic correlations in two ways: 1) it reduces the density of magnetic moments, and 2) the motion of the hole can stir up the spins.  Here we will consider the dilution effect.

One can study the dilution effect directly by preparing samples with a fraction $z$ of the Cu ions replaced by Zn and/or Mg.\footnote{The substituted ions are not exactly the same size as the Cu, and the associated strains can limit the amount of a given ion that can be substituted.  Empirically, it if found that one can achieve greater levels of substitution using a combination of Zn and Mg \cite{vajk02}.}   For a classical 2D AF, it is known that AF order is destroyed at the percolation limit of $z\approx0.41$ \cite{newm00}.  Given the large quantum fluctuations for $S={\scriptstyle \frac12}$, it was suspected that the loss of magnetic order might occur at a lower threshold.

\begin{figure}[t]
  \includegraphics[width=0.4\textwidth]{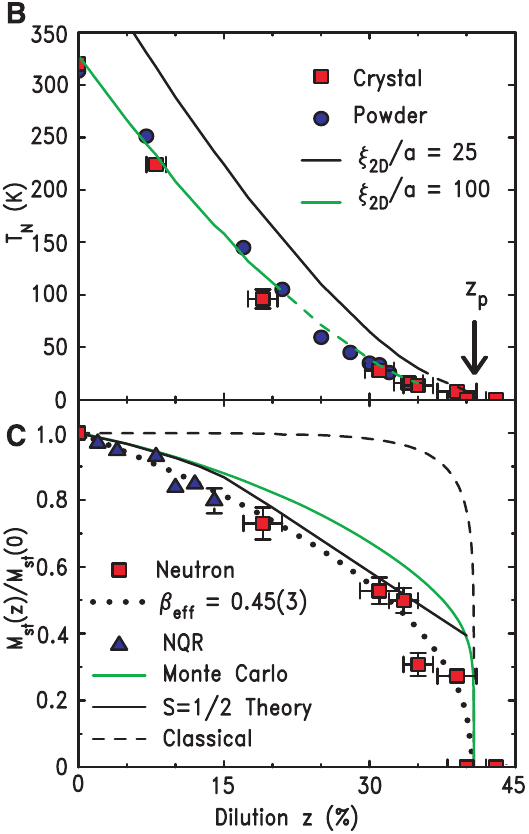}
  \caption{Impact of non-magnetic dilution on antiferromagnetic order in La$_2$Cu$_{1-z}$(Zn,Mg)$_z$O$_4$ from the work of Vajk {\it et al.} \cite{vajk02}.  Reprinted with permission from AAAS.  (B)  $T_{\rm N}$ vs.\ $z$ measured by neutron diffraction and magnetometry on crystals (squares) and by magnetometry on powders (circles).  Solid lines are from Monte Carlo calculations. (C) Staggered moment per Cu, normalized to $z=0$, from neutron diffraction (squares) and nuclear quadrupole resonance (NQR, triangles).  Lines represent various models. }
  \label{fg:lczo}
\end{figure}

Figure~\ref{fg:lczo} shows the results reported by Vajk {\it et al.} \cite{vajk02}.  They find that the destruction of long-range order due to dilution occurs at the classical percolation limit.   At the same time, there is some reduction of the ordered moment compared to the classical prediction, though the difference is largely captured by models that take account of the quantum fluctuations.

It is clear from these results that the impact of hole doping in CuO$_2$ layers is much greater than the impact of dilution alone.

\subsection{Local magnetism survives}

\begin{figure}[t]
  \includegraphics[width=0.4\textwidth]{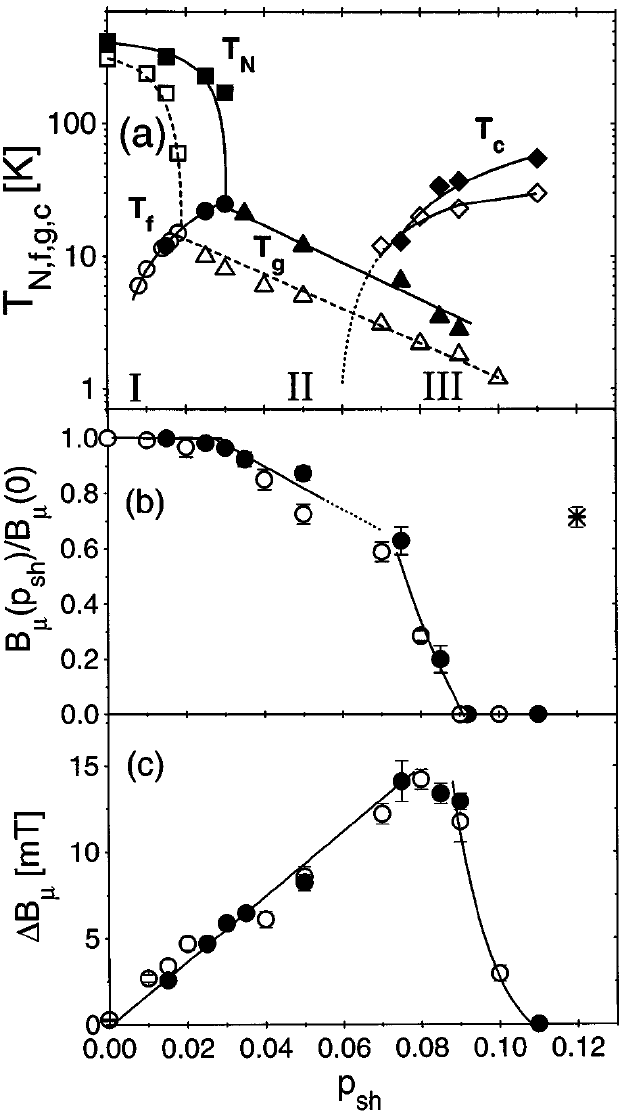}
  \caption{Results of a $\mu$SR study of ordering in La$_{2-x}$Sr$_x$CuO$_4$ (open symbols) and Y$_{1-2x}$Ca$_{2x}$Ba$_2$Cu$_3$O$_{6.02}$ (filled symbols; with hole density per sheet $p_{\rm sh}=x$) from Niedermayer {\it et al.} \cite{nied98}.  (a) Ordering phase diagram, indicating N\'eel temperature $T_{\rm N}$, freezing temperature $T_f$, spin-glass temperature $T_g$, and superconducting transition $T_c$.  (b)  Average magnetic field at the muon site measured at $T<1$~K, normalized to the antiferromagnet.  (c) Width of the magnetic field distribution.}
  \label{fg:nied}
\end{figure}

While long-range AF order is destroyed by a small density of holes ($\sim0.02$ in La$_{2-x}$Sr$_x$CuO$_4$), early studies with local probes, such as nuclear quadrupole resonance (NQR) and nuclear magnetic resonance (NMR), demonstrated that local magnetic order can survive, at least at low temperatures.   A particularly useful illustration of this is given by Fig.~\ref{fg:nied}, which presents the results of a $\mu$SR study on two different cuprate families.  Even before long-range order is destroyed, finite hole doping leads to a second magnetic transition at low temperature associated with localization of the holes.  Beyond that, the magnetic ordering has a spin-glass character, as indicated by bulk susceptibility studies \cite{chou95}.  Some local magnetic order can even coexist with superconductivity.

From these results it is clear that holes frustrate antiferromagnetic order; however, they do not kill local magnetic correlations.  To get a better view of the surviving correlations, we need to turn to neutron scattering studies.

\subsection{Magnetic excitation spectrum in doped cuprates}

\begin{figure}[t]
  \includegraphics[width=1\textwidth]{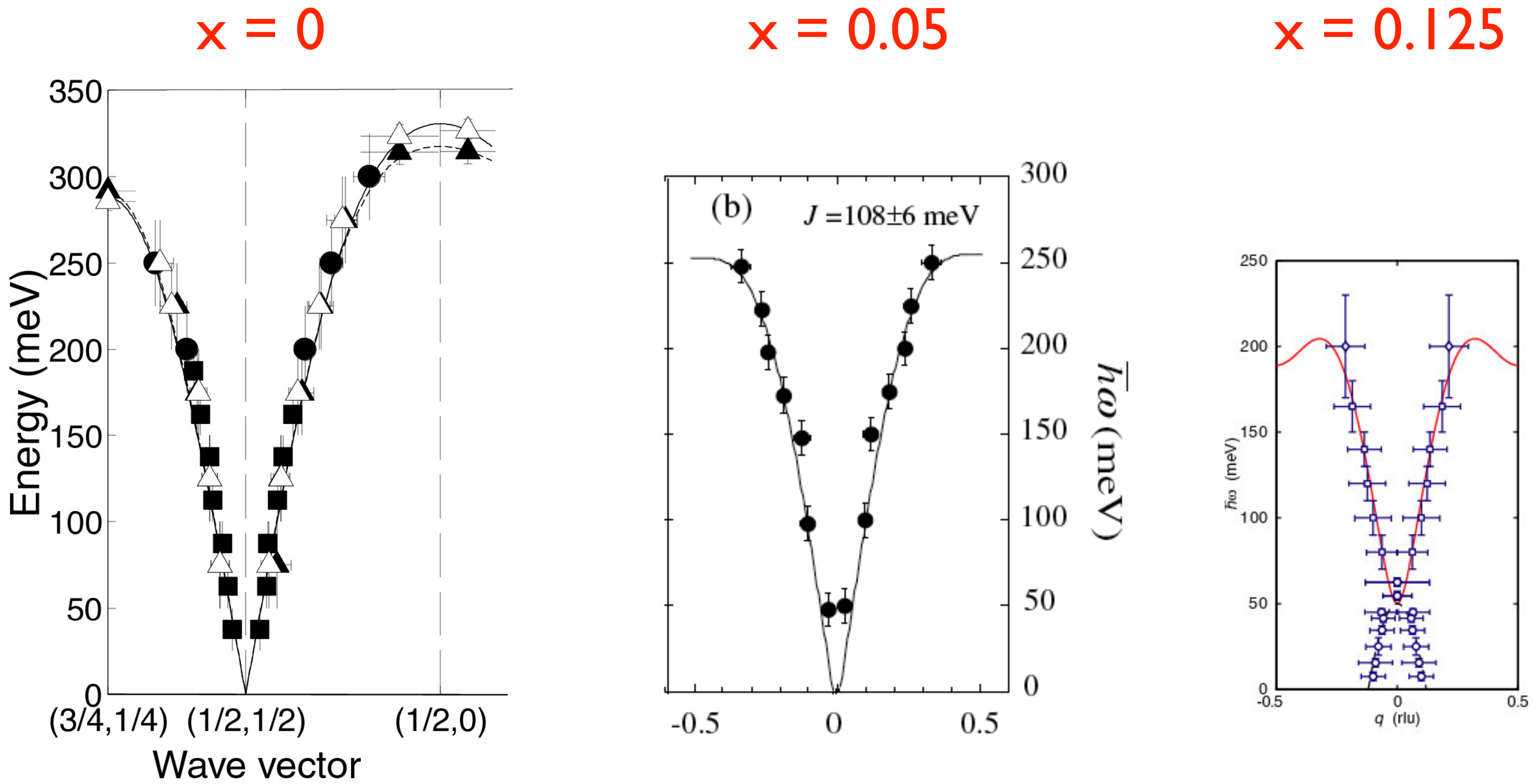}
  \caption{Comparison of effective magnetic dispersions measured by inelastic neutron scattering for: (left) La$_2$CuO$_4$ from Coldea {\it et al.} \cite{cold01}; (middle) La$_{1.95}$Sr$_{0.05}$CuO$_4$ from Goka {\it et al.} \cite{goka03} (reprinted with permission from Elsevier); (right) La$_{1.875}$Ba$_{0.125}$CuO$_4$ \cite{tran04}.  In each case the dispersion is measured about the AF wave vector.}
  \label{fg:compare}
\end{figure}

Another perspective on the magnetic correlations of the hole-doped planes is given by measurements of the magnetic excitations obtained by inelastic neutron scattering.  Some examples, measured about the AF wave vector, ${\bf Q}_{\rm AF}$,  are shown in Fig.~\ref{fg:compare} for La$_{2-x}$(Sr,Ba)$_x$CuO$_4$.  At the higher energies, one can see that AF-like spin waves survive, though the effective strength of the superexchange is renormalized downwards.  Note that the effective bandwidth of the excitations corresponds to $\sim2J_{\rm eff}$, and this estimate of $J_{\rm eff}$ is consistent with the analysis of the magnitude of the bulk spin susceptibility in Fig.~\ref{fg:lsbco_chi}.

One change that is apparent in the right-hand panel of Fig.~\ref{fg:compare} is that, below an energy scale $E_{\rm cross}$, the magnetic dispersion is associated with wave vectors split incommensurately about ${\bf Q}_{\rm AF}$.  Studies have shown that both the direction and magnitude of the incommensurability are doping dependent.  

\begin{figure}[t]
  {\includegraphics[width=0.4\textwidth]{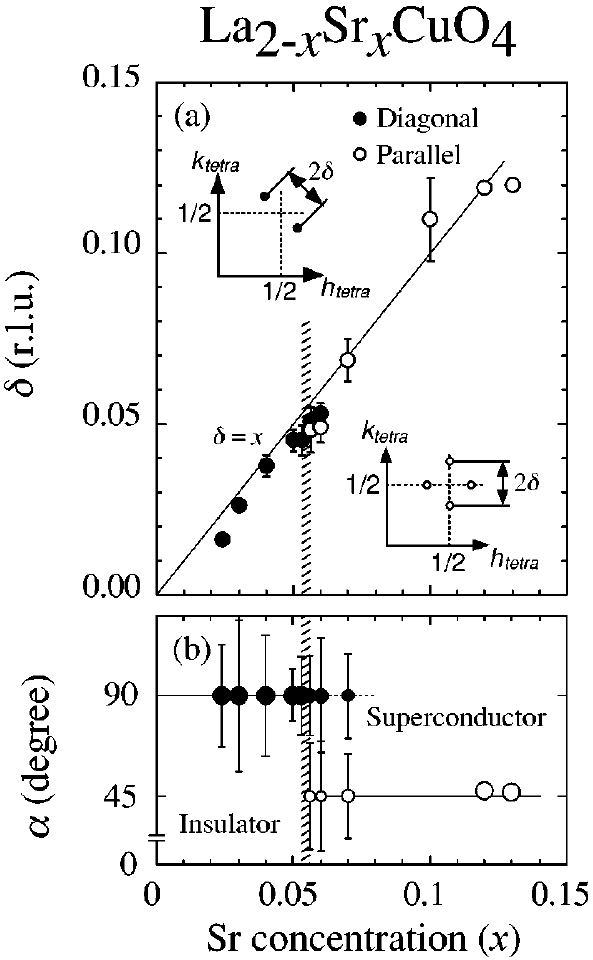}\hskip50pt
  \includegraphics[width=0.4\textwidth]{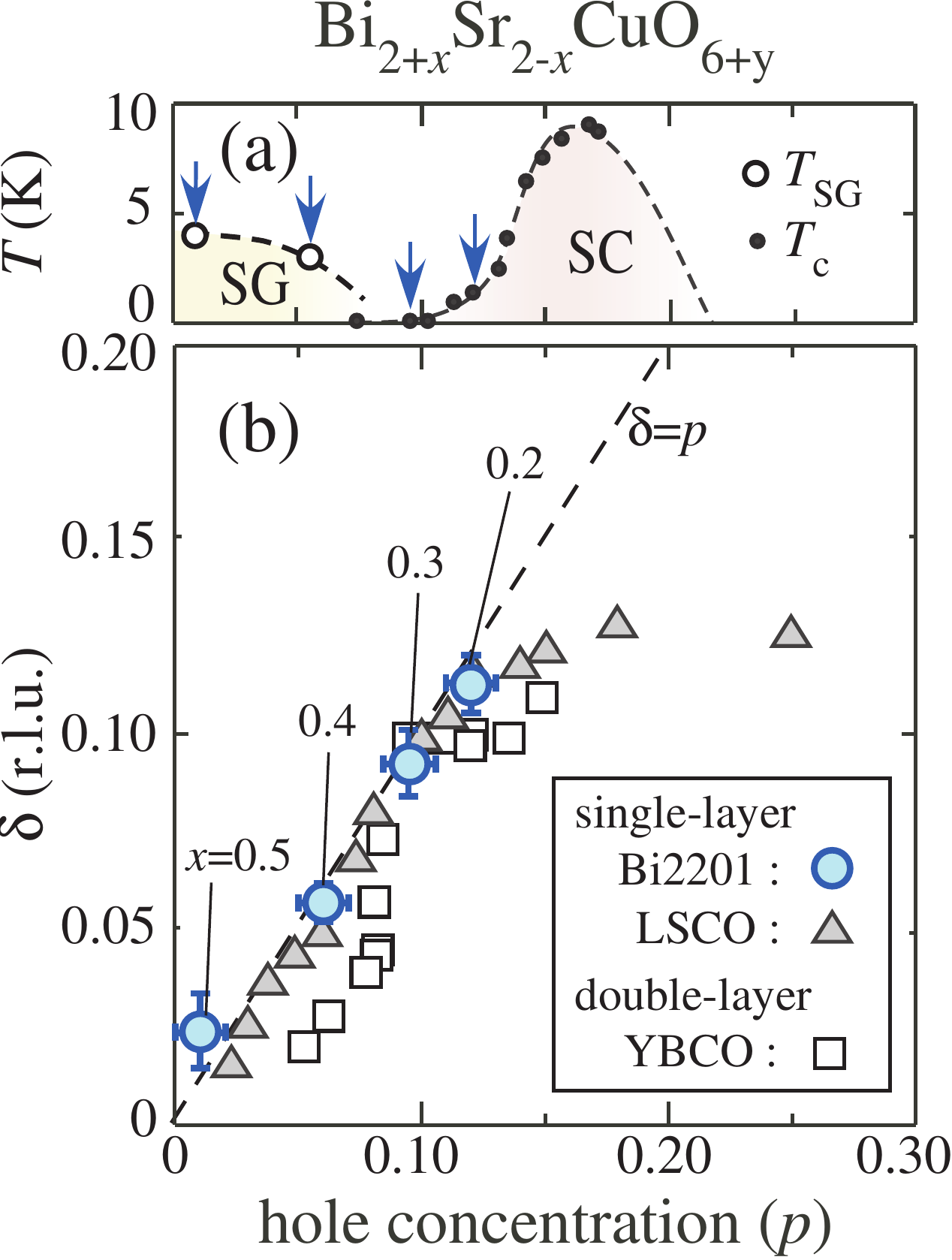}}
  \caption{(Left) (a) Plot of incommensurability $\delta$ vs.\ $x$ for La$_{2-x}$Sr$_x$CuO$_4$ (LSCO), from Fujita {\it et al.} \cite{fuji02c}.  Insets indicate orientation of incommensurate peaks for $x\lesssim0.055$ (upper left) and for $x\gtrsim0.055$ (lower right). (b) Orientation angle of the incommensurate peaks.  (Right) (a) Phase diagram of Bi$_{2+x}$Sr$_{2-x}$CuO$_{6+y}$ (Bi2201), showing spin glass (SG) and superconducting (SC) phases.  (b) Incommensurability vs.\ hole concentration for Bi2201 (circles), LSCO (triangles), and YBa$_2$Cu$_3$O$_{6+x}$ (YBCO) (squares).  The orientation of the incommensurate peaks in Bi2201 rotates from the SG to the SC phase just as in LSCO.  From Enoki {\it et al.} \cite{enok13}.}
  \label{fg:mag_inc}
\end{figure}

The nature of the incommensurability has been studied in detail for La$_{2-x}$Sr$_x$CuO$_4$ by neutron scattering \cite{fuji02c}. The left-hand side of Fig.~\ref{fg:mag_inc} summarizes the results.  For $x\lesssim0.055$, the system is insulating at low temperature, and elastic scattering is observed at incommensurate peaks split about ${\bf Q}_{\rm AF}$ along a diagonal direction, as indicated by the upper inset in (a); the modulation is uniquely along the orthorhombic $b$ axis.  As the doping level is increased to $x\gtrsim0.055$, the system becomes superconducting, the elastic scattering strongly weakens, and the low-energy spin excitations are oriented parallel to the Cu-O bond directions, as shown in the lower-right inset of (a).  While the orientation of the peaks rotates with doping, the magnitude $\delta$ of the incommensurability remains very close to $x$ across the transition.

Evidence that this trend is universal is presented on the right-hand side of Fig.~\ref{fg:mag_inc}.  The incommensurability measured vs.\ hole concentration in Bi$_{2+x}$Sr$_{2-x}$CuO$_{6+y}$ is quantitatively identical to that in La$_{2-x}$Sr$_x$CuO$_4$, and the behavior in YBa$_2$Cu$_3$O$_{6+x}$ is qualitatively quite similar.

\begin{figure}[t]
 { \includegraphics[width=0.35\textwidth]{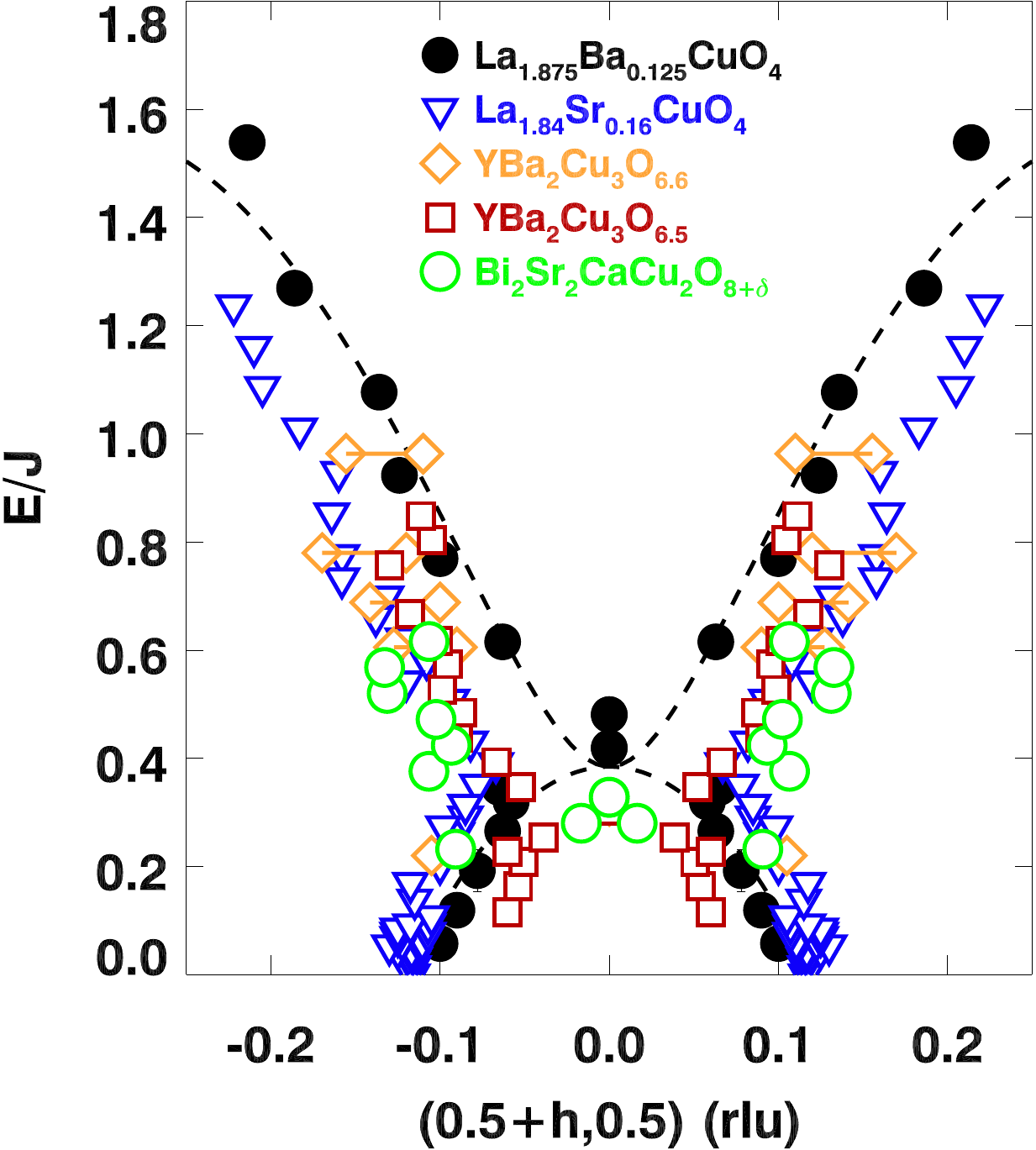}\hskip20pt
  \includegraphics[width=0.6\textwidth]{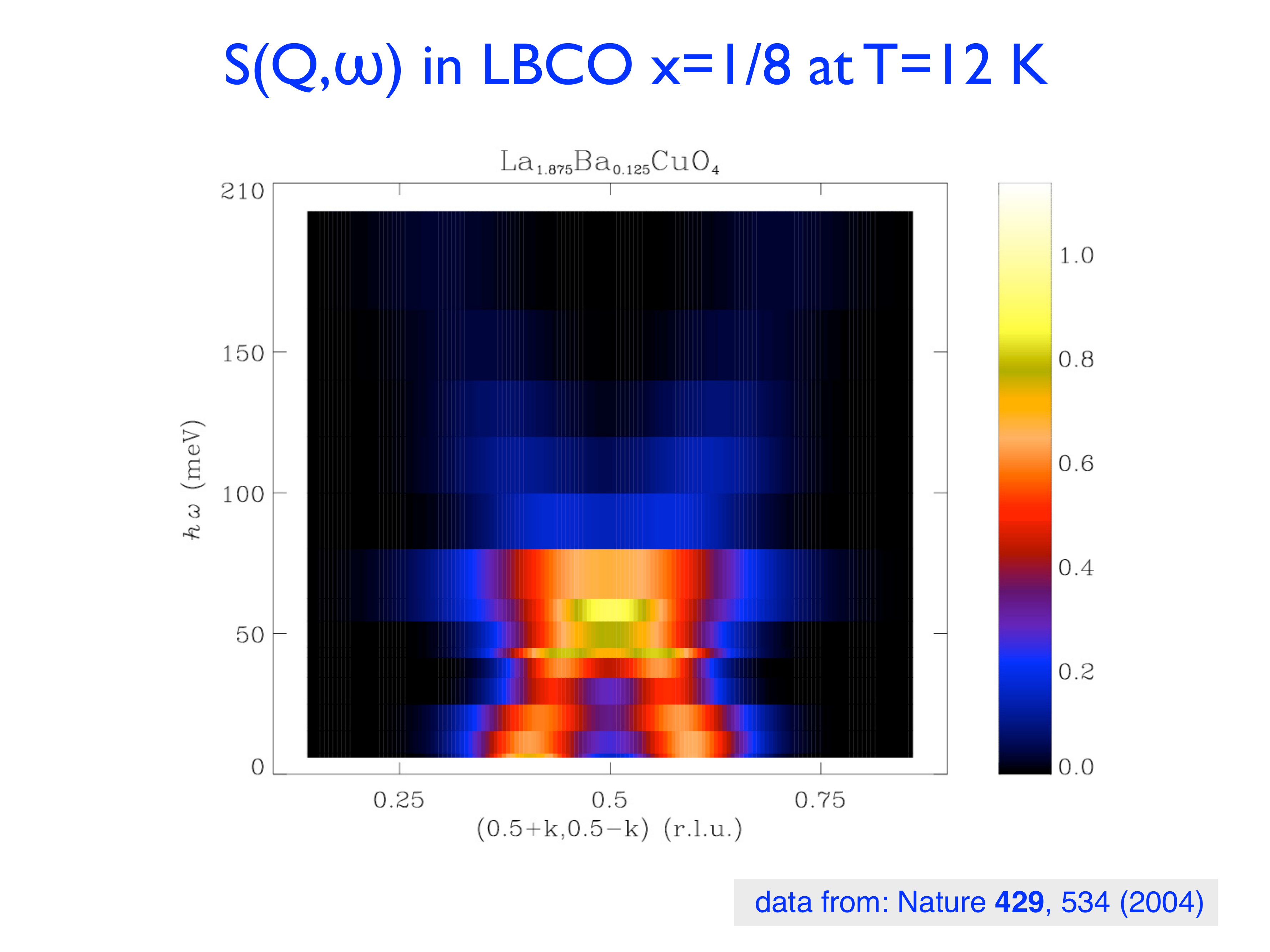}}
  \caption{(Left) Comparison of magnetic dispersion along $(0.5+h, 0.5, 0)$ in several cuprate families using data from the literature \cite{tran04,vign07,hayd04,stoc05,stoc10,xu09}, with the energy scale normalized to $J$ for the appropriate parent compound \cite{tran07,suga03}. From Fujita {\it et al.} \cite{fuji12a}.  (Right) Plot of magnetic ${\cal S}({\bf Q},\omega)$ reconstructed from fits to neutron scattering measurements on La$_{1.875}$Ba$_{0.125}$CuO$_4$ \cite{tran04,xu07}, courtesy of Guangyong Xu.}
  \label{fg:hourglass}
\end{figure}

The hourglass-like nature of the dispersion at higher-energies is also rather universal.  The left-hand side of Fig.~\ref{fg:hourglass} compares magnetic dispersions for several cuprate families, all of which are quite similar when one normalizes the energy scale to the superexchange $J$ for the appropriate parent compound.

Of course, plotting the effective dispersion can be a bit misleading, as the observed scattering is quite broad in {\bf Q} and the intensity varies greatly with energy.  The right-hand side of Fig.~\ref{fg:hourglass} presents an approximate version of the experimental ${\cal S}({\bf Q},\omega)$ for magnetic excitations in La$_{1.875}$Ba$_{0.125}$CuO$_4$ \cite{tran04,xu07}.  As one can see, there is a lot of weight near $E_{\rm cross} \sim 50$~meV.  Because of the outward dispersion at higher energies, the peak signal falls off much faster than the integrated intensity.  The weight at $E < E_{\rm cross}$ is also reduced.

\begin{figure}[t]
  \includegraphics[width=0.4\textwidth]{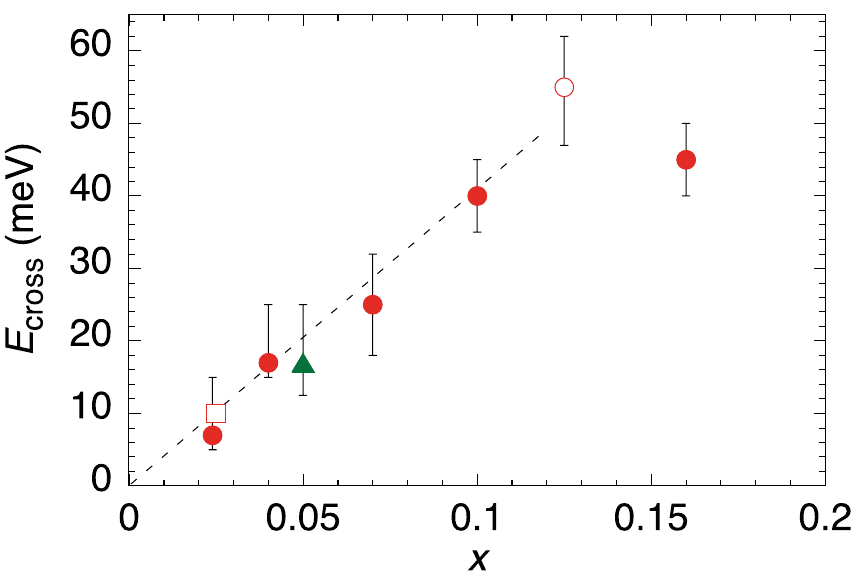}
  \caption{Plot of $E_{\rm cross}$ vs. $x$ in LSCO (filled circles and open square), Zn-doped LSCO (filled triangle), and La$_{2-x}$Ba$_x$CuO$_4$ (open circle) from Matsuda {\it et al.} \cite{mats11}. The dashed line is a guide for the eye.}
  \label{fg:ecross}
\end{figure}

Given that the low-energy incommensurability is doping dependent, one might expect that $E_{\rm cross}$ should be doping dependent, as well.  This is indeed the case, as shown in Fig.~\ref{fg:ecross}.  $E_{\rm cross}$ grows linearly with doping until one approaches optimal doping.

\subsection{Magnetic spectral weight}

\begin{figure}[t]
  {\includegraphics[width=0.47\textwidth]{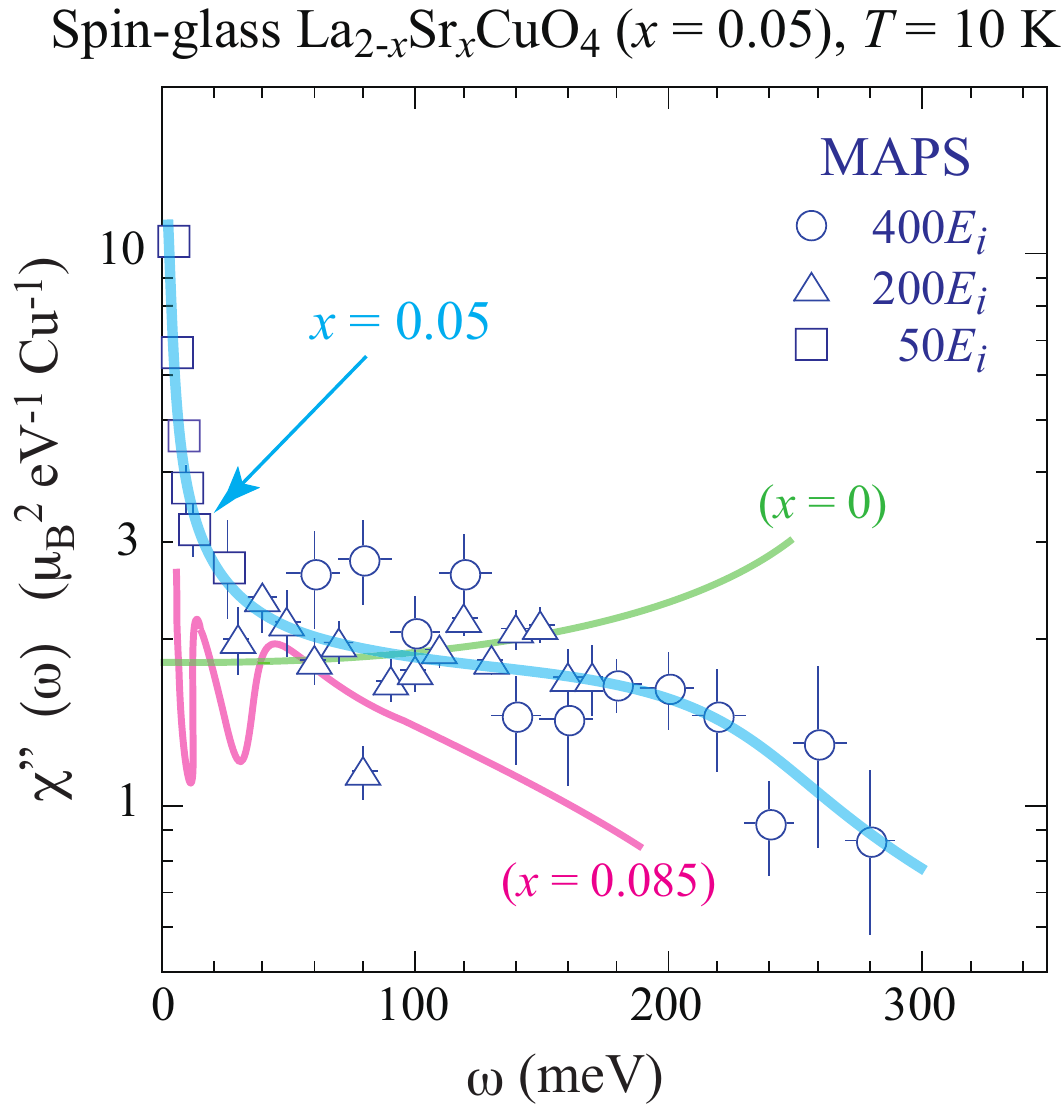}\hskip20pt
   \includegraphics[width=0.5\textwidth]{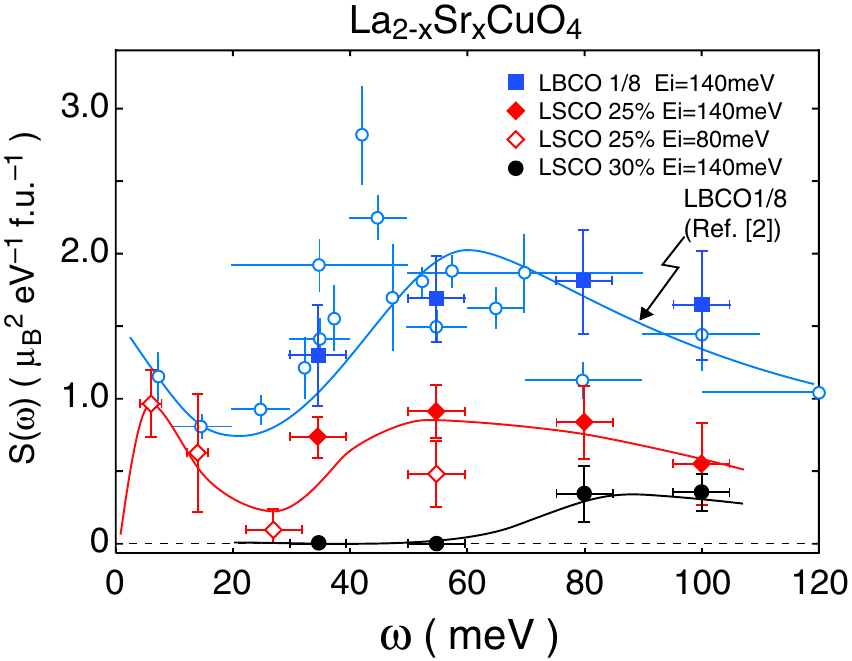}}
  \caption{(Left) Plot of magnetic spectral weight for La$_{2-x}$Sr$_x$CuO$_4$; symbols are results of $x=0.05$, with line as a guide to the eye; curves for $x=0$ and $x=0.085$ are from other work as discussed by Fujita {\it et al.} \cite{fuji12a}. (Right) Magnetic spectral weight for overdoped samples, with $x=0.25$ and 0.30, compared with LBCO ($x=1/8$), with lines as guides to the eye, from Wakimoto {\it et al.} \cite{waki07b}.  All results are at low temperature.}
  \label{fg:mag_wt}
\end{figure}

Besides dispersion, one can also consider how the magnetic spectral weight evolves with doping.  It has become common to express this in terms of the imaginary part of the local spin susceptibility, $\chi''(\omega)$, obtained by averaging $\chi''({\bf Q},\omega)$ over a Brillouin zone.   Examples of $\chi''(\omega)$ for several doping levels of LSCO are shown on the left-hand side of Fig.~\ref{fg:mag_wt}.   The result for La$_2$CuO$_4$ shown there includes only the weight from the spin waves, and does not include the signal associated with elastic scattering in the antiferromagnetic Bragg peak.  For $x=0.05$, the peak at low frequency can be viewed as a quasi-elastic version of the weight in the Bragg peak at $x=0$.  Doping into the spin-glass regime frustrates the commensurate order, but the spins are still close to ordering.  The magnitude of $\chi''(\omega)$ at 100 meV is very similar to that in the antiferromagnet, but it eventually falls away at higher doping.  The curve for $x=0.085$ \cite{lips09}, again, is close to the antiferromagnet at lower energies, and then falls away.

The fall of of spectral weight is especially significant for overdoped samples.  The right-hand side of Fig.~\ref{fg:mag_wt} shows the loss of low-energy spectral weight in highly overdoped LSCO compared to LBCO with $x=1/8$. 

\begin{figure}[t]
  \includegraphics[width=0.4\textwidth]{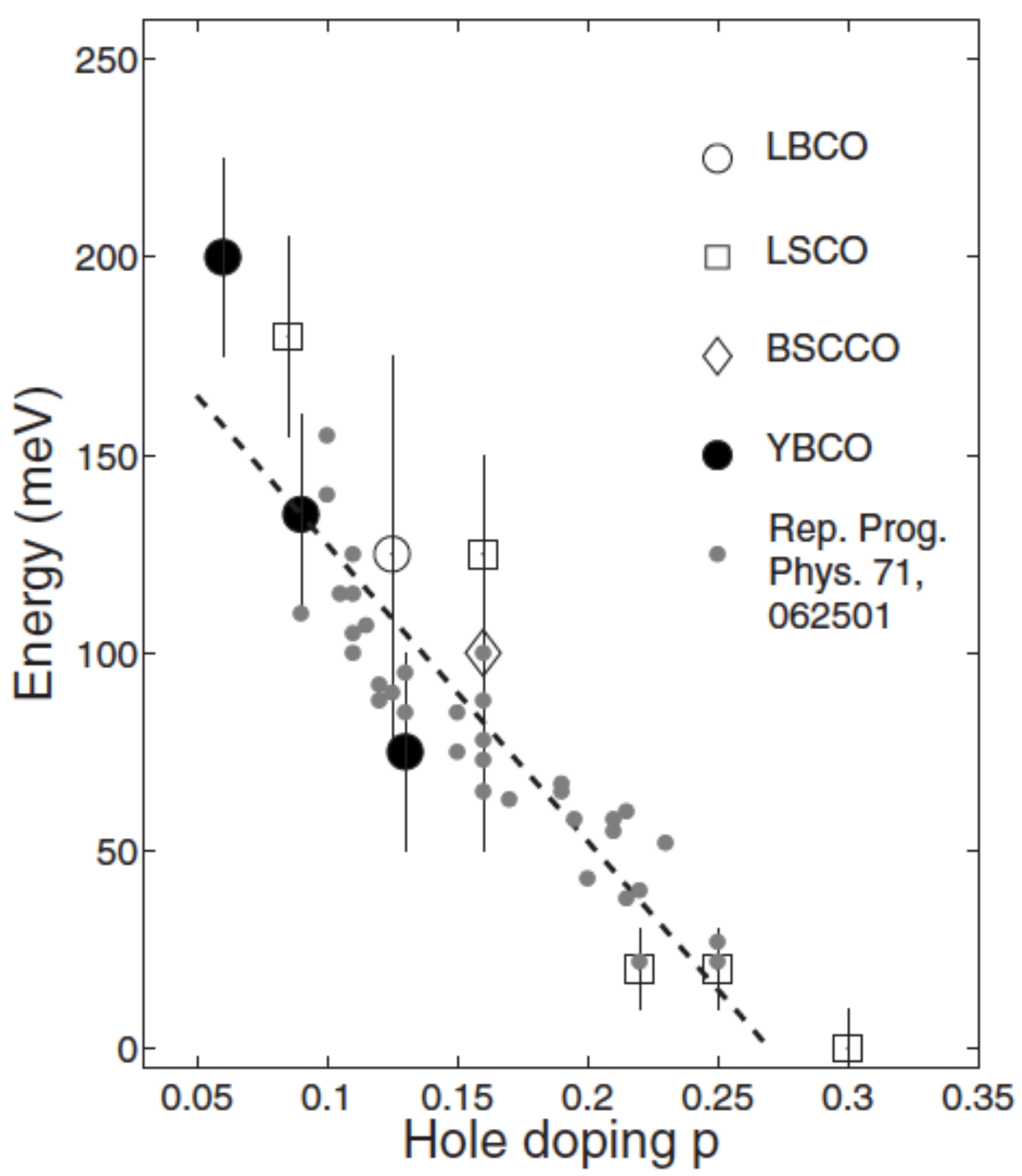}
  \caption{Energy scale at which magnetic spectral weight falls below half that of the antiferromagnetic parent, evaluated from neutron scattering results on a number of cuprate superconductors by Stock {\it et al.} \cite{stoc10}.  The dots are from \cite{hufn08}.}
  \label{fg:Stock}
\end{figure}

Stock and workers \cite{stoc10} were the first to address this trend.  Looking over all available data, they attempted to identify, for each sample, the energy at which the magnetic spectral weight falls to half of that for the antiferromagnet.  The results are indicated by the large symbols in Fig.~\ref{fg:Stock}.  Note that this plot includes results from 3 different families of superconductors.  The small points are various electronic spectroscopic measures (from ARPES, Raman scattering, etc.) of the pseudogap energy from \cite{hufn08}.   This plot identifies an intriguing competition between antiferromagnetic and electronic excitations.  The antiferromagnetic correlations are strong {\it below} the pseudogap energy, but rapidly lose weight above it.  In contrast, the electronic excitations are strong {\it above} the pseudogap energy and weak below it.

\subsection{2-Magnon Raman scattering}

Another measure of magnetic correlations is provided by 2-magnon Raman scattering.  Raman scattering is a resonant process, involving virtual absorption of the incident photon associated with an electron-hole excitation and reemission of a photon from a de-excitation process.\footnote{For a review, see Devereaux and Hackl \cite{deve07}.}  If the sample is left in a different state than it started in, then the scattering process is inelastic.  The scattering cannot couple to a single spin-flip, but it can couple to one spin-flip in the absorption process and another on the de-excitation process.  The Raman cross section is sensitive to the orientations of the polarizations of the incident and scattering photons with respect to crystal axes.  For an antiferromagnet, most of the 2-magnon response appears in the channel with $B_{1g}$ symmetry, with the peak intensity occurring at $\sim3J$.  The nature of the 2-magnon process could change when antiferromagnetic CuO$_2$ layers are doped, but the energy of the peak intensity provides a measure of the effective $J$ describing the magnetic correlations.

\begin{figure}[t]
  \includegraphics[width=0.8\textwidth]{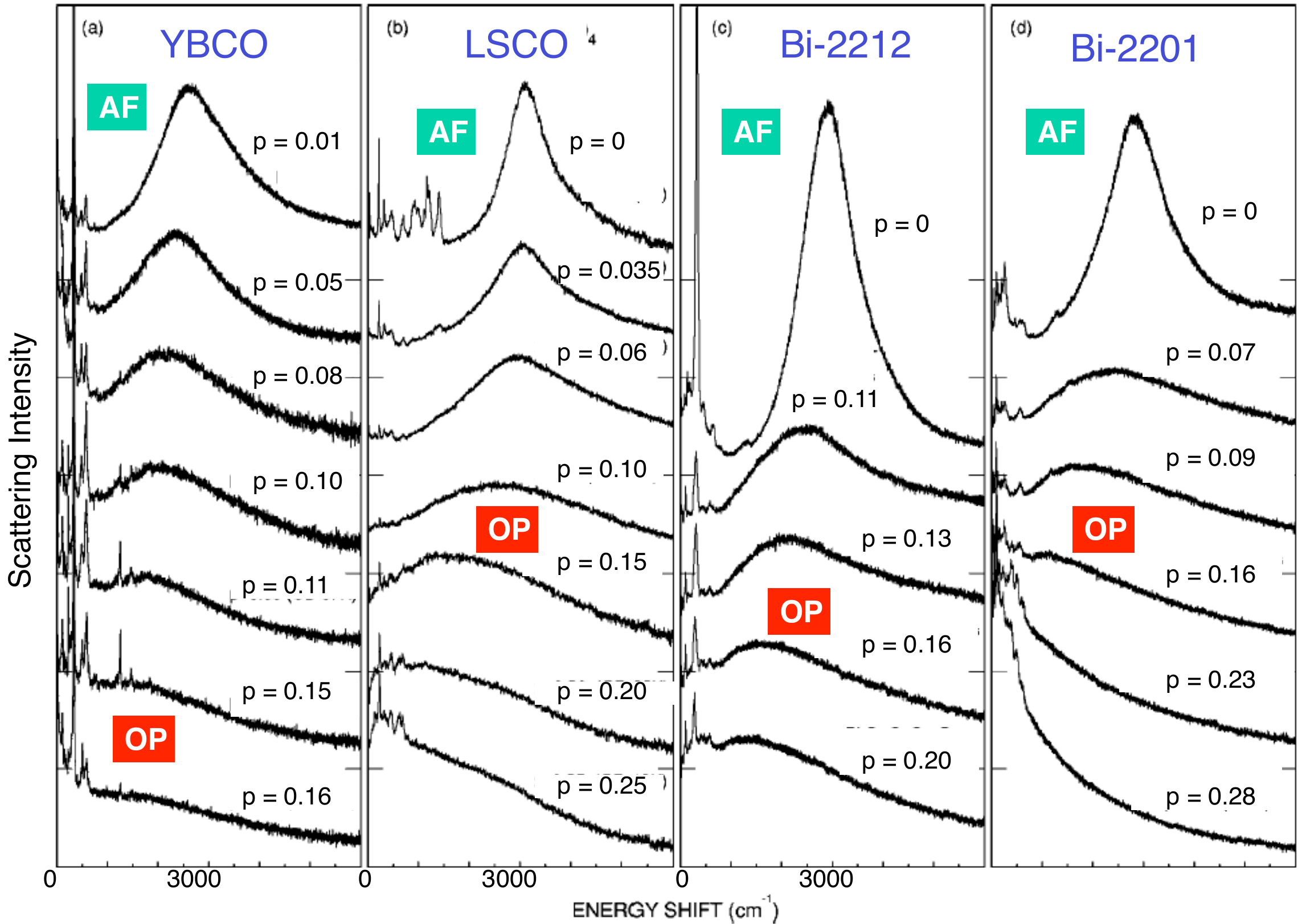}
  \caption{$B_{1g}$ Raman spectra for several cuprate families as a function of doping, from Sugai {\it et al.} \cite{suga03}.  One can see the evolution of the 2-magnon peak; sharp peaks at small energy are from phonons.}
  \label{fg:raman}
\end{figure}

Figure~\ref{fg:raman} shows $B_{1g}$-symmetry Raman spectra for four cuprate families as a function of doping, from Sugai {\it et al.} \cite{suga03}.  For each of the antiferromagnetic parent compounds, one can see a strong 2-magnon peak centered at 2500--3000 cm$^{-1} \approx 310$--380~meV.   With doping the peak shape broadens and the peak energy decreases.  By optimal doping, a peak is just resolvable, and with overdoping the peak is completely overdamped.   This trend is quite consistent with the previously-discussed neutron scattering results in Fig.~\ref{fg:Stock}.

The clear trend is that the magnetic correlations present in the superconducting cuprates are related to the antiferromagnetism present in the parent materials, which are correlated insulators.  The mobile carriers frustrate commensurate order, and they lead to damping of the high-energy magnetic excitations.   We have metallic materials that retain symptoms ({\it i.e.}, antiferromagnetic correlations) of correlated-insulator character.  The struggle to understand this state has been one of the major challenges of the field.

\subsection{Spin resonance}

\begin{figure}[t]
  \includegraphics[width=0.4\textwidth]{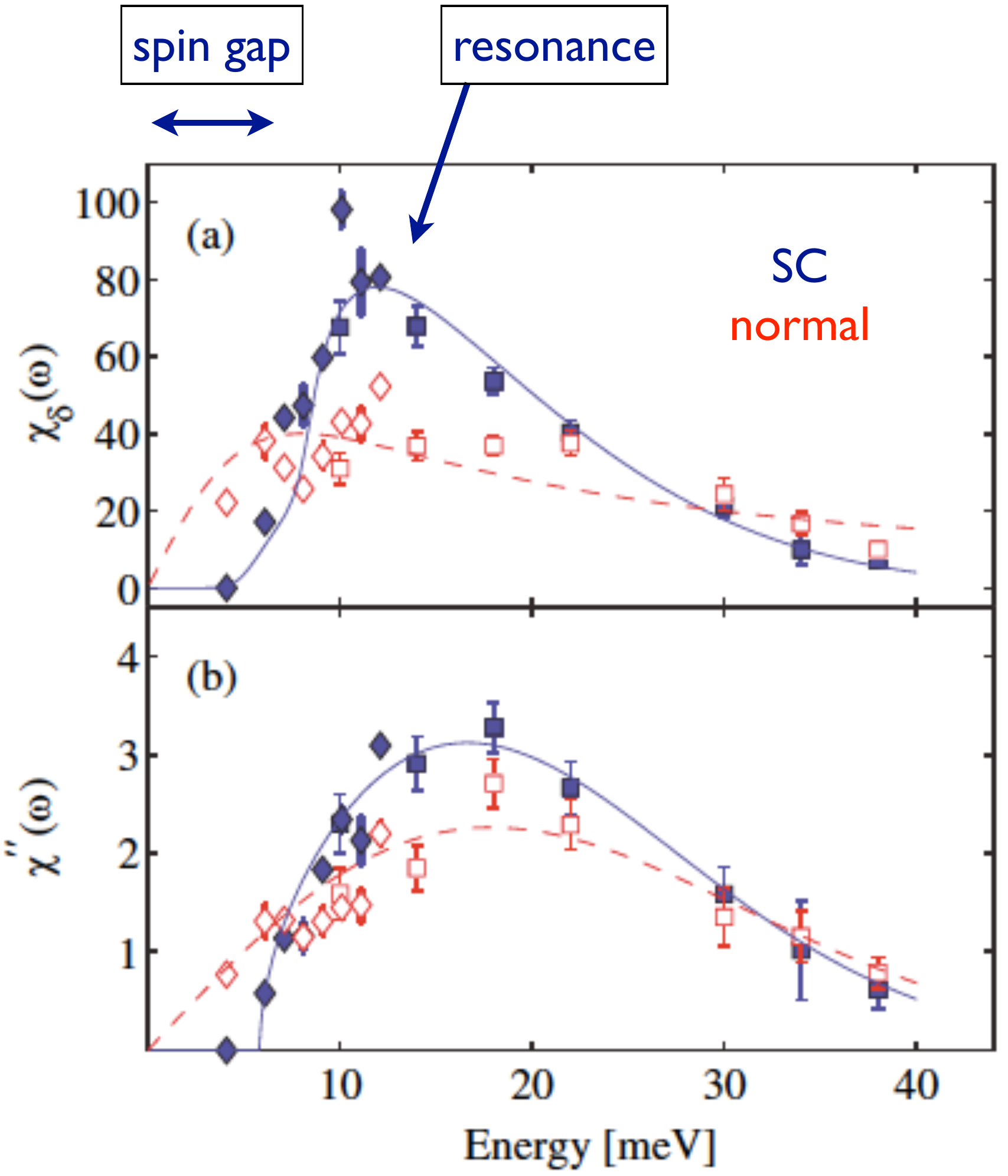}
  \caption{Neutron scattering results for LSCO with $x=0.16$ ($T_c=38.5$~K) in the normal (open symbols) and superconducting (filled symbols) states, from Christensen {\it et al.} \cite{chri04}. (a) $\chi''({\bf Q},\omega)$ at the incommensurate peak positions; (b) local susceptibility $\chi''(\omega)$.  Lines are guides to the eye.}
  \label{fg:reso}
\end{figure}

Given the substantial interaction between the doped holes and the antiferromagnetic spin correlations that we have already seen, it should not be surprising that the magnetic spectrum is modified when a sample is cooled into the superconducting state.  At least for optimally-doped cuprates, with $T<T_c$ one observes a gap in the spin excitations and a pile up of weight into a spin resonance peak above the spin gap.  An example is shown in Fig.~\ref{fg:reso} for LSCO with $x=0.16$ from Christensen {\it et al.} \cite{chri04}.

\begin{figure}[t]
  \includegraphics[width=0.4\textwidth]{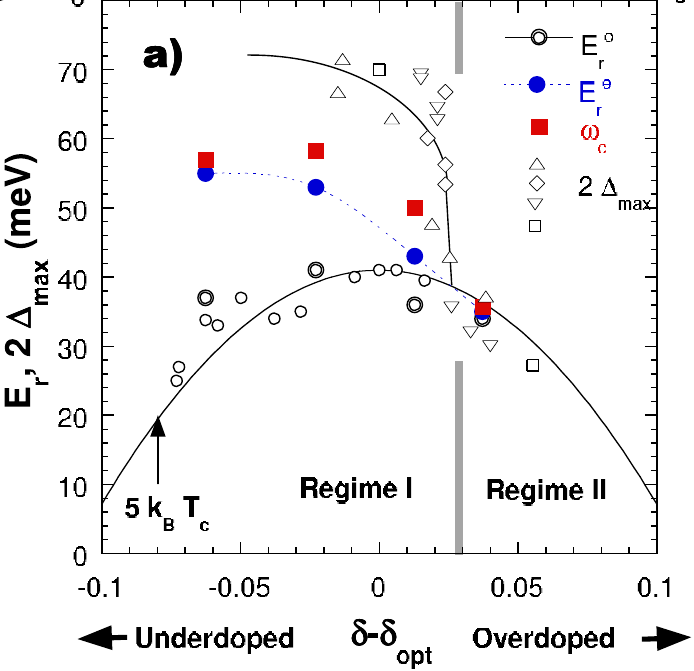}
  \caption{Plot of $E_r$ (labelled $E_r^o$ here) for YBCO from neutron scattering (open circles), $2\Delta_{\rm max}$ from electronic Raman scattering (open triangles, diamonds, squares), and $5kT_c$ (indicated by arrow) in YBCO as a function of doping, relative to $n_{\rm opt}\equiv0.16$,  from Pailh\`es {\it et al.} \cite{pail06}. See original paper for definitions of $E_r^e$ and $\omega_c$.}
  \label{fg:resovp}
\end{figure}

The spin resonance was first identified in optimally-doped YBCO by Rossat-Mignod {\it et al.} \cite{ross91c}, and is also apparent in underdoped YBCO \cite{dai99}.   Figure~\ref{fg:resovp} compares measurements of the spin-resonance energy, $E_r$, with twice the superconducting gap maximum, $2\Delta_{\rm max}$, as a function of doping in YBCO \cite{pail06}.\footnote{For a plot including results for Bi$_2$Sr$_2$CaCu$_2$O$_{8+\delta}$, see Sidis {\it et al.} \cite{sidi04}.}  The trend is that $E_r/kT_c\sim 5$.  $E_r$ is always less than $2\Delta_{\rm max}$, but the ratio varies with doping.  In contrast, Yu {\it et al.} \cite{yu09} have argued for a universal ratio, $E_r/2\Delta_{\rm max}=0.64$, for a broad range of superconductors, including heavy-fermion and Fe-based superconductors.

\begin{figure}[t]
  \includegraphics[width=0.6\textwidth]{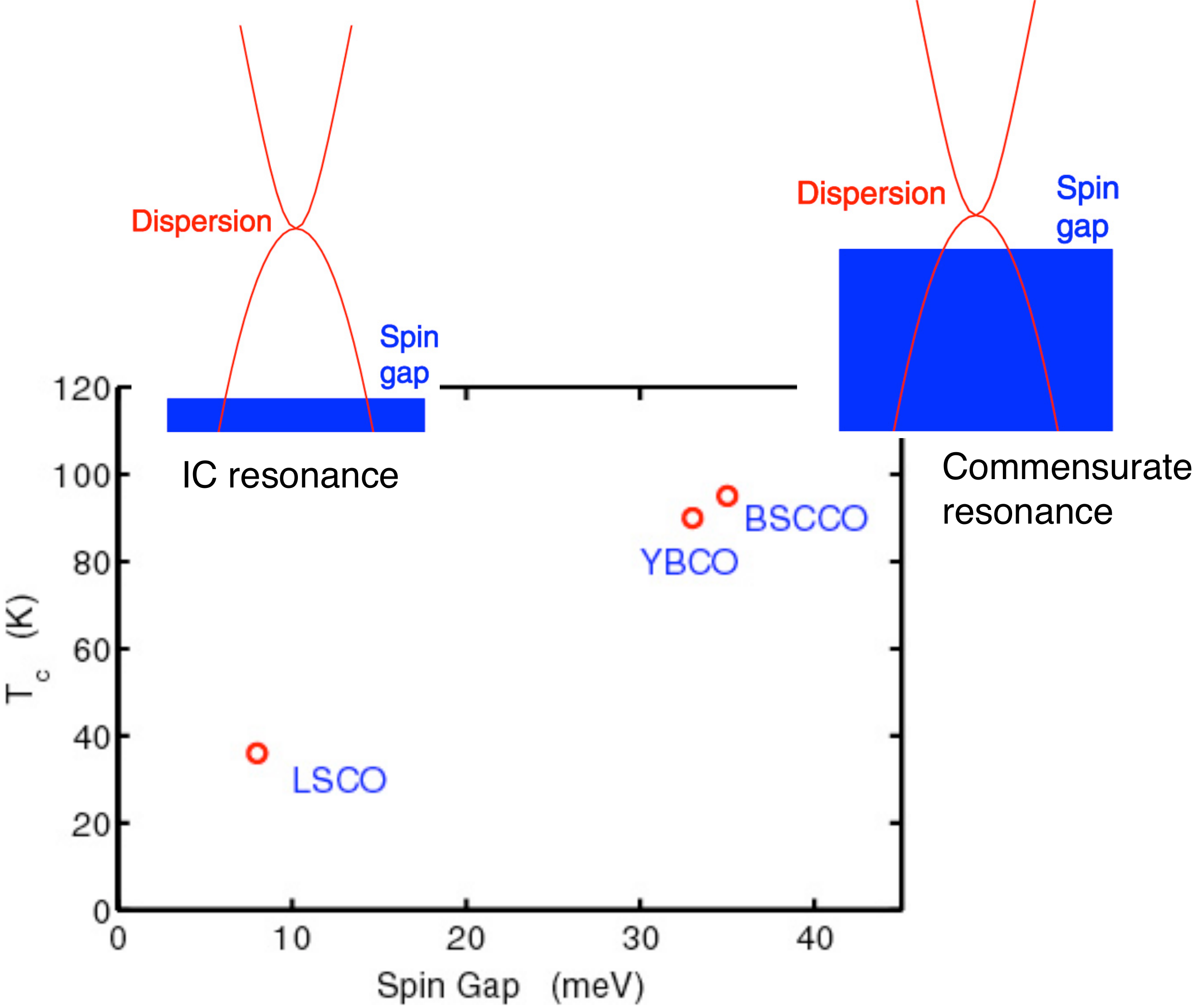}
  \caption{Plot of $T_c$ vs.\ spin gap for three optimally-doped cuprates, from \cite{tran07}.  Left inset indicates the case of LSCO, where the spin gap energy is much less than $E_{\rm cross}$, while the right inset indicates the situation for YBCO and BSCCO, where the two scales are much closer in energy, and $E_r\approx E_{\rm cross}$.}
  \label{fg:tc_vs_gap}
\end{figure}

It is also of interest to consider the magnitude of the spin gap and its relationship to $E_{\rm cross}$.  Figure~\ref{fg:tc_vs_gap} shows $T_c$ as a function of spin gap energy for three optimally doped cuprates.  The relationship is roughly linear.  In contrast, $E_{\rm cross}$ is in the range of 40--50~meV for all three compounds.  As a consequence, the appearance of the spin resonance can vary significantly.  For YBCO and BSCCO, $E_r\approx E_{\rm cross}$, so that the resonance is centered commensurately at ${\bf Q}_{\rm AF}$, whereas $E_r\ll E_{\rm cross}$ for LSCO, so that the resonance appears only at incommensurate wave vectors.

\section{Spin and charge stripe order in layered transition-metal oxides}

We have seen that doping holes into the CuO$_2$ planes frustrates long-range AF order while allowing dynamic AF spin correlations to survive.  In this section, we will consider how spatial segregation of holes and spins, in terms of periodic stripes, can explain such behavior.

\subsection{Note on models vs.\ reality}

Since conventional density-functional calculations failed to capture the strong correlation effects in lightly-doped LSCO, one must turn to model Hamiltonians, such as the Hubbard, to more directly address the impact of strong, poorly-screened Coulomb interactions.  Given the analysis of Zaanen, Sawatzky, and Allen \cite{zaan85}, it would appear that a 3-band Hubbard model (including one Cu $3d_{x^2-y^2}$ and two O $2p_\sigma$ orbitals) would be the minimum necessary to describe the charge-transfer character of the optical gap in La$_2$CuO$_4$ and the O $2p$ character of the doped holes seen by x-ray spectroscopy.  While such a model has been advocated by some, the challenges of performing calculations beyond mean field on anything more than a very small cluster have provided the motivation to consider further simplifications.

For example, it is common to start with a 1-band Hubbard model, focusing just on the Cu $d$ state and hoping that the impact of ignoring the oxygen states does not qualitatively change the physics.  For those who believe that the large $U$ limit is the relevant parameter range, it is common to make a further simplification, moving to the $t$-$J$ model, which keeps the kinetic energy term, adds a Heisenberg interaction between neighboring spins, and takes account of Coulomb repulsion by projecting out any states with double occupancy.

All of these models are challenging to solve, as a glance at the very large literature will quickly confirm.  Analysis of these models has provided various insights, and progress with treating any one model can benefit the field of many-body theory.  At the same time, care is needed when comparing calculations on models to experiments on real materials.  

A matter of continuing controversy is the character of the minimal model required to capture the essence of high-temperature superconductivity.  The debate began early, with Zhang and Rice \cite{zhan88} suggesting that a doped hole can be treated as forming a singlet state with a Cu $d$ hole, and Emery and Reiter \cite{emer88} countering that a 3-band model allows an oxygen hole to create a triplet between neighboring Cu $d$ states.  This debate continues today.  M\"oller, Sawatzky, and Berciu \cite{moll12a,moll12b} have recently performed an interesting analysis of a one-dimensional model in which they compare one-band and two-band (in 1D, one gives up one of the oxygen bands) models.  To allow exact solutions, they consider the energy of two carriers in a ferromagnetic background, and find evidence for pair binding in two-band, but not in one-band, models.  They expect that their results should also have relevance to the case of an AF background.  Their results lead them to "question the ability of simple one-band models to accurately describe the low-energy physics of cuprate layers" \cite{moll12b}.  This is clearly not the final word on this problem, but it does motivate one to maintain an open and skeptical mind.  It is reasonable to see what insights are provided by one-band models, but one should not dismiss the possibility that qualitatively different outcomes might be possible in models that contain more of the degrees of freedom present in real materials.

\subsection{Impact of hole motion and stripe concept}

\begin{figure}[t]
  {\includegraphics[width=0.4\textwidth]{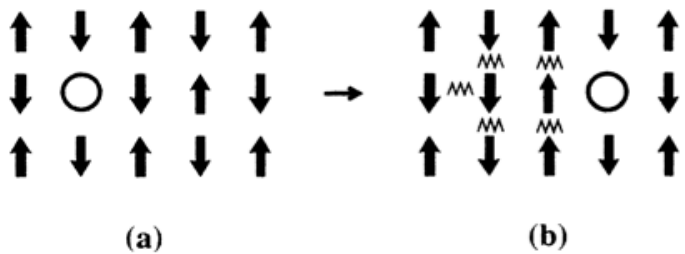}\hskip40pt
  \includegraphics[width=0.4\textwidth]{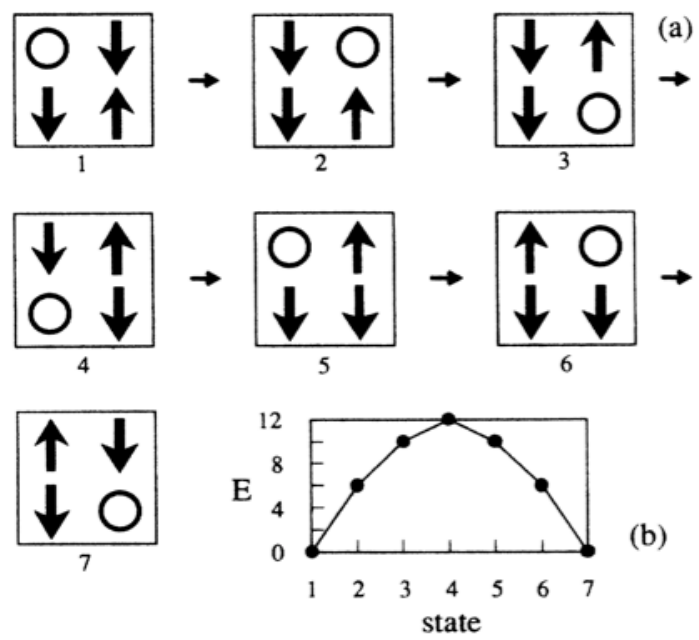}}
  \caption{Analysis of motion of one hole in an antiferromagnet described by a single-band Hubbard model with only nearest-neighbor hopping, from Trugman \cite{trug88}.  (Left) Starting with (a), moving the hole along a row of spins to the configuration (b) causes spins to be flipped, thus resulting in energetically-costly ferromagnetic bonds, indicates by squiggles.  (Right) (a) Starting at configuration 1, moving the hole around the square 1.5 times gets one to 7, which is degenerate with 1.  (b) Plot of relative energy of each configuration in units of $J$.}
  \label{fg:trug}
 \end{figure}

The impact of hole motion in an antiferromagnetic background  was analyzed early on by Trugman \cite{trug88} for a 1-band Hubbard model (with only nearest-neigbor hopping).  As shown on the left-hand side of Fig.~\ref{fg:trug}, hole motion along a row of sites creates ferromagnetic spin correlations, which cost exchange energy $J$.  In contrast, the right-hand side shows that a hole can effectively move in a diagonal direction by rotating around a square 1.5 times.  The first time around, the displacement of spins creates ferromagnetic correlations with respect to the AF background, but these are repaired the second time around.

\begin{figure}[t]
  {\includegraphics[width=0.3\textwidth]{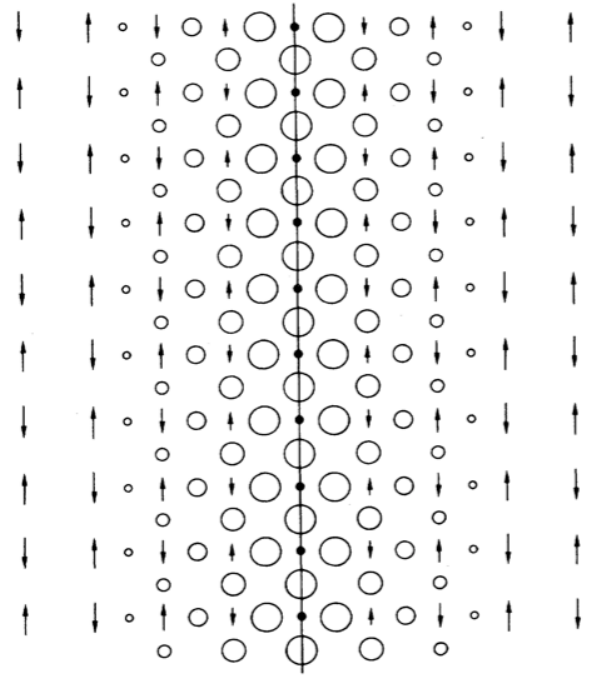}\hskip40pt
  \includegraphics[width=0.5\textwidth]{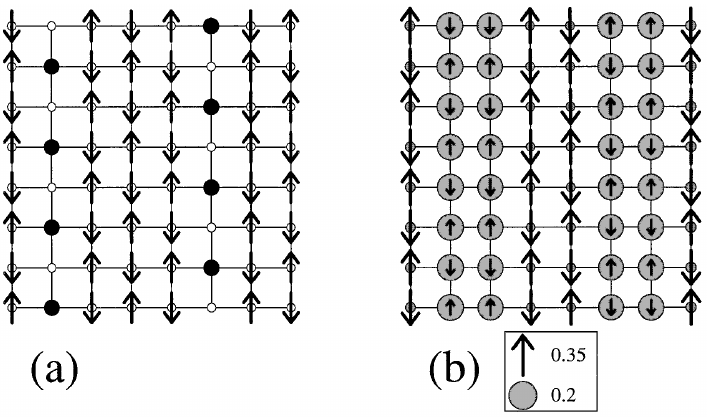}}
  \caption{(Left) Hartree-Fock calculation of a stripe in a 3-band Hubbard model by Zaanen and Gunnarsson \cite{zaan89}.  (Right) Density matrix renormalization group (DMRG) calculations of stripes in a $t$--$J$ model at $p=1/8$, with (a) site-centered stripe, (b) bond-centered stripe, from White and Scalapino \cite{whit98a}.}
  \label{fg:calc_stripe}
 \end{figure}

The first suggestion that a hole-doped cuprate layer might develop an inhomogeneous state was provided by Zaanen and Gunnarsson \cite{zaan89}, who performed a Hartree-Fock calculation on a 3-band Hubbard model.\footnote{The stripe state was discovered independently by Machida \cite{mach89}.  It was also confirmed by many other groups.}  Their stripe solution is shown on the left-hand side of Fig.~\ref{fg:calc_stripe}.  A key feature is that the phase of the AF background shifts by $\pi$ on crossing the charge stripe.  Hence, charge stripes destroy long-range AF order, but periodic stripes can lead to incommensurate spin order.  A problem with this particular solution is that the charge stripe has an effective hole density of one per Cu site along the stripe.  This corresponds to an insulating state, and it also gives a spin incommensurability that is a factor of two smaller than the incommensurability observed in early inelastic neutron scattering measurements of LSCO \cite{cheo91}.

Later studies have found stripe states that have a hole density of one-half per Cu site along the charge stripe, consistent with neutron and transport experiments.  For example, the right-hand side of Fig.~\ref{fg:calc_stripe} shows the correlations found by White and Scalapino \cite{whit98a,scal12b} using the density matrix renormalization group technique on a $t$-$J$ model at an average hole density of 1/8.  There have also been extensive calculations of stripes using the Gutzwiller approximation by Seibold {\it et al.} \cite{seib12a}; these are just a few of many calculations that have been done.

\begin{figure}[t]
  \includegraphics[width=0.6\textwidth]{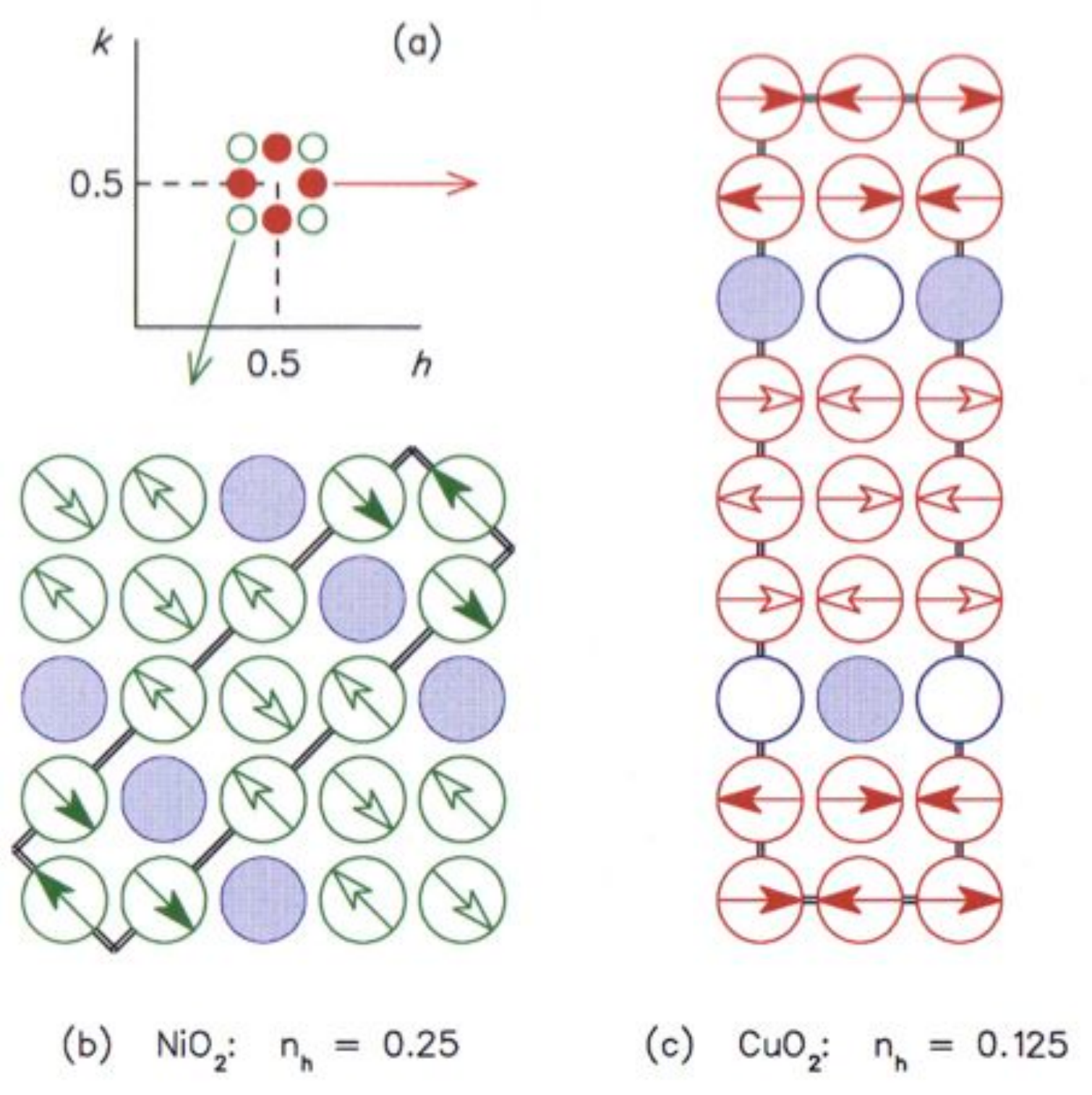}
  \caption{(a) Orientation of magnetic superlattice peaks for diagonal (open circles) and bond-parallel (filled circles) stripes.  (b) Cartoon of stripe order found in nickelates; hole density is one per Ni along a charge stripe.  (c) Cartoon of stripe order detected in cuprates; hole density is approximately one for every two Cu sites along a charge stripe.  The double lines in (b) and (c) indicate the magnetic unit cell, which is twice the size of the cell describing charge order.  From \cite{tran95a}.}
  \label{fg:diag_horiz}
 \end{figure}

Spin and charge inhomogeneity have also been proposed from another direction.  Based on early calculations for a $t$-$J$ model indicating that doped holes would tend to phase separate \cite{emer90}, Emery and Kivelson \cite{emer93} argued that inclusion of extended Coulomb interactions should frustrate the phase separation, resulting in spatially-modulated structures such as striped and checkerboard states \cite{low94}.  The density of the holes in the stripes would not be fixed, but should correspond to whatever value minimizes the free energy associated with the competing short- and long-range interactions.

Experimentally, charge and spin stripe order were first identified in an oxide with copper replaced by nickel, La$_2$NiO$_{4.125}$ \cite{tran94a}.  The stripes in that case run diagonally, as indicated in Fig.~\ref{fg:diag_horiz}(b).  The discovery of stripe order in a cuprate, La$_{1.48}$Nd$_{0.4}$Sr$_{0.12}$CuO$_4$, followed soon after \cite{tran95a}; there the stripes run parallel to the Cu-O bonds, as shown schematically in Fig.~\ref{fg:diag_horiz}(c).

We will eventually discuss the evidence for stripes in cuprates; however, stripe order is rather rare in cuprates, and the relationship with superconductivity is fairly subtle.  To establish some of the phenomenology of stripes, and the fact that they are rather common in layered transition-metal oxides, we will start off with the case of nickelates, and then briefly mention a few other examples.

\subsection{Nickelates}

\begin{figure}[t]
  {\includegraphics[width=0.45\textwidth]{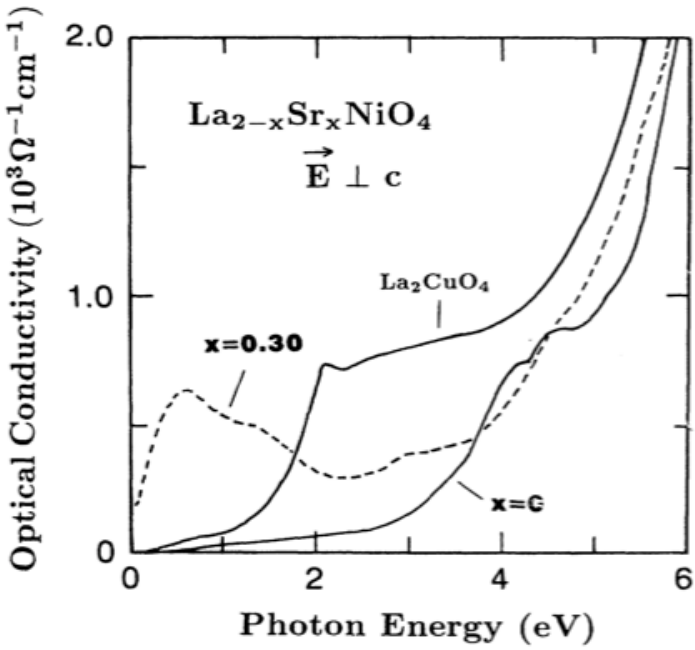}\hskip30pt
  \includegraphics[width=0.45\textwidth]{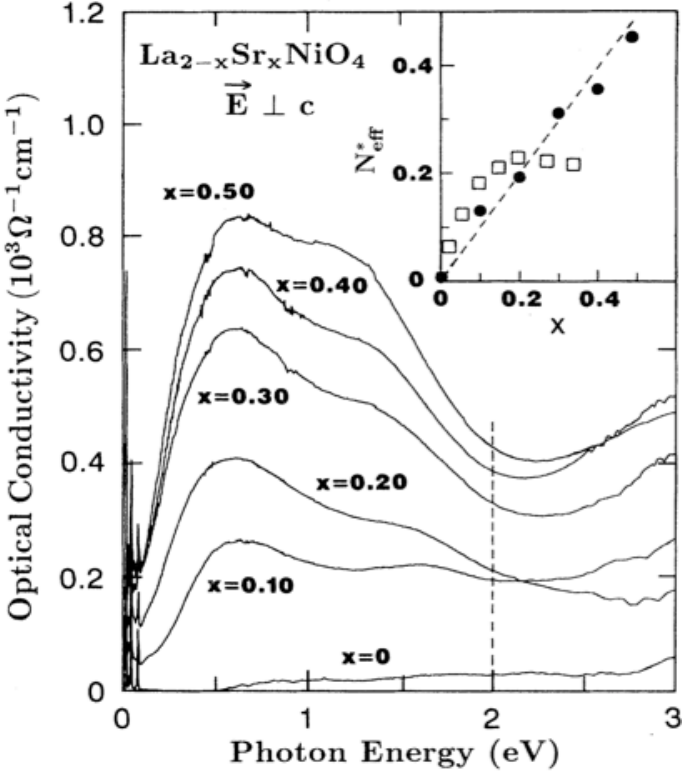}}
  \caption{(Left) In-plane optical conductivity for La$_{2-x}$Sr$_x$NiO$_4$ with $x=0$ and 0.30, compared with La$_2$CuO$_4$.  (Right) Room-temperature spectra as a function of doping; inset shows effective carrier density obtained by integrating the optical conductivity up to 2 eV for LSNO (filled circles) and up to 1 eV for LSCO (open squares).  Both figures are from Ido {\it et al.} \cite{ido91}.}
  \label{fg:lsno_opt}
 \end{figure}

As with cuprates, one can dope La$_2$NiO$_4$ either by substituting Sr for La or by adding excess oxygen.  It is useful to consider both types of samples.  Strontium substitution provides a continuous range of doping but at the cost of a disordered dopant potential.  In contrast, the ordering of oxygen interstitials at particular concentrations leads to stripe-ordered states with long-range order, from which one can gain considerable information.   We will jump back and forth between both types of systems.

From the comparison of optical conductivity on the left-hand side of Fig.~\ref{fg:lsno_opt}, one can see that the charge-transfer gap in La$_2$NiO$_4$ is $\sim4$ eV, about twice that of La$_2$CuO$_4$.  Introducing holes through Sr substitution creates states within the gap, as indicated by the dashed line \cite{ido91}.  If this substantial doping had created a metallic state, then one would expect to see the optical conductivity peak at zero energy; the fact that the peak is at $\sim0.6$~eV indicates that the doped charges are rather localized.  As shown on the right-hand side of Fig.~\ref{fg:lsno_opt}, the mid-gap peak induced by doping simply grows in magnitude in proportion to doping, suggesting that the nature of the charge localization does not change greatly over a very large doping range. 

\begin{figure}[t]
  \includegraphics[width=0.7\textwidth]{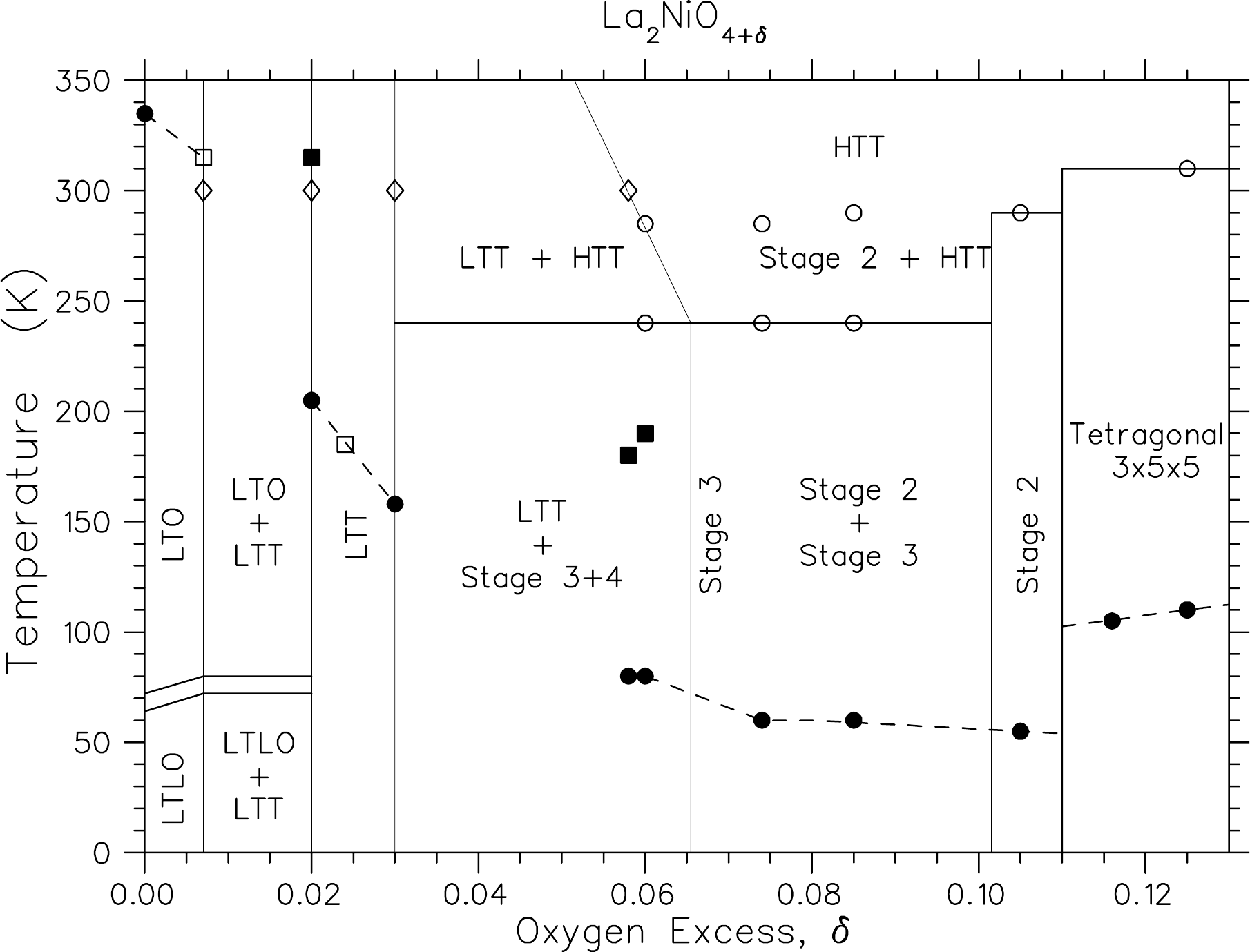}
  \caption{Phase diagram for La$_2$NiO$_{4+\delta}$ from \cite{tran94b}.  Open circles (diamonds) indicate phase boundaries determined by neutron single-crystal (x-ray powder) diffraction.  Solid circles denote N\'eel temperatures of primary phase, solid squares indicate N\'eel temperatures of secondary phases, and open squares indicate these transitions translated to the appropriate values of $\delta$.  Charge and spin stripe order occur in the phase with highest doping, where the oxygen interstitials have a complex 3D order \cite{tran95b}.}
  \label{fg:lno_pd}
 \end{figure}

Turning to oxygen-doped samples, Figure~\ref{fg:lno_pd} shows the phase diagram worked out experimentally for La$_2$NiO$_{4+\delta}$ \cite{tran94b}.   The intermediate phases involve stage-ordering of oxygen, as indicated in Fig.~\ref{fg:stage}.   Within this region, commensurate AF order is observed, but with a depressed $T_{\rm N}$.  Doping beyond stage-2 leads to a phase with 3D interstitial order as well as 3D ordering of charge and spin stripes.  Analysis of the structure indicates that it corresponds to an ideal oxygen excess of $\delta=2/15$ \cite{woch98}.

\begin{figure}[t]
  \includegraphics[width=0.8\textwidth]{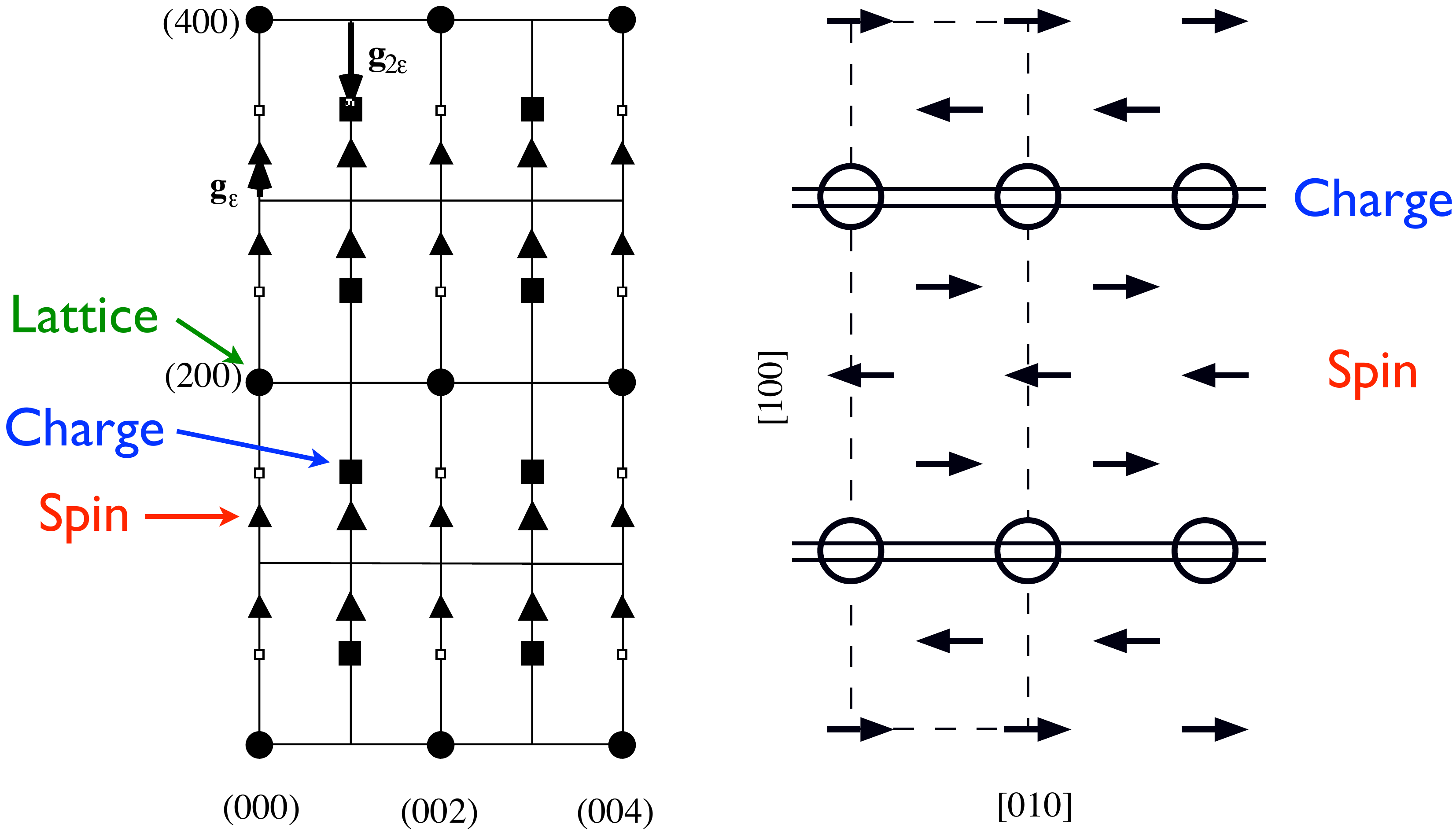}
  \caption{(a) Diagram showing the positions of various diffraction peaks in the $(h0l)$ zone, using $a=5.46$~\AA.  Solid circles, fundamental Bragg peaks; solid triangles, magnetic superlattice peaks; solid squares, nuclear superlattice peaks corresponding to charge order; open squares, allowed but unobserved charge-order peak positions.  Oxygen-ordering superlattice peaks are excluded.  (b) Cartoon of the stripe order; dashed line denotes a unit cell, and double lines mark antiphase domain walls for the spins.  From \cite{woch98}.}
  \label{fg:lno_super}
 \end{figure}

The 3D order in La$_2$NiO$_{4.133}$ makes it possible to measure a large array of superlattice peaks having differing characters, as indicated in Fig.~\ref{fg:lno_super}(a).  The magnetic peaks are split about the AF wave vectors of the type $(1,0,0)$ by $g_\epsilon=(\epsilon,0,0)$, while the charge order peaks are split about positions of the type $(2n,0,2m+1)$ ($n,\,m = $ integers) by $g_{2\epsilon}=(2\epsilon,0,0)$.  

\begin{figure}[t]
  {\includegraphics[width=0.43\textwidth]{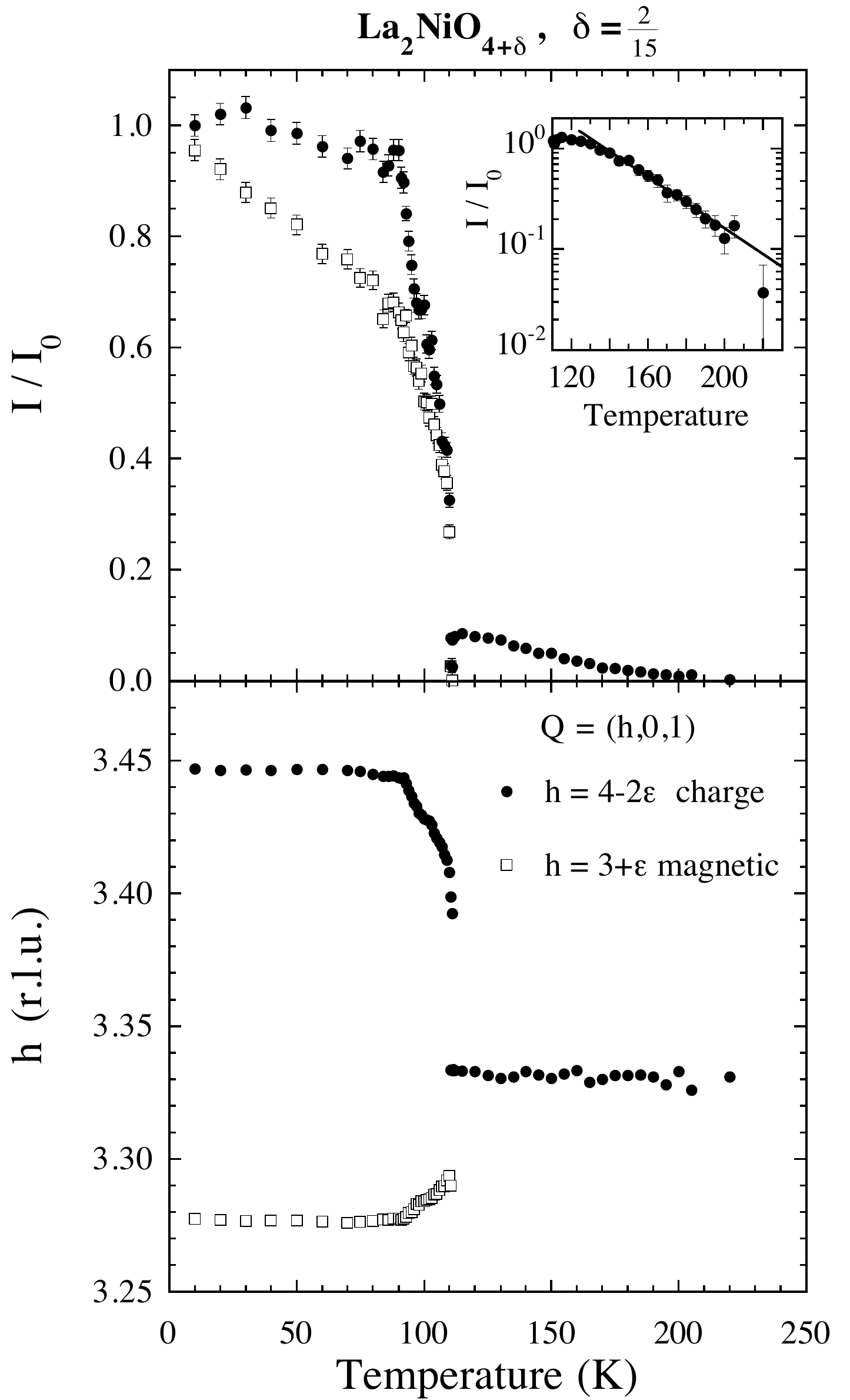}\hskip30pt
  \includegraphics[width=0.49\textwidth]{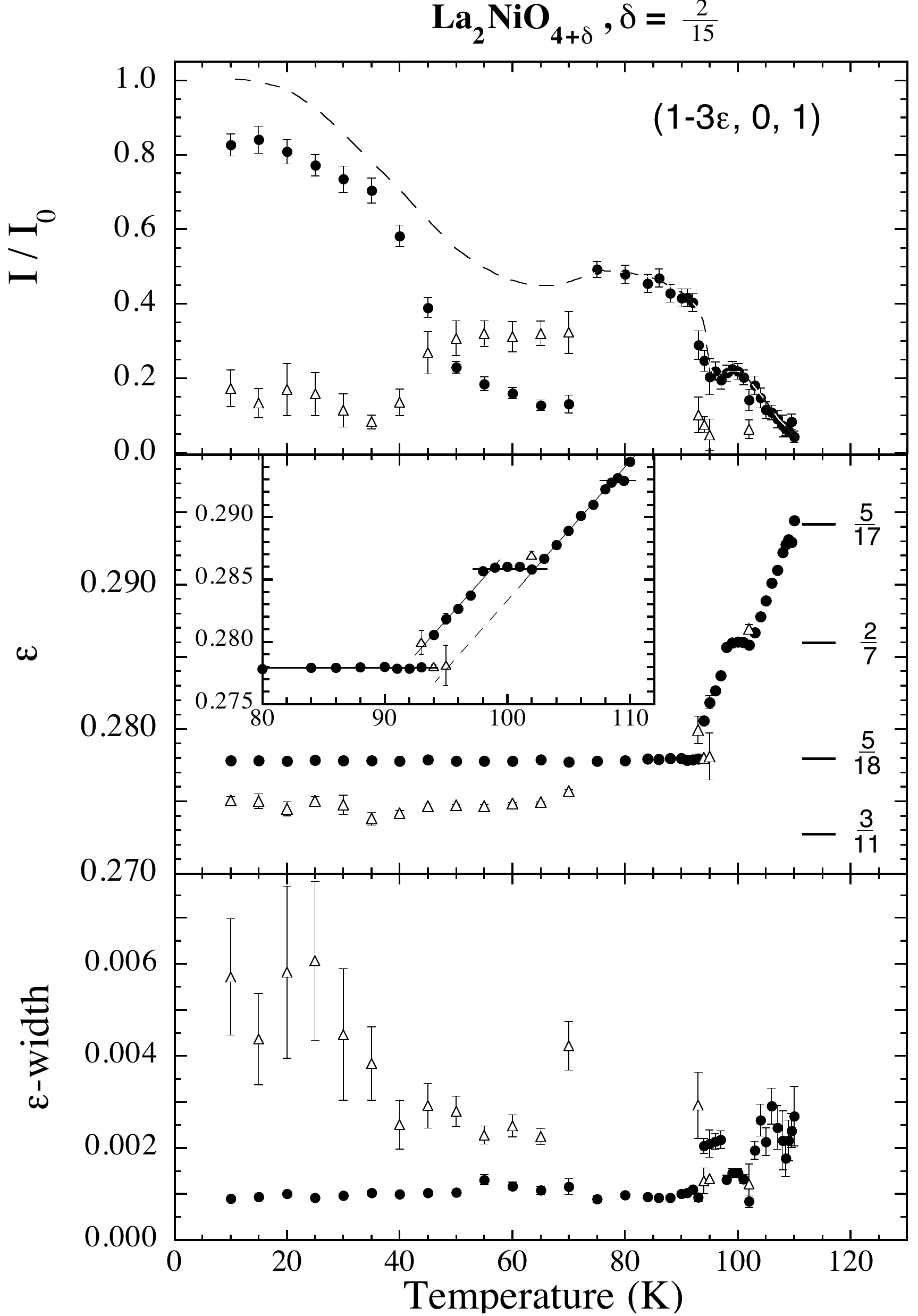}}
  \caption{(Left) Temperature dependence of   (top) intensities and (bottom) wave vectors associated with charge (filled circles) and spin (open squares) order in La$_2$NiO$_{4.133}$.  (Right) Temperature dependences of third harmonic magnetic peaks at
$(1-3\epsilon,0,1)$. At most temperatures two components with different
$\epsilon$ coexist. In all panels,  the  dominant
component is represented by the filled circles and the secondary  by open
triangles. The top panel shows the integrated intensity normalized to the total
intensity from both components at 10~K. The middle panel and inset show
$\epsilon$ and the bottom panel the  peak width, divided by 3 (in units of
$2\pi/a$).  From Wochner {\it et al.} \cite{woch98}.}
  \label{fg:lno_tdep}
 \end{figure}

The temperature dependence of the diffraction associated with representative charge and magnetic peaks is presented on the left-hand side of Fig.~\ref{fg:lno_tdep}; the wave vectors are 
plotted at the bottom, with the corresponding intensities (normalized at low temperature) shown at the top.  Above 110~K, there is only charge order, characterized by $\epsilon=0.33$.  Intriguingly, there is no obvious critical behavior that might be associated with a well-defined onset of charge order; instead, the intensity falls off exponentially with temperature, much like the behavior of a Debye-Waller factor.   On cooling below 110~K, there is a discontinuous jump in $\epsilon$ to a smaller value when the magnetic order appears.  Together with the onset of the spin stripe order, there is a large jump in the charge order intensity.  The incommensurability continues to vary with cooling, until it appears to saturate.

Better resolution of $\epsilon$ and its temperature dependence can be obtained by measuring a third-harmonic magnetic peak.   Results for $\epsilon$ and associated peak intensities and widths are presented on the right-hand side of Fig.~\ref{fg:lno_tdep}.  As one can see in the middle panel, $\epsilon$ appears to follow a devil's staircase, locking in at certain rational fractions, such as ${\scriptstyle \frac{5}{17}}$, ${\scriptstyle \frac27}$, and ${\scriptstyle \frac{5}{18}}$.  To describe these various periodicities, one can consider a mixture of both Ni-centered and O-centered stripes.   By analyzing the intensities of higher harmonics, it has been possible to sort out the most likely configurations.  The conclusion is that there is a shift in character of the stripes with temperature.

\begin{figure}[t]
  \includegraphics[width=0.7\textwidth]{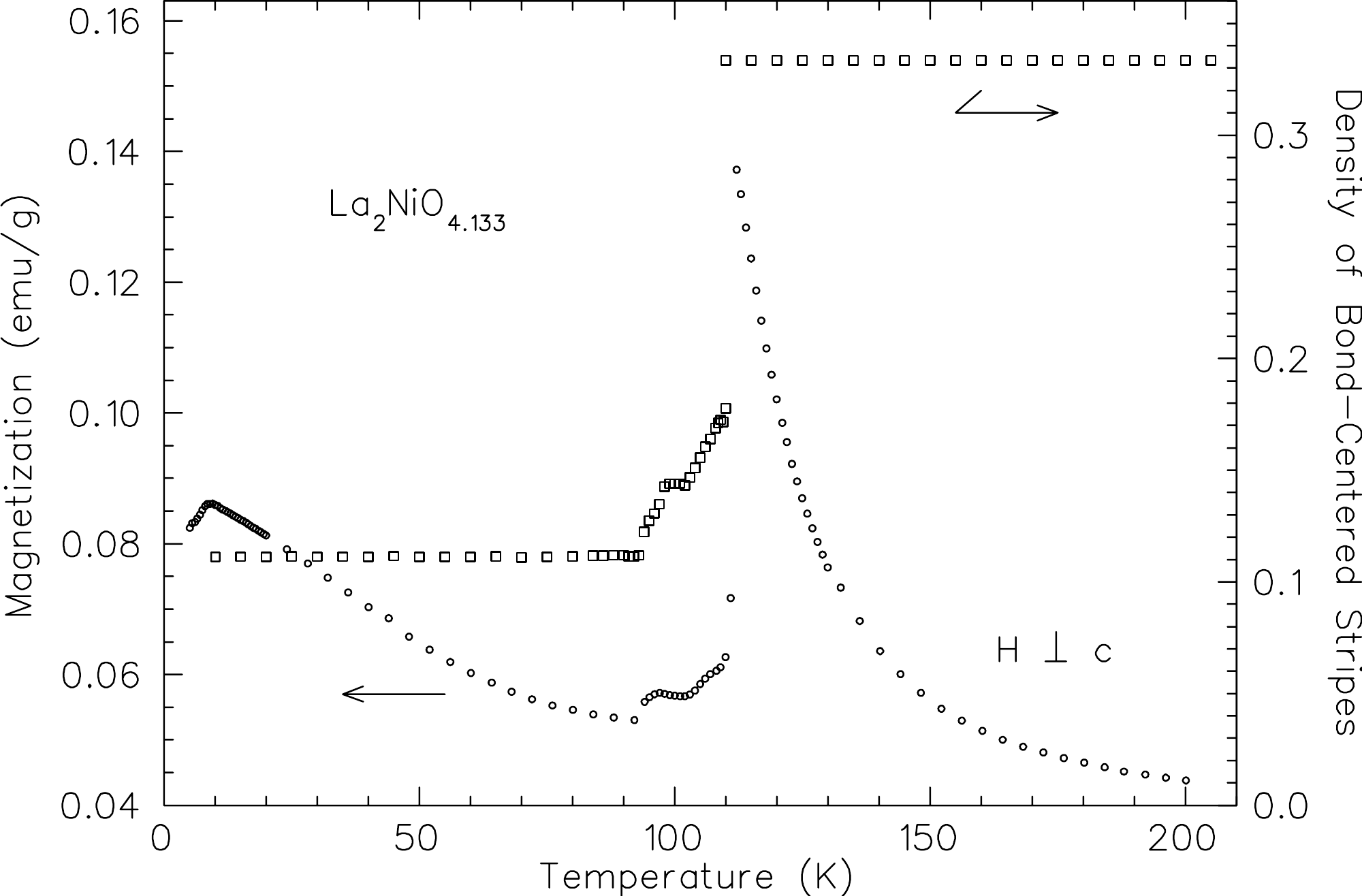}
  \caption{Bulk magnetization measured (on warming, after zero-field cooling) with an applied field of 1 T aligned parallel to the NiO$_2$ planes (circles), and density of bond-centered stripes (squares), which is equal to $4\epsilon-1$.  From \cite{tran97b}.}
  \label{fg:lno_m}
 \end{figure}

To give a simple description of the stripe order, it is convenient to consider just a few ordering wave vectors.  We start with one that is not quite reached in the present case, $\epsilon={\scriptstyle \frac14}$.  We denote the order along the in-plane direction, perpendicular to the stripes, with $\uparrow(\downarrow)$ indicating a Ni spin, $\cdot$ indicating an O site, and $\circ$ indicating a hole ({\it i.e.}, the center of a charge stripe).  For $\epsilon ={\scriptstyle\frac14}$, which would be characteristic of the low-temperature case with Ni-centered stripes, the configuration looks like 
 \begin{eqnarray} \label {Eq:Ni1_4}
 \circ\cdot
 \underbrace{\uparrow\cdot\downarrow\cdot\uparrow}_{\displaystyle{\uparrow}}
 \cdot\circ\cdot
\underbrace{\downarrow\cdot\uparrow\cdot\downarrow}_{\displaystyle{\downarrow}}
 \cdot\circ
\end{eqnarray} 
where we indicate below the configuration that the net moments of neighboring spin stripes are antiferromagnetically correlated.  For $\epsilon = {\scriptstyle\frac27}$, we get equal numbers of Ni-centered and O-centered stripes:
\begin{eqnarray}
  \label {Eq:Ni_O_2_7}
  \circ\cdot\uparrow\cdot\downarrow\cdot
  \underbrace{\uparrow\circ\uparrow}_{\displaystyle{\Uparrow}}
  \cdot\downarrow\cdot\uparrow\cdot\circ\cdot\downarrow\cdot\uparrow\cdot
  \underbrace{\downarrow\circ\downarrow}_{\displaystyle{\Downarrow}}
  \cdot\uparrow\cdot\downarrow\cdot\circ
\end{eqnarray} 
At high temperature, $\epsilon = {\scriptstyle\frac13}$ has only O-centered stripes:
\begin{eqnarray}
  \downarrow\cdot\underbrace{\uparrow\circ\uparrow}_{\displaystyle{\Uparrow}}
  \cdot\downarrow\cdot
  \underbrace{\uparrow\circ\uparrow}_{\displaystyle{\Uparrow}}
  \cdot\downarrow
\end{eqnarray}
Experimentally, we see $\epsilon={\scriptstyle\frac13}$ with only charge order; however, if the spins ordered, one might expect to see a ferrimagnetic response, as the net moments across charge stripes do not cancel out.  Indeed, it has been shown that application of a magnetic field can induce spin stripe order for $T\gtrsim 110$~K \cite{tran97b}.  The ferrimagnetic correlations are also evident in the temperature dependence of the magnetization, as shown in Fig.~\ref{fg:lno_m}.

\begin{figure}[t]
  \includegraphics[width=0.4\textwidth]{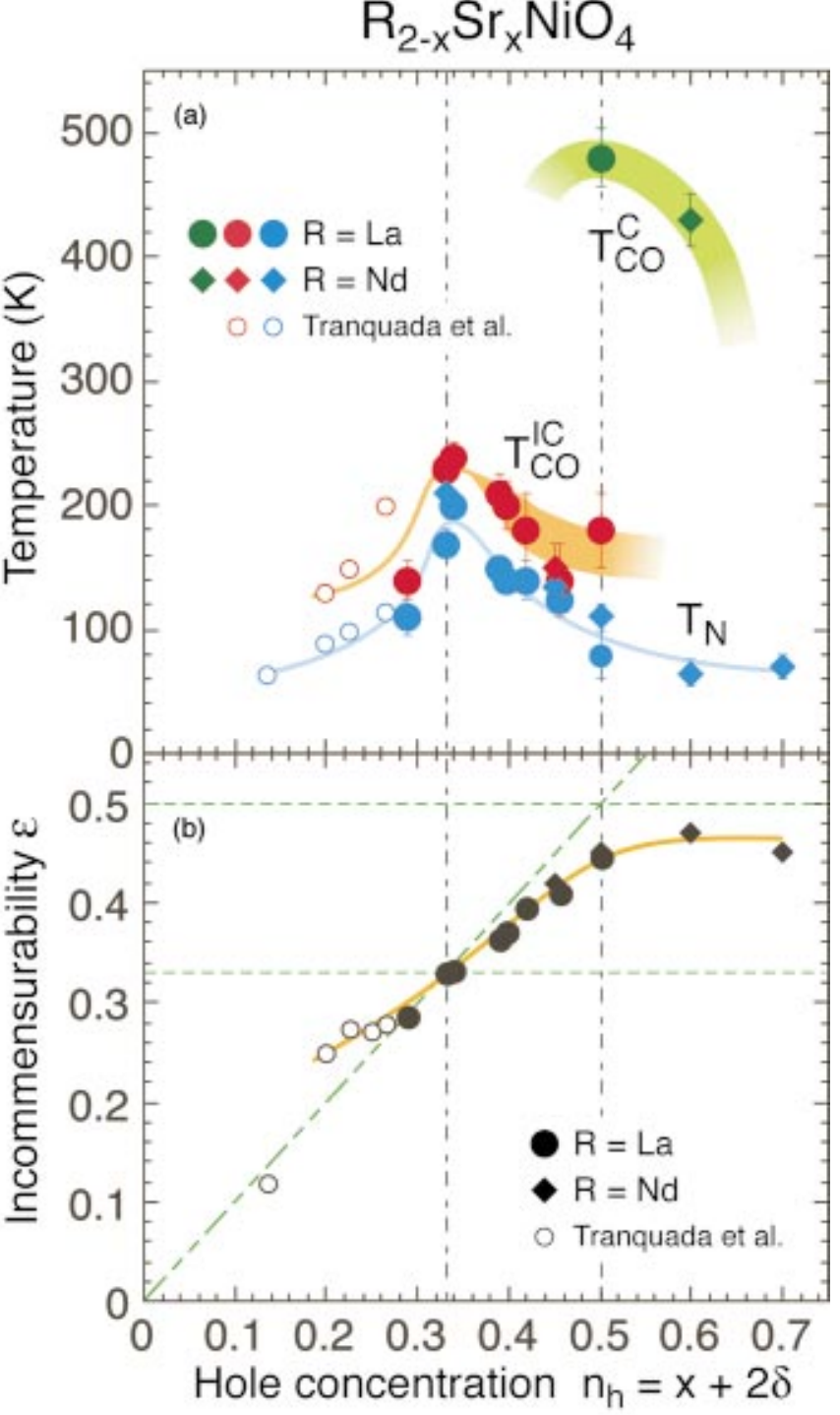}
  \caption{Phase diagram of R$_{2-x}$Sr$_x$NiO$_4$: (a) Hole concentration dependence of the transition temperatures, for incommensurate and commensurate charge order $T_{\rm co}^{\rm IC}$, $T_{\rm co}^{\rm C}$, and for spin stripe order $T_{\rm N}$ . (b) Hole concentration dependence of the incommensurability $\epsilon$. Circles denote the data for La$_{2-x}$Sr$_x$NiO$_4$, while diamonds are for Nd$_{2-x}$Sr$_x$NiO$_4$.  From Kajimoto {\it et al.} \cite{kaji03}.}
  \label{fg:lsno_pd}
 \end{figure}

To consider the doping dependence of stripe order, we return to La$_{2-x}$Sr$_x$NiO$_4$.  Figure~\ref{fg:lsno_pd}(a) shows the phase diagram obtained from neutron scattering studies \cite{kaji03}, including results for Nd$_{2-x}$Sr$_x$NiO$_4$.\footnote{At higher Sr levels, successful crystal growth requires replacement of La by smaller trivalent ions, such as Nd.}  It is consistently observed that the charge stripe order appears at a higher temperature than the spin stripe order, with the maximum ordering temperatures occurring at $x=1/3$ \cite{lee97}.  For $x\approx0.5$, a checkerboard pattern of charge order develops at quite high temperature; some competing stripe order appears at lower temperatures.

The incommensurability obtained at low temperature as a function of doping is shown in Fig.~\ref{fg:lsno_pd}(b).  For much of the range, it stays close to $\epsilon=x$, which would correspond to keeping the charge density in the stripes fixed and simply varying the stripe density with doping.  As a function of temperature, it is generally found that $\epsilon$ heads toward $1/3$ as the disordered state is approached \cite{ishi04}.

\begin{figure}[t]
  \includegraphics[width=0.5\textwidth]{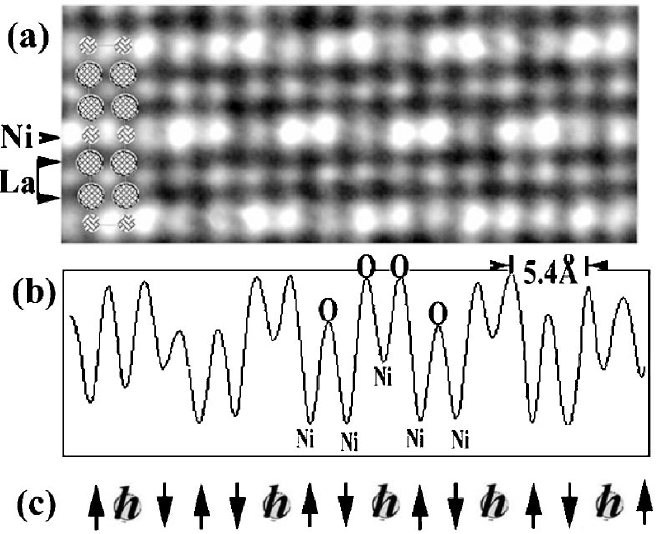}
  \caption{(a) High-resolution TEM image of the $a$-$c$ plane ($c$ vertical) illustrating the modulation at low temperature. The bright segments arising from charge modulation within the NiO$_2$ layers can be recognized. (b) An intensity profile showing contrast variation within the bottom NiO$_2$ layer of (a). This curve clearly indicates that the brighter segments in this region are Ni centered. (c) Schematic model of spin and charge ordering corresponding to the curve in (b).  From Li {\it et al.} \cite{li03}.}
  \label{fg:lsno_tem}
 \end{figure}

A real-space image of stripes in La$_{1.725}$Sr$_{0.275}$NiO$_4$ obtained by transmission electron microscopy \cite{li03} is shown in Fig.~\ref{fg:lsno_tem}(a).  Here we see the stripes edge on, in planes that run horizontally.  The image shows scattering strength along columns of atoms in a very thin sample.  The period of the contrast modulation is consistent with the charge-order period from neutron scattering measurements \cite{lee02}.   Unfortunately, it is not known whether the brighter or darker contrast corresponds to the hole location.  (This can only be determined by a simulation, which is challenging in the present case.)

 Another feature apparent in the image is the shift in stripe registry on moving along the $c$ axis from one layer to the next.  This staggered arrangement had already been inferred in the neutron scattering studies from analysis of the dependence of the structure factor on momentum perpendicular to the planes.  It is consistent with minimizing the Coulomb energy of the stripes, together with preferential registry with the NiO$_2$ lattice.

Both spin and charge order in La$_{1.8}$Sr$_{0.2}$NiO$_4$ have also been detected by the technique of resonant soft-x-ray scattering \cite{schu05}.  Here, one tunes the x-ray energy to that of the Ni $L_3$ edge (850 eV) in order to get a very strong enhancement of the superlattice peaks.  The results are also sensitive to the polarization of the incident photons.

One might ask whether it is possible to distinguish experimentally between a model in which stripes locally run in only one direction, and one in which there is a superposition of stripes in orthogonal directions.  To make this distinction, one needs a sample in which the two diagonal directions in the NiO$_2$ planes are inequivalent.  This can be achieved by studying Nd$_{2-x}$Sr$_x$NiO$_4$, which has the same low-temperature orthorhombic structure as La$_2$CuO$_4$, with inequivalent lattice parameters along the diagonals.  A careful x-ray diffraction study by H\"ucker {\it et al.} \cite{huck06} on crystals with $x=0.33$ demonstrated that the charge stripes run uniquely along the shorter $a$ axis.

\begin{figure}[t]
  \includegraphics[width=0.8\textwidth]{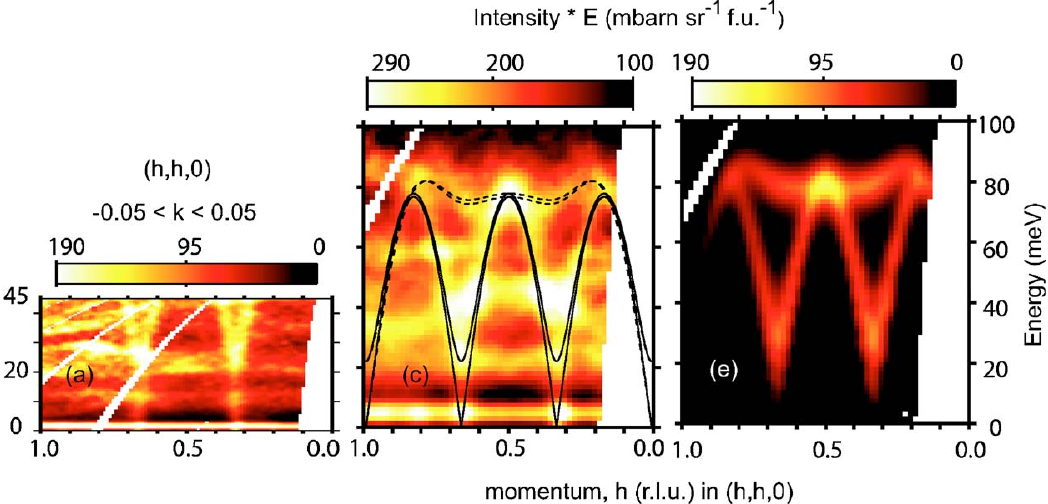}
  \caption{(a,c) Slices through inelastic neutron scattering measurements for ${\bf Q}$ along the spin stripe modulation direction of La$_{1.67}$Sr$_{0.33}$NiO$_4$.  The steep dispersion of the spin waves is superimposed on phonons, which have a flatter dispersion.  The fitted spin-wave dispersion is plotted as lines in (c) and is simulated in (e).  From Woo {\it et al.} \cite{woo05}.}
  \label{fg:lsno_sw}
 \end{figure}

Besides static order, it is also possible to measure the spin dynamics.  Figure~\ref{fg:lsno_sw} shows measurements of the spin waves (and a simulation of the fit) for La$_{1.67}$Sr$_{0.33}$NiO$_4$ \cite{woo05}.  It is possible to get a satisfactory description of the dispersion starting with a Heisenberg model, with different exchange parameters within and between the AF stripe domains.  The magnitude of $J$ within an AF domain is equal to 90\%\ of that in the undoped AF.  This is a remarkably small change considering the very large doping, and it is consistent with a strong localization of the holes to the charge stripes.

It is important to keep in mind that each Ni$^{2+}$ has $S=1$.  It follows that, even with one doped hole per Ni site along a charge stripe, there should still be at least a $S={\scriptstyle\frac12}$ per Ni there.  Indeed, Boothroyd {\it et al.} \cite{boot03b} discovered spin waves with a 1D dispersion that are consistent with excitations within the charge stripes.  While they are readily detected at low temperature, the intensity falls off rapidly on warming towards 100~K.

\begin{figure}[t]
 { \includegraphics[width=0.55\textwidth]{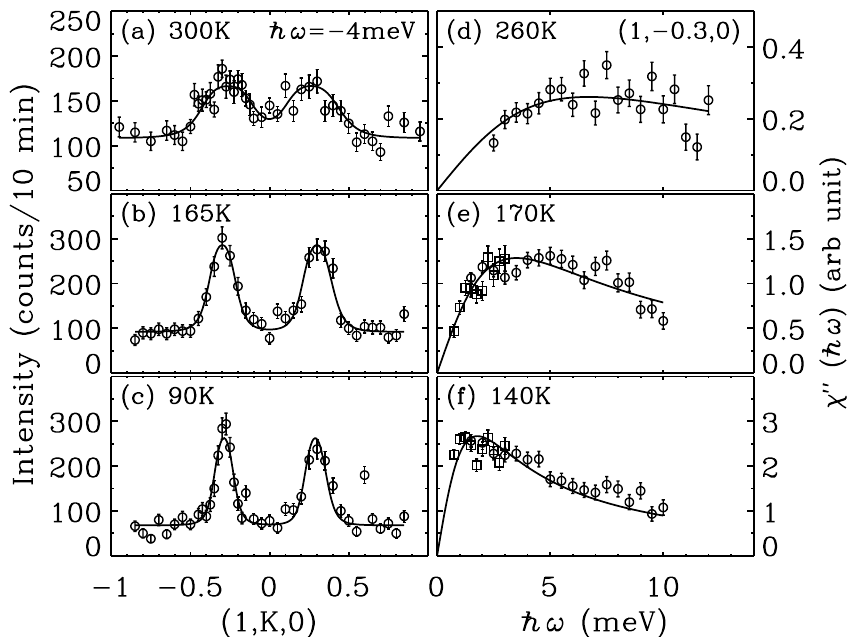}\hskip30pt
   \includegraphics[width=0.4\textwidth]{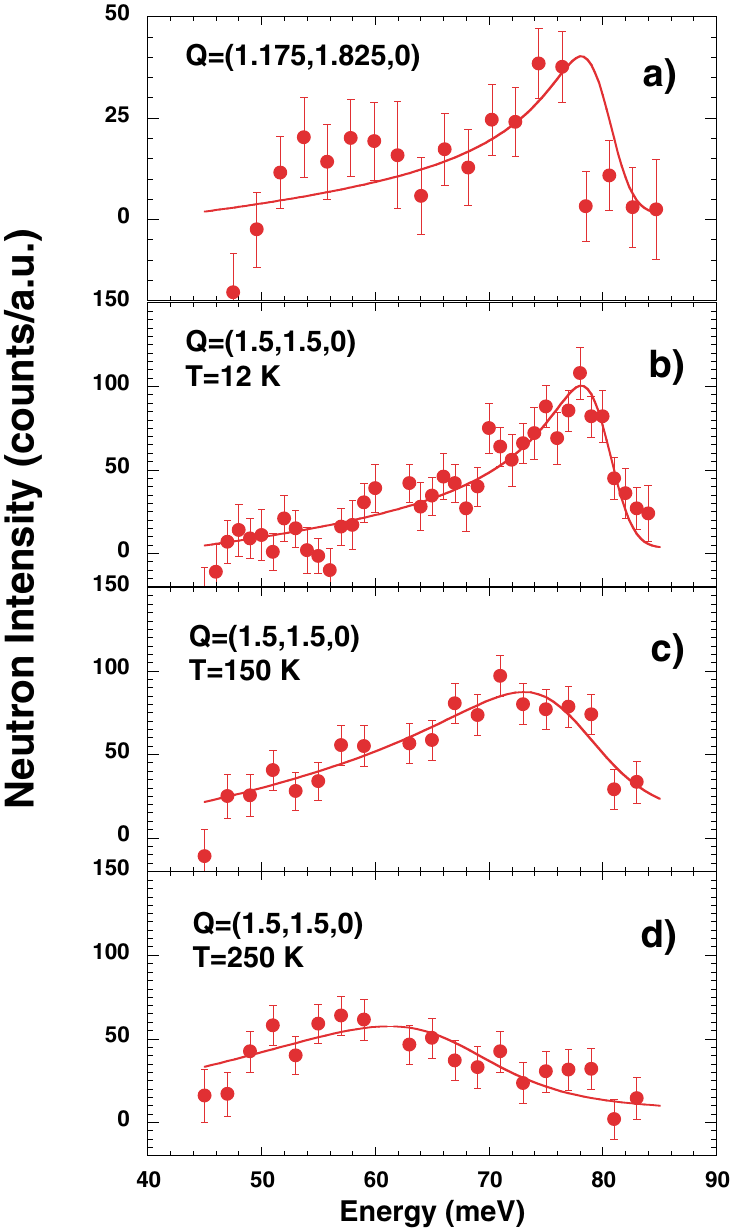}}
  \caption{(Left) (a-c) Constant-energy scans for $\hbar\omega=-4$~meV in LSNO with $x=0.275$ for three different conditions: (a) 300 K, no static order; (b) 165 K, charge but no spin order; (c) 90 K, spin and charge order.  (d-f) Measurements of $\chi''({\bf Q}_{\rm peak},\omega)$ for temperatures (d) 260 K, (e) 170 K, (f) 140 K.  From Lee {\it et al.} \cite{lee02}.  (Right)  (b-d) Measurements of spin-fluctuation intensity vs.\ energy through the dispersion maximum (at ${\bf Q}_{\rm AF}$) for LSNO with $x=0.31$ at low temperature (b), near the spin ordering temperature (c), and just above the charge ordering temperature (d).   From Bourges {\it et al.} \cite{bour03}.}
  \label{fg:lsno_tdep}
 \end{figure}

It is especially interesting to consider the temperature dependence of the spin fluctuations.  The left-hand side of Fig.~\ref{fg:lsno_tdep} demonstrates that incommensurate excitations survive at temperatures above the charge-ordering transition, $T_{\rm co}$.  The right-hand side shows that the high-energy spin fluctuations become strongly damped when charge order is lost.

\begin{figure}[t]
 { \includegraphics[width=0.43\textwidth]{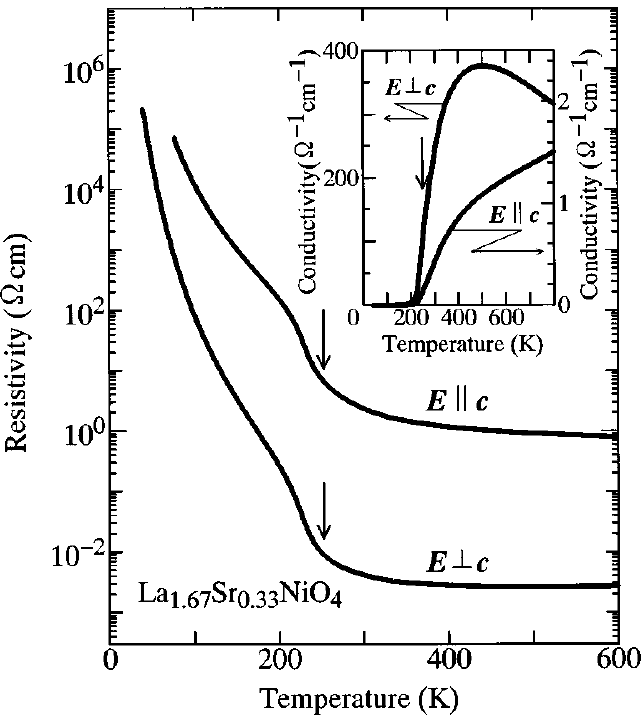}\hskip30pt
   \includegraphics[width=0.52\textwidth]{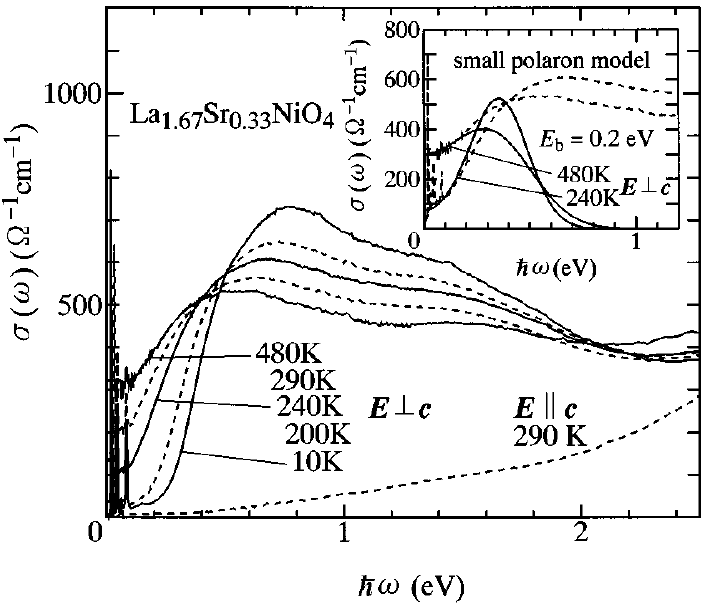}}
  \caption{Results for La$_{1.67}$Sr$_{0.33}$NiO$_4$ from Katsufuji {\it et al.} \cite{kats96}.  (Left) Resistivity vs.\ temperature measured both parallel to the planes and parallel to the $c$ axis.  (Right) Optical conductivity vs.\ energy for temperatures spanning the charge order transition at 240 K.}
  \label{fg:lsno_transport}
 \end{figure}

We can also consider the thermal evolution of the resistivity and optical conductivity, as shown in Fig.~\ref{fg:lsno_transport}.  The in-plane resistivity drops significantly upon warming across $T_{\rm co}$ (indicated by the arrow); however, the magnitude of the resistivity in this disordered state corresponds to a very poor metal.  Examination of the optical conductivity shows why: much of the charge remains quasi-localized, with a peak in $\sigma(\omega)$ at a substantial energy even at a temperature of $2T_{\rm co}$.

Considering the high-temperature spin and charge responses together, we see that there is evidence that stripe correlations survive in the disordered state.  When the charge becomes mobile, the high-energy spin fluctuations become strongly damped.  That behavior is reminiscent of the trend in cuprates presented in Fig.~\ref{fg:Stock}.

To summarize this section, we have seen that:
\begin{itemize}
\item Doped holes frustrate antiferromagnetism;
\item The balance between kinetic, Coulomb, and superexchange energies can result in stripe order;
\item Dynamic stripe correlations survive above $T_{\rm co}$;
\item Mobile holes damp the high-energy spin fluctuations.
\end{itemize}

\subsection{Cobaltates and Manganites}

Charge and spin order are also common in other layered transition-metal oxides.  We will just briefly touch on these materials here in order to reinforce the case that electronic inhomogeneity is a common phenomenon.  For more details, one can turn to the review by Ulbrich and Braden \cite{ulbr12b}.

Magnetic and charge ordering have been investigated by neutron scattering in La$_{2-x}$Sr$_x$CoO$_4$ \cite{zali00,cwik09} and in Pr$_{2-x}$Ca$_x$CoO$_4$ \cite{saki08}.   One has to dope to a higher level ($x>0.3$) than in nickelates to obtain incommensurate spin order.  Short-range checkerboard charge order is especially stable at $x=0.5$, with ordering at $\sim 825$~K \cite{zali01b}.  An extra feature in this system is that Co$^{3+}$ ions can have a low-spin state with $S=0$, as well as intermediate and high spin states.  The size of the ion is correlated with the spin state, being smaller in the low spin state \cite{asai94}.  This size effect can couple to lattice strain and influence the charge ordering.

The spin excitation spectrum has been measured in La$_{1.67}$Sr$_{0.33}$CoO$_4$ by Boothroyd {\it et al.} \cite{boot11}.   In contrast to the nickelate, this system exhibits an hourglass dispersion, reminiscent of the cuprates.

In manganites, the interesting behavior occurs with mixtures of Mn$^{3+}$ and Mn$^{4+}$ ions, both of which have magnetic moments; in addition, one must take account of ordering of $e_g$ orbitals on Mn$^{3+}$ ions.   For example, La$_{0.5}$Sr$_{1.5}$MnO$_4$ exhibits an ordered structure with a large unit cell involving ferromagnetic zigzag chains of ions, with neighboring magnetic chains having a relative antiferromagnetic alignment \cite{ster96}; this structure was first predicted by Goodenough \cite{good55} in the 1950's.  By adding more Sr, it is possible induce extra rows of Mn$^{4+}$ ions, changing the periodicity of the structure \cite{ulbr11}.  Again, an hourglass magnetic dispersion has been reported for Nd$_{0.33}$Sr$_{1.67}$MnO$_4$ \cite{ulbr12a}.

\section{Stripes and superconductivity in the cuprates}

We have seen that charge and spin order, especially in the form of stripes, are quite common in layered transition-metal oxides.   Such states tend to be insulating, though thermally-disordered stripes can have a bad-metal character.   Given such a background, it should not be surprising that stripes might play a role in cuprates.   

In considering the more metallic character of the cuprates, it is valuable to consider the difference between having Cu$^{2+}$ with $S={\scriptstyle\frac12}$ rather than, say, Ni$^{2+}$ with $S=1$.  In a cuprate, a hole moving past a Cu ion will leave it in a final state that is equivalent to its initial state, except possibly for spin direction.  This is no longer true in a nickelate.  Ni$^{2+}$ has two half-filled $3d$ orbitals, $x^2-y^2$ and $3z^2-r^2$.  If one flips the spin in the $x^2-y^2$ orbital, then it will become antiparallel with $3z^2-r^2$ electron, which is not equivalent to the initial state.  This latter effect can contribute to a tendency towards charge localization that is not relevant to the cuprates.

\subsection{The 1/8 anomaly}

\begin{figure}[t]
 { \includegraphics[width=0.65\textwidth]{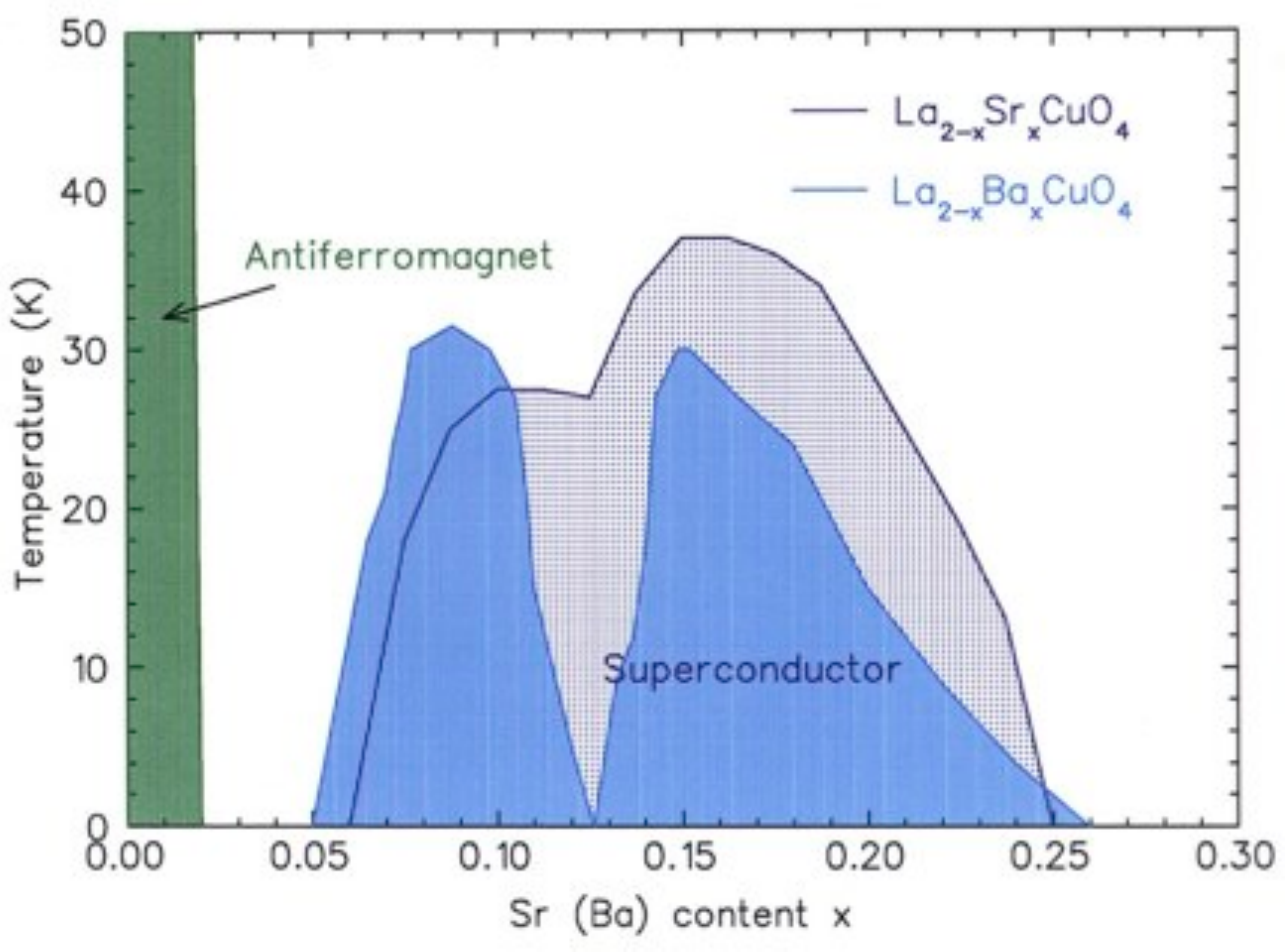}\hskip40pt
 \includegraphics[width=0.25\textwidth]{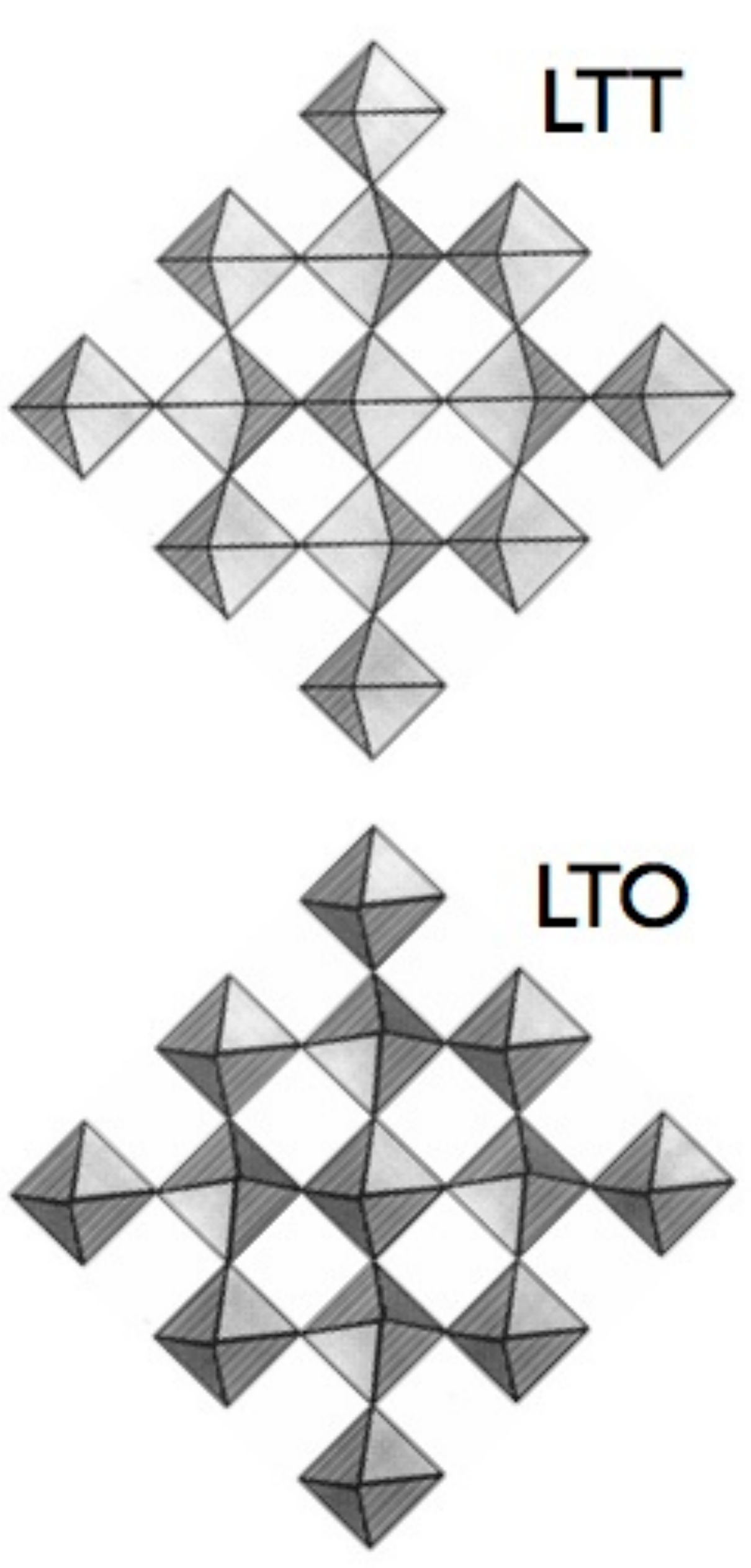}}
  \caption{(Left)  Comparison of superconducting $T_c$ vs.\ $x$ in LBCO and LSCO, illustrating the anomaly at  $x\approx{\scriptstyle\frac18}$ found by Moodenbaugh {\it et al.} \cite{mood88}. (Right) View from the top of CuO$_2$ planes comparing the pattern of CuO$_6$ octahedral tilts in the (top) low-temperature-tetragonal (LTT) phase, and (bottom) low-temperature-orthorhombic (LTO) phase.  Both LBCO and LSCO exhibit the LTO structure, but  Axe {\it et al.} \cite{axe89} discovered a transition from LTO to LTT on LBCO.}
  \label{fg:lbco_mood}
 \end{figure}

Not long after the discovery of  high temperature superconductivity, a careful study of the transition temperature as a function of doping in La$_{2-x}$Ba$_x$CuO$_4$ revealed a surprising anomaly \cite{mood88}.  As shown on the left-hand side of Fig.~\ref{fg:lbco_mood}, $T_c$ shows a remarkable dip at $x\approx{\scriptstyle\frac18}$; in contrast, $T_c(x)$ for La$_{2-x}$Sr$_x$CuO$_4$ shows only a slight kink at this doping level.  Electronically, one expects these materials to be virtually identical, as Ba and Sr are both divalent ions.   The dip in $T_c$ at a special value of $x$ suggests the presence of some type of competing order.

A distinction between these compounds was revealed in an x-ray diffraction study by Axe {\it et al.} \cite{axe89}.   It was already known that the CuO$_6$ octahedra that make up the CuO$_2$ planes are unstable to tilt distortions.  At high temperature, there are no average tilts, but on cooling there is a second-order phase transition to the low-temperature-orthorhombic (LTO) phase, in which octahedra rotate around $[1\bar{1}0]$ axes, with corner-sharing neighbors tilting in opposite directions, as illustrated on the right-hand side of Fig.~\ref{fg:lbco_mood}.   While this tilt pattern makes the diagonal directions of the Cu-O plaquettes inequivalent, it provides no distinction between the Cu-O bonds in orthogonal directions.  The x-ray study showed that LBCO can, at even lower temperature, undergo a second transition to the low-temperature-tetragonal (LTT) phase, with the tilt axis changed to [010], as indicated on the right-hand side of Fig.~\ref{fg:lbco_mood}; the orientation of the tilt axis rotates $90^\circ$ between neighboring layers along the $c$ axis.  In the LTT phase, orthogonal Cu-O bonds are no longer equivalent.  In a given CuO$_2$ layer, the Cu-O-Cu bonds along one direction are perfectly straight, while those at $90^\circ$ are bent.  This anisotropy is key to the pinning of the electronic order that competes with the bulk superconductivity.

\subsection{Diffraction evidence for stripe order}

We have already seen that antiferromagnetic spin correlations are a characteristic feature of superconducting cuprates.  Characterization of these correlations by neutron scattering requires large single-crystal samples.  Early measurements revealed the incommensurate character of low-energy spin fluctuations in LSCO \cite{cheo91}.   There are challenges to growing LBCO crystals that were not overcome for more than a decade after that; however, it was discovered that partial substitution of Nd for La in LSCO results in a system, such as La$_{1.6-x}$Nd$_{0.4}$Sr$_x$CuO$_4$ (LNSCO), that exhibits the same structural phases as LBCO and also has the ${\scriptstyle\frac18}$-anomaly in $T_c(x)$ \cite{craw91,naka92}.   A neutron diffraction study on a crystal of LNSCO with $x=0.12$ found separate superlattice peaks for charge and spin order that were interpreted as evidence for the bond-parallel stripe order indicated in Fig.~\ref{fg:diag_horiz} \cite{tran95a}.   When crystals of LBCO with $x=0.125$ were eventually grown, stripe order was confirmed there, as well \cite{fuji04}.

\begin{figure}[t]
 \includegraphics[width=0.7\textwidth]{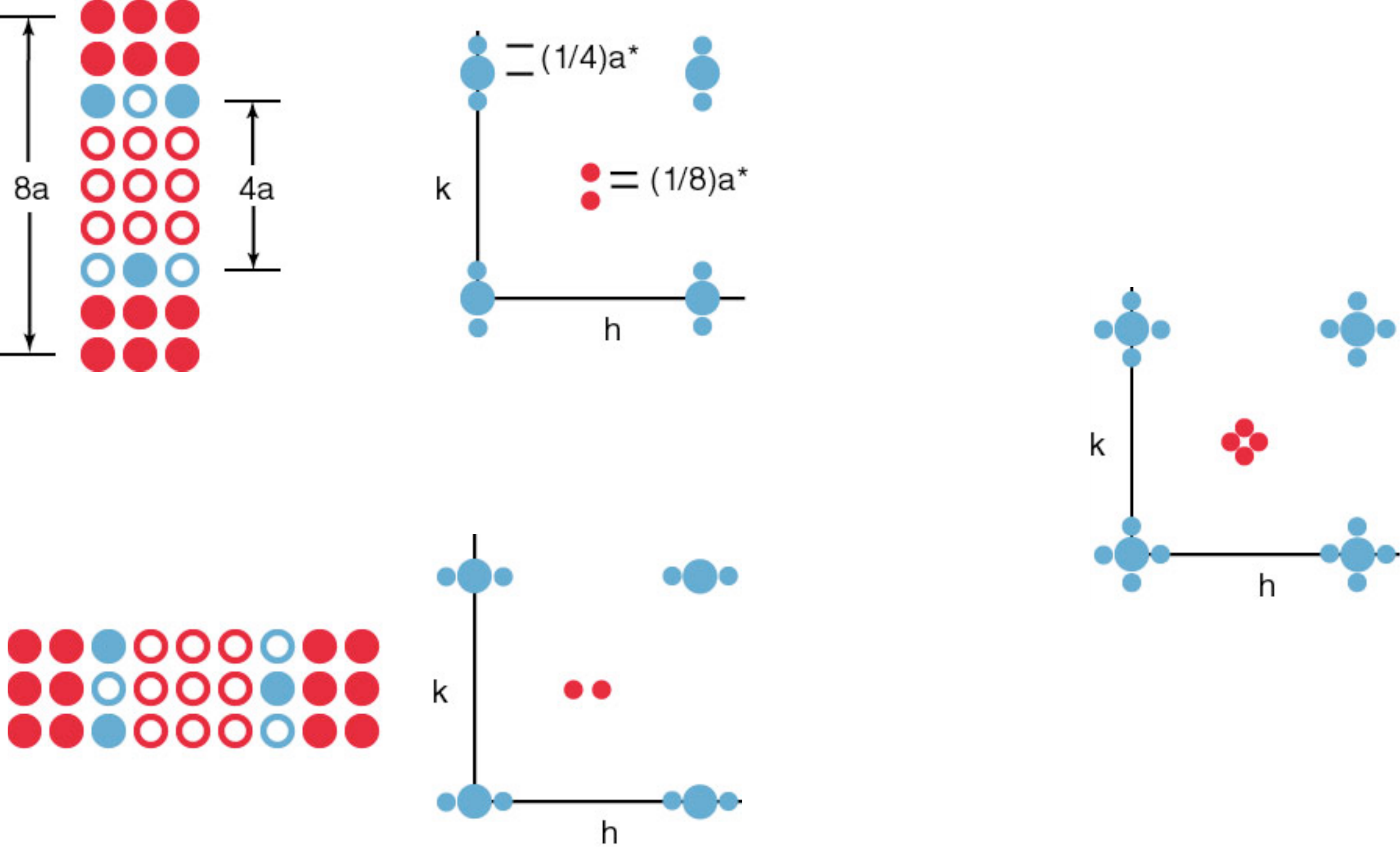}
  \caption{(Upper left) Illustration of a domain of spin and charge stripe order for $x={\scriptstyle\frac18}$, indicating the periods of the charge ($4a$) and spin ($8a$) modulations, and the positions of corresponding superlattice peaks in reciprocal space.  For simplicity, the antiphase AF stripes are indicated by all filled or all open circles.  (Lower left) An orthogonal stripe domain and associated superlattice peaks.  (Right) Pattern of superlattice peaks when both orientations of stripe domains are present }
  \label{fg:diff}
 \end{figure}
 
 The relationship between stripe order, domain orientation, and superlattice peaks is illustrated in Fig.~\ref{fg:diff}.  The magnetic peaks are split around ${\bf Q}_{\rm AF}$ by an amount inversely proportional to the magnetic period in real space.  A magnetic period of $8a$ corresponds to an incommensurate splitting of ${\scriptstyle\frac18}$ in reciprocal lattice units.  Similarly, a charge modulation of $4a$ leads to superlattice peaks split by ${\scriptstyle\frac14}$; since the diffraction signal is due to the atomic displacements associated with charge modulation, the superlattice peaks are centered around fundamental Bragg peaks.  
 
 Since the LTT structure is key to the occurrence of stripe order, it is reasonable to assume that it is the anisotropy between orthogonal Cu-O bonds that pins the stripes.  Hence, each CuO$_2$ should pin stripes with a unique orientation.  Since a unit cell of the LTT structure contains two CuO$_2$ layers with the anisotropy rotated by $90^\circ$ from one to the next,  a diffraction experiment must sample equal numbers of the two domain orientations shown on the left of Fig.~\ref{fg:diff}, resulting in the net superlattice pattern shown on the right.
 
\begin{figure}[t]
 \includegraphics[width=0.7\textwidth]{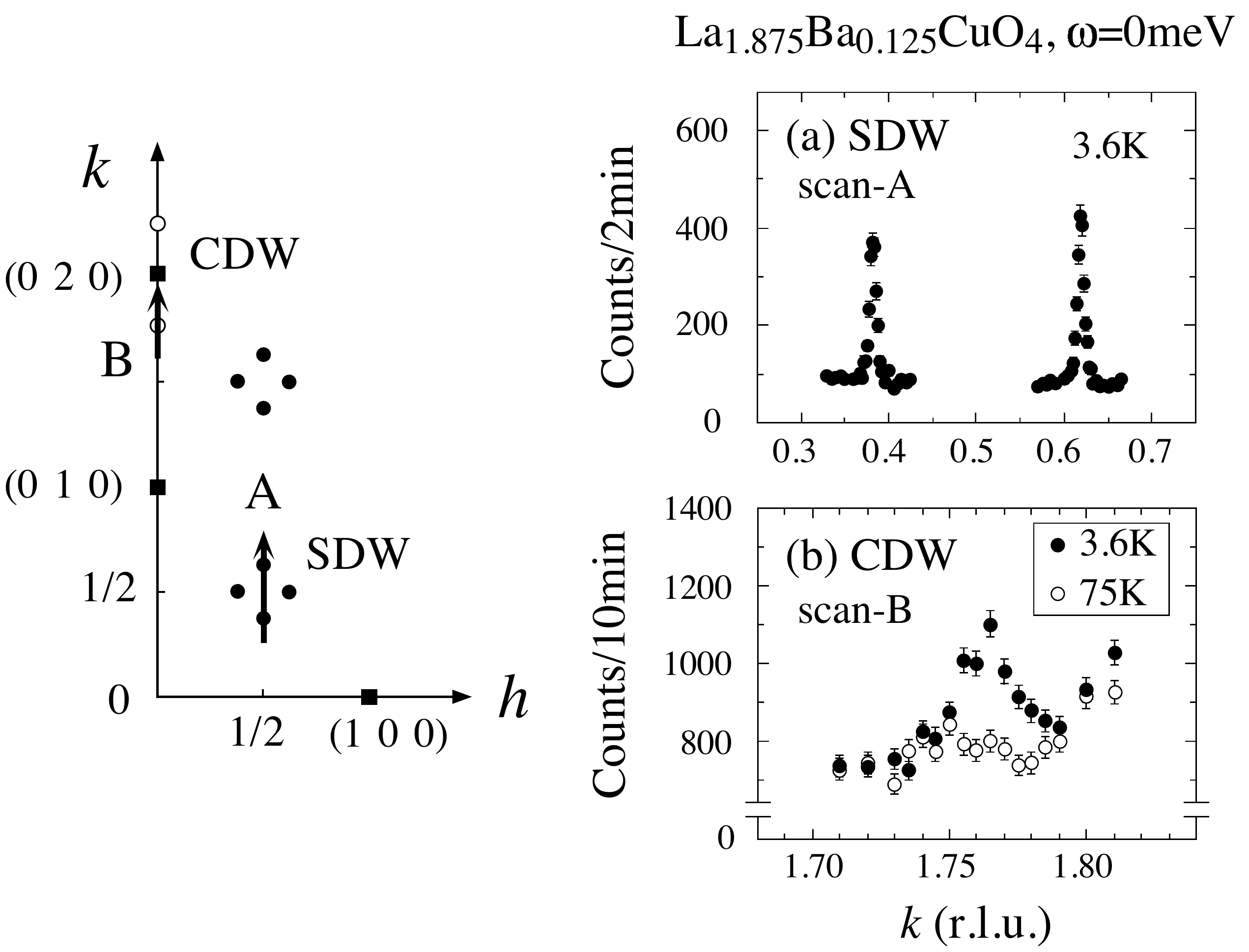}
  \caption{(Left) Reciprocal space map indicating locations of superlattice peaks, with arrows indicating paths of neutron diffraction scans A and B.  (Right) (a) Magnetic superlattice peaks measured along scan A in LBCO with $x=0.125$ at $T=3.6$ K.  (b) Charge-order superlattice peak measured along scan B at 3.6~K (filled circles), with background measured at 75~K (open circles). From Fujita {\it et al.} \cite{fuji04}. }
  \label{fg:fuji}
 \end{figure}

Figure~\ref{fg:fuji} shows examples of superlattice peak measurements on LBCO with $x={\scriptstyle\frac18}$ \cite{fuji04}.  The peak positions are close, but not precisely equal, to the values suggested in Fig.~\ref{fg:diff}; rather, the spin incommensurability is $\delta=0.118$ and the charge peaks appear at $2\delta=0.236$; it should also be noted that the peaks have widths greater then resolution, indicating finite correlation lengths.  The fact that $\delta$ is not a rational fraction can be reconciled with a model having local commensurability of stripes with the lattice, by taking account of disordered variations in the stripe spacing \cite{tran99a}.

\begin{figure}[t]
 {\includegraphics[width=0.4\textwidth]{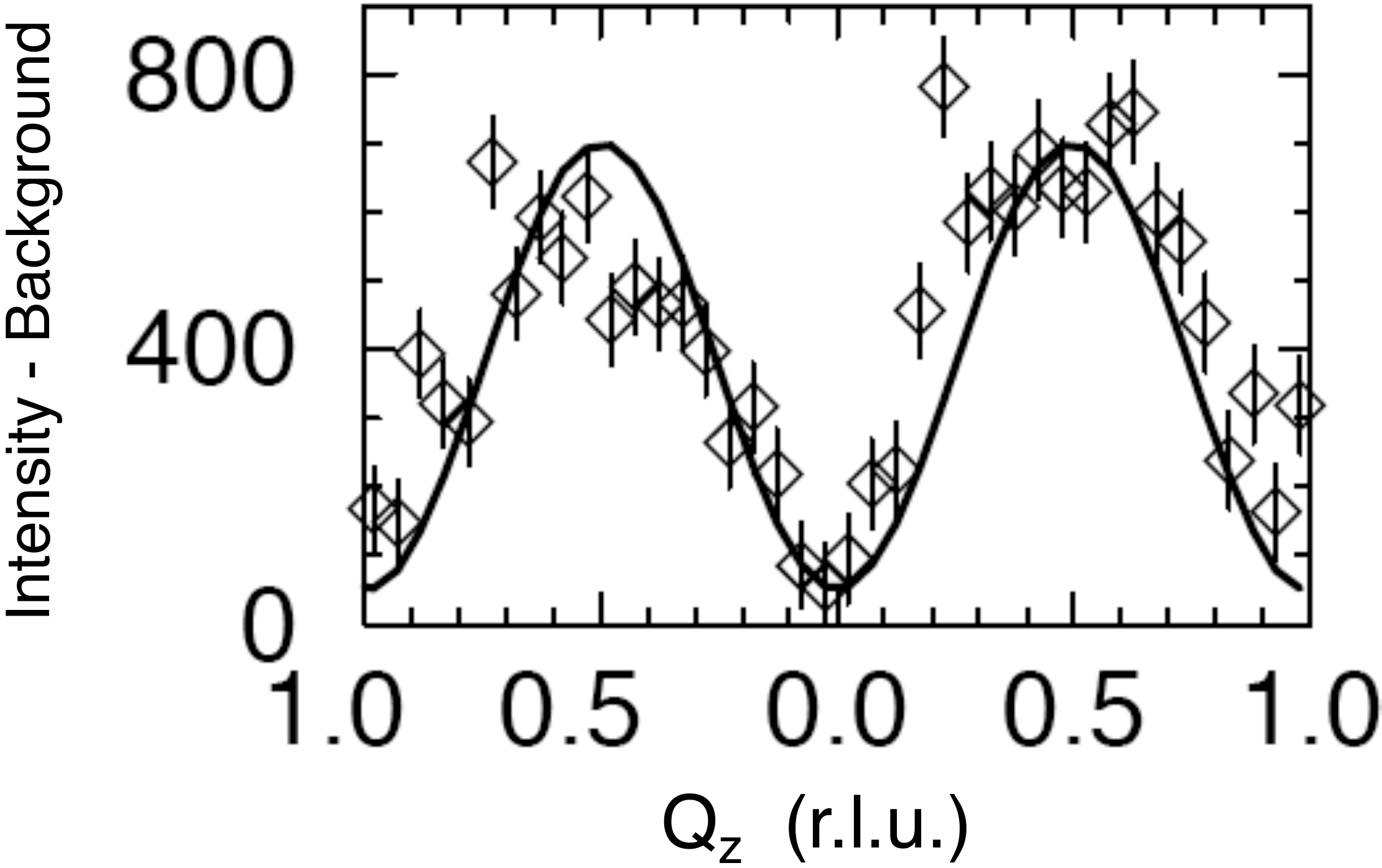}\hskip50pt
 \includegraphics[width=0.45\textwidth]{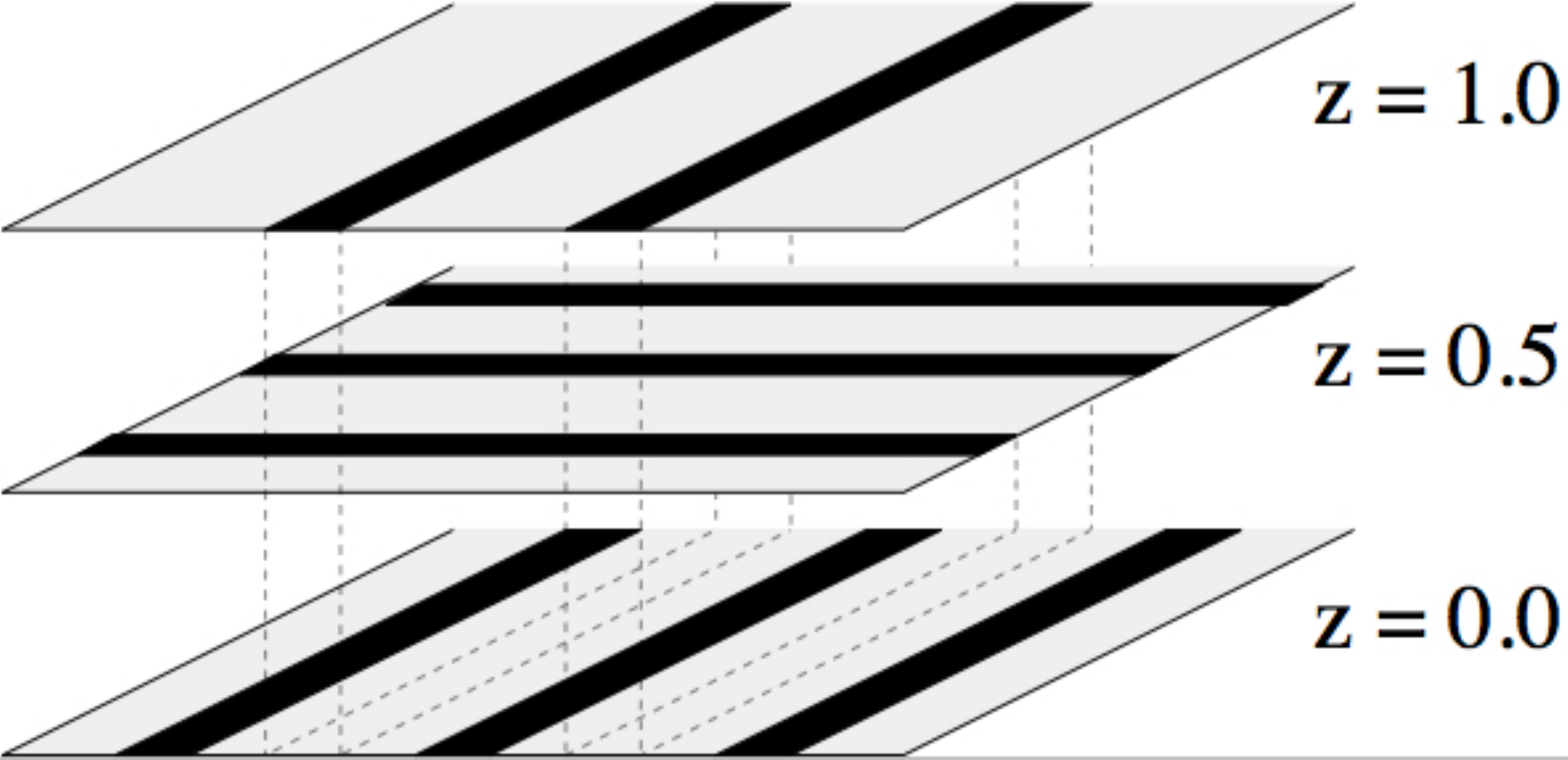}}
  \caption{(Left) Plot of x-ray intensity along ${\bf Q}=(2-2\delta,0,Q_z)$ (after background subtraction) for a charge order peak in LNSCO with $x=0.12$, from Zimmermann {\it et al.} \cite{vonz98}.  (Right) Stacking arrangement for charge stripe order consistent with the unit-cell doubling along the $c$ axis indicated by the x-ray measurement.}
  \label{fg:qz}
 \end{figure}

Besides considering the order within the planes, one can also measure the correlations between the planes.  An x-ray diffraction study of LNSCO with $x=0.12$ demonstrated that, for charge order, the superlattice intensity along $Q_z$ is peaked at half-integer values, indicating a doubling of the unit cell along the $c$-axis, as indicated on the left-hand side of Fig.~\ref{fg:qz}.  A stacking model consistent with the scattering result is shown on the right-hand side.  The chemical unit cell contains two CuO$_2$ layers, and these should have orthogonal stripe domains.   A centered arrangement of stripes in equivalent layers, as might occur due to Coulomb repulsion, then yields a doubling of the cell.  The very broad width of the scattering along $Q_z$ indicates that the correlation length in that direction is quite short.

\begin{figure}[t]
 \includegraphics[width=0.7\textwidth]{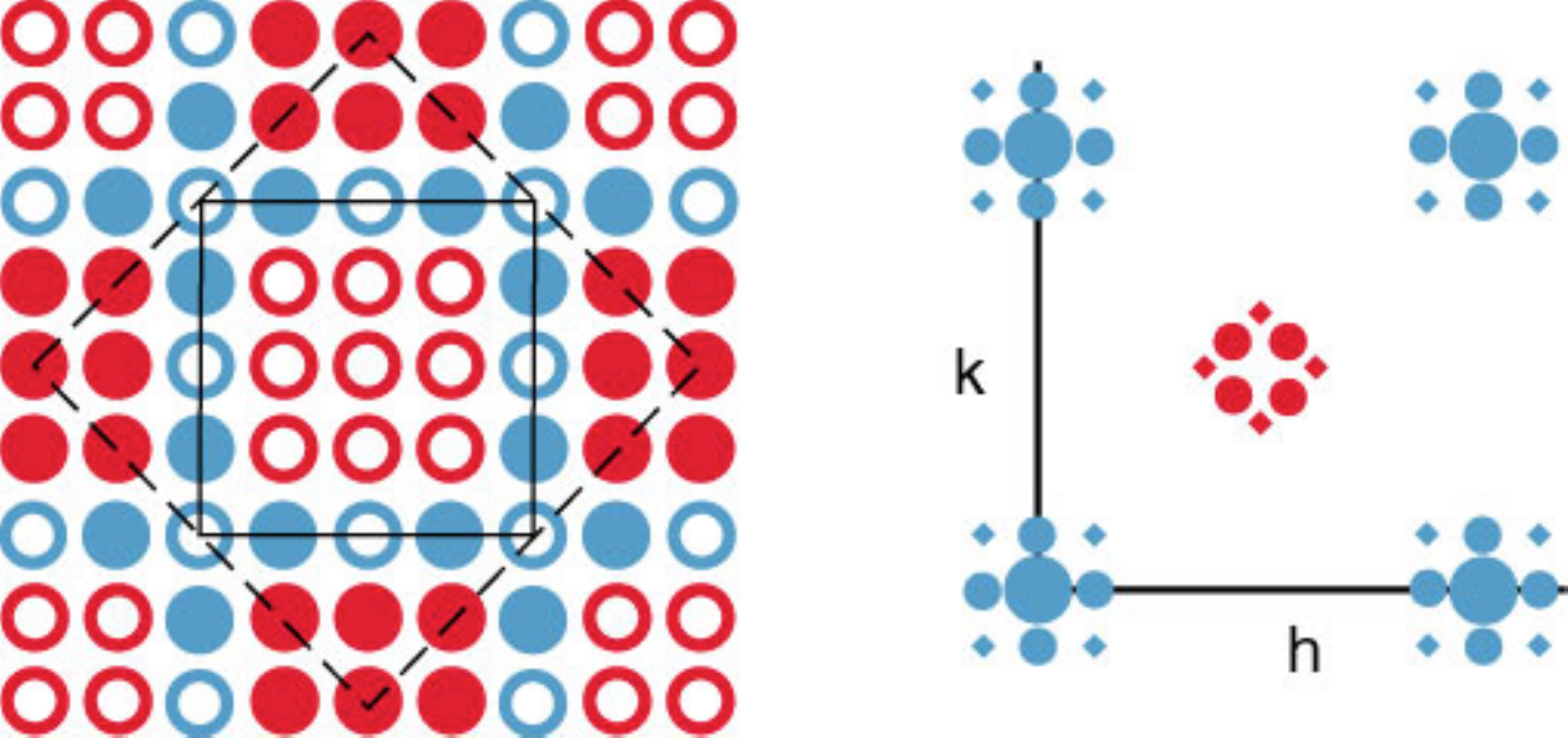}
  \caption{(Left) Real-space model of checkerboard order from superimposed stripe domains.  (Right) Corresponding superlattice peaks in reciprocal space.  Note that the lowest order magnetic and charge peaks are now rotated by $45^\circ$ with respect to each other.}
  \label{fg:grid}
 \end{figure}

It has been proposed \cite{fine04} that, instead of unidirectional stripes, one might have a two-dimensional pattern, such as that shown on the left-hand side of Fig.~\ref{fg:grid}.  It turns out that there are difficulties in describing the experimental observations with such a model.  For one thing, if one relies on the charge stripes to act as antiphase domain walls for the spins, then one finds that the  orientation of the first order magnetic peaks is rotated $45^\circ$ relative to the charge order peaks, as indicated on the right-hand side of Fig.~\ref{fg:grid}.  Secondly, a 2D grid pattern would allow second order superlattice peaks (not present for unidirectional stripes) that have not been detected in diffraction experiments, despite attempts to measure them.

The charge stripe order has also been confirmed in LBCO with $x={\scriptstyle\frac18}$ by resonant soft-x-ray diffraction by Abbamonte {\it et al.} \cite{abba05}.  Measuring with x-ray energies near the O $K$ edge, superlattice intensity is seen only when the energy is resonant with electronic excitations into the O $2p$ hole states (or the $3d$ hole).  This result is consistent with spatial modulation of the O $2p$ hole density.\footnote{For some recent alternative interpretations, see \cite{achk13,benj13}.} 

\subsection{Temperature dependence and dynamics}

\begin{figure}[t]
 {\includegraphics[width=0.35\textwidth]{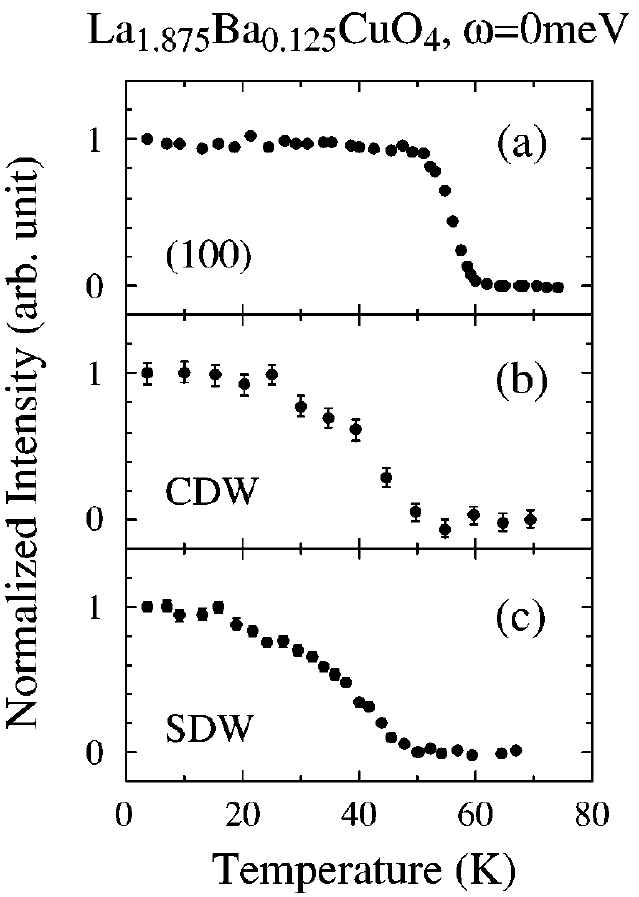}\hskip40pt
  \includegraphics[width=0.35\textwidth]{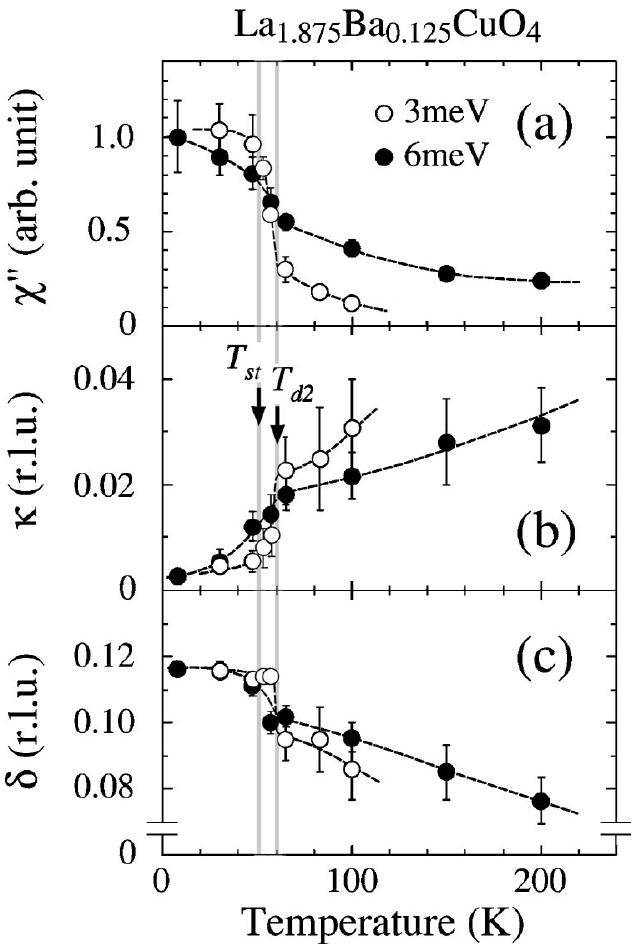}}
  \caption{(Left) Temperature dependence of superlattice peak intensities associated with (a) LTT phase, (b) charge stripe order, and (c) spin stripe order, measured by neutron diffraction on LBCO with $x=0.125$.  (Right) Temperature dependence of parameters characterizing inelastic neutron scattering measurements of spin fluctuations at $\hbar\omega =3$~meV (open circles) and 6~meV (filled circles): (a) local spin susceptibility, $\chi''(\omega)$, (b) peak width, and (c) peak incommensurability, $\delta$.  From Fujita {\it et al.} \cite{fuji04}. }
  \label{fg:cdw_tdep}
 \end{figure}

As a function of temperature, charge stripe order develops as soon as the crystal structure transforms to the LTT phase, as shown on the left-hand side of Fig.~\ref{fg:cdw_tdep} \cite{fuji04}.  Spin stripe order tends to develop at a slightly lower temperature, though gapless spin excitations are detected as soon as charge order appears \cite{tran08}.  The development of charge order before spin order is  similar to the situation in nickelates.

We have already looked at the magnetic spectrum associated with spin order, as the dispersion shown for LBCO with $x={\scriptstyle\frac18}$ shown in Fig.~\ref{fg:hourglass} was measured in the stripe-ordered phase.  The thermal evolution of parameters related to low-energy spin fluctuations is shown on the right-hand side of Fig.~\ref{fg:cdw_tdep}.  The low-energy magnetic spectral weight changes relatively little within the stripe-ordered state, but falls off substantially in the LTO phase, where there is no stripe order.  The incommensurability is also fixed within the stripe-ordered state, but shows a significant jump and decrease at higher temperatures.

\subsection{Doping dependence}

\begin{figure}[t]
 \includegraphics[width=0.6\textwidth]{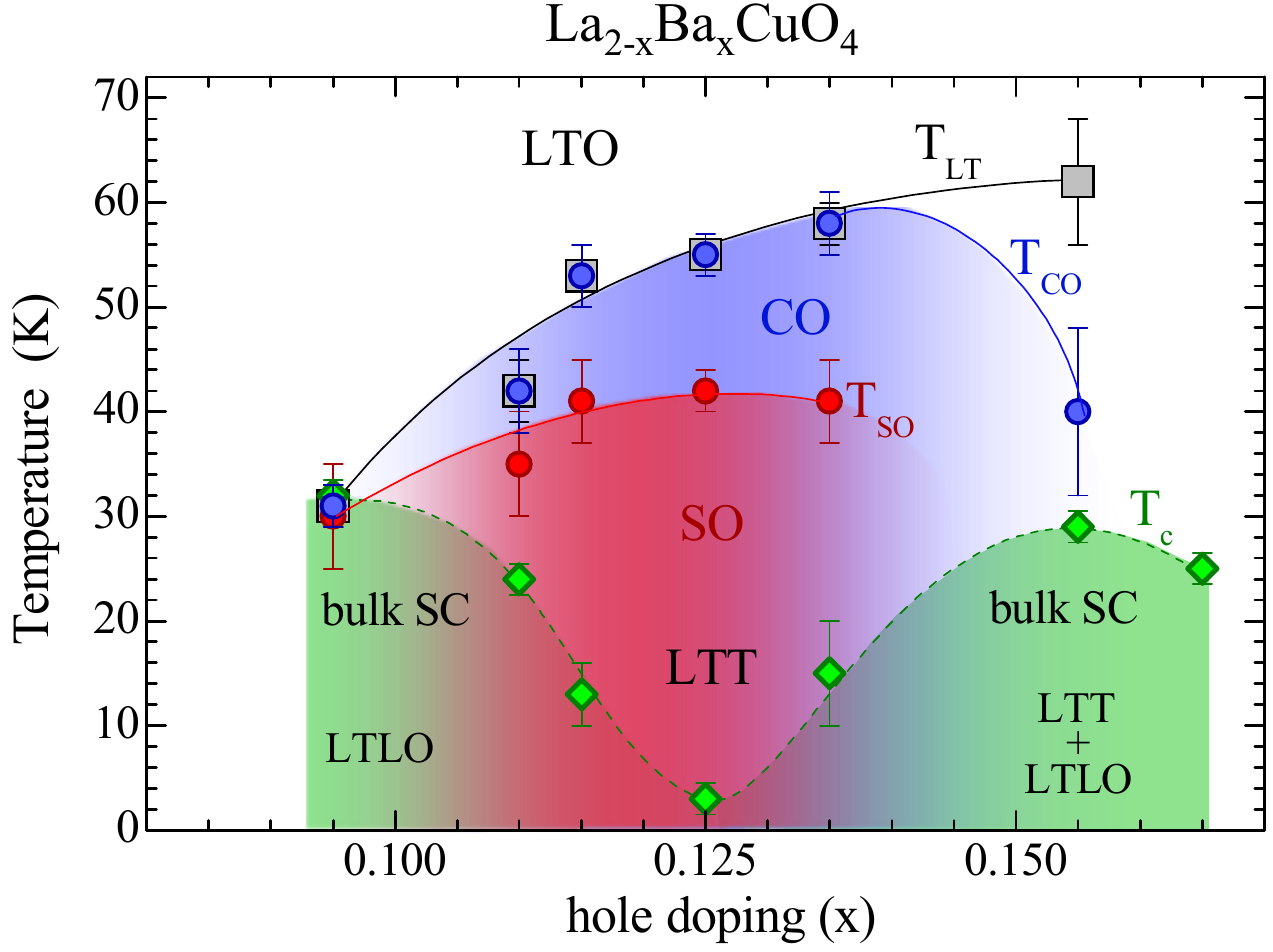}
  \caption{Temperature vs.\ hole-doping phase diagram of La$_{2-x}$Ba$_x$CuO$_4$ single crystals determined by x-ray diffraction, neutron diffraction, and magnetic susceptibility measurements, from H\"ucker {\it et al.} \cite{huck11}. Onset temperatures: $T_c$ of bulk superconductivity (SC) (diamonds), $T_{\rm co}$ of charge stripe order (CO) (circles), $T_{\rm so}$ of spin stripe order (SO) (circles), and $T_{\rm LT}$ of the low-temperature structural phases LTT and LTLO (squares). At base temperature CO, SO, and SC coexist, at least in the crystals with $0.095\le x\le 0.135$. }
  \label{fg:lbco_pd}
 \end{figure}

The phase diagram for structural and stripe order as a function of doping in LBCO is shown in Fig.~\ref{fg:lbco_pd}.   Charge order only appears below the structural transition to the low-temperature-tetragonal phase.\footnote{Besides the LTT phase, samples further away from $x={\scriptstyle\frac18}$ tend to show a low-temperature less-orthorhombic phase, as discussed in \cite{huck11}.}  The maximum spin stripe ordering temperature occurs at $x={\scriptstyle\frac18}$, where the bulk $T_c$ is a minimum.  The amplitudes of the spin and charge order found at low temperature decrease as one moves away from $x={\scriptstyle\frac18}$.

\begin{figure}[t]
 \includegraphics[width=0.45\textwidth]{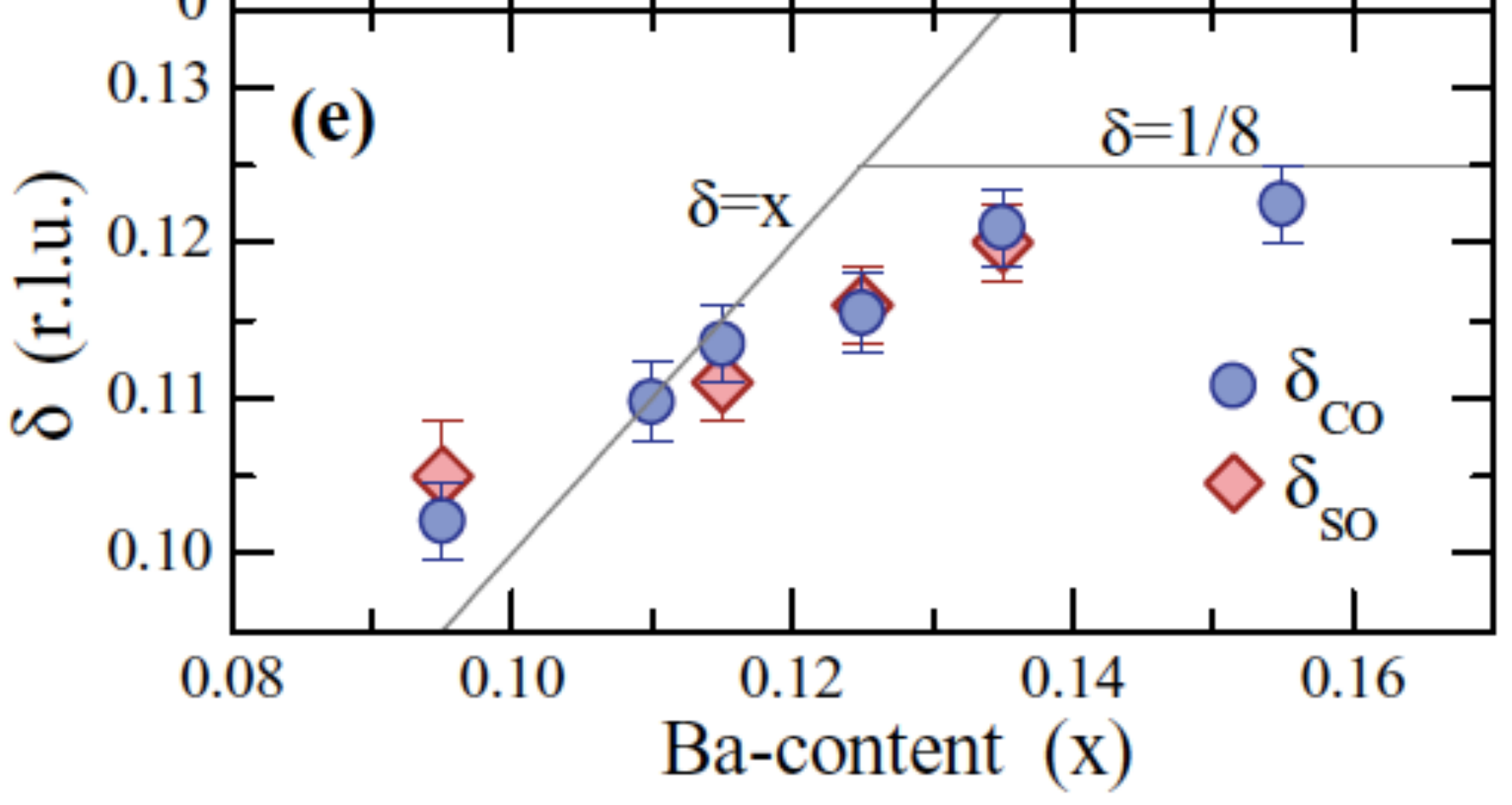}
  \caption{Incommensurability $\delta$ extracted by x-ray diffraction from the CO-peaks and by neutron diffraction from the SO-peaks.  From H\"ucker {\it et al.} \cite{huck11}. }
  \label{fg:icvsx}
 \end{figure}

The doping dependence of the incommensurability obtained from spin and charge order peaks is shown in Fig.~\ref{fg:icvsx}.  The charge and spin modulations are locked together everywhere they are both seen.

 Similar doping dependences of stripe order have been reported for La$_{1.6-x}$Nd$_{0.4}$Sr$_x$CuO$_4$ \cite{ichi00} and La$_{1.8-x}$Eu$_{0.2}$Sr$_x$CuO$_4$ \cite{fink11}.  The overall scale of the superconducting $T_c$ is depressed as one goes from LBCO to LNSCO to LESCO.

\subsection{Stripes and superconductivity}

\begin{figure}[t]
 \includegraphics[width=0.5\textwidth]{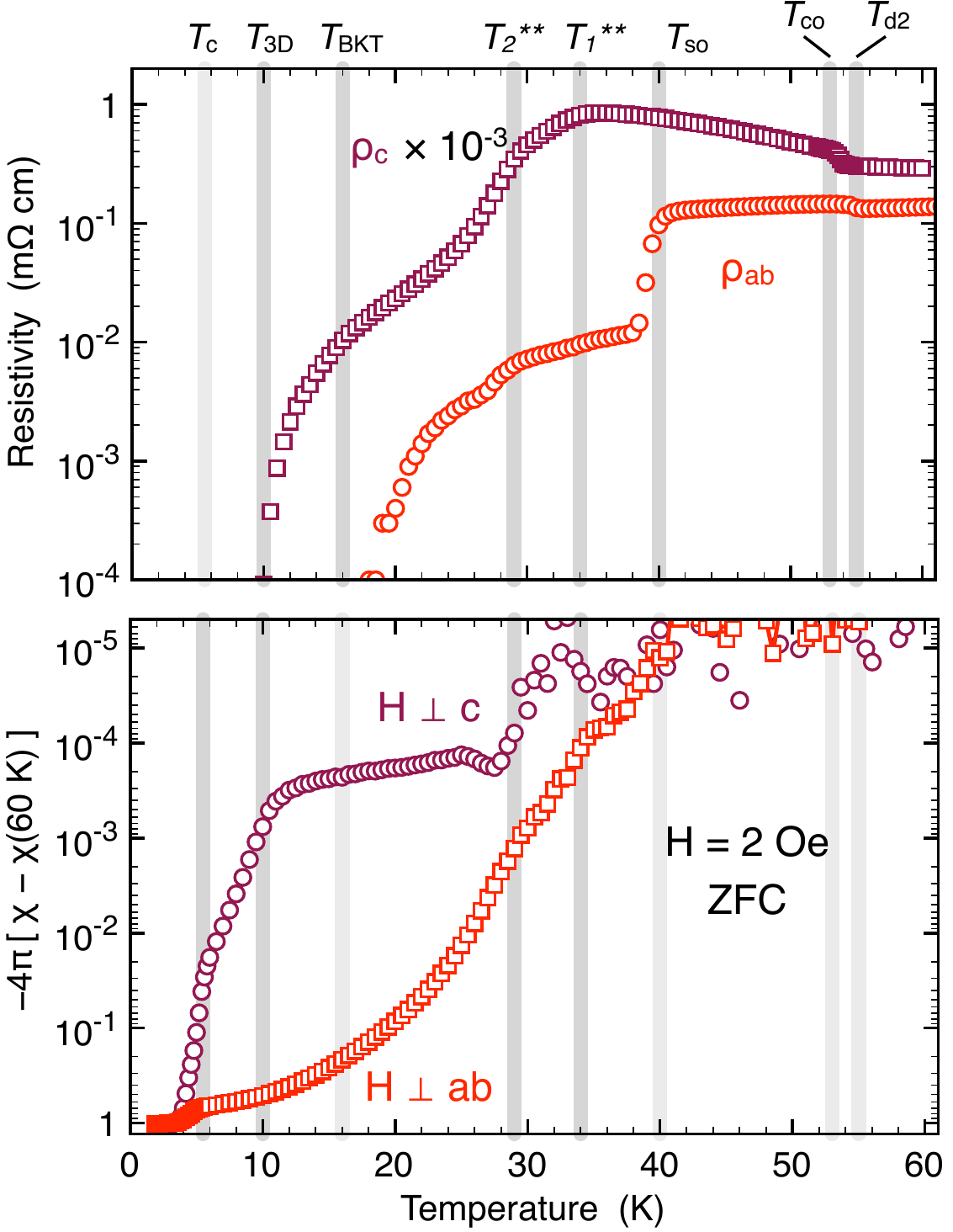}
  \caption{(Top) In-plane resistivity, $\rho_{\rm ab}$ (circles), and resistivity perpendicular to the planes, $\rho_{\rm c}$, divided by $10^3$ (squares). (Bottom) Magnetic susceptibility measured with the field perpendicular to the planes (squares) and parallel to the planes (circles) for an applied field of 2 Oe (after cooling in zero field). $\chi$ has been corrected for shape anisotropy.  Vertical gray lines denote relevant temperatures, with labels at the top. From  \cite{tran08}. }
  \label{fg:rho_chi}
 \end{figure}

One might conclude from Fig.~\ref{fg:lbco_pd} that stripes and superconductivity are mutually incompatible types of order; however, the situation turns out to be rather subtle and considerably more interesting.  Careful measurements of the anisotropic properties of LBCO with $x={\scriptstyle\frac18}$ have provided clear evidence that 2D superconducting correlations onset together with spin stripe order \cite{li07}.  For example, the in-plane resistivity, $\rho_{\rm ab}$, drops by an order of magnitude at 40~K, while the $c$-axis resisitivity, $\rho_{\rm c}$, shows no change, as illustrated in Fig.~\ref{fg:rho_chi} \cite{tran08}.  That figure also shows that there is an onset of weak 2D diamagnetism that coincides with the drop in $\rho_{\rm ab}$.  

\begin{figure}[t]
 \includegraphics[width=0.5\textwidth]{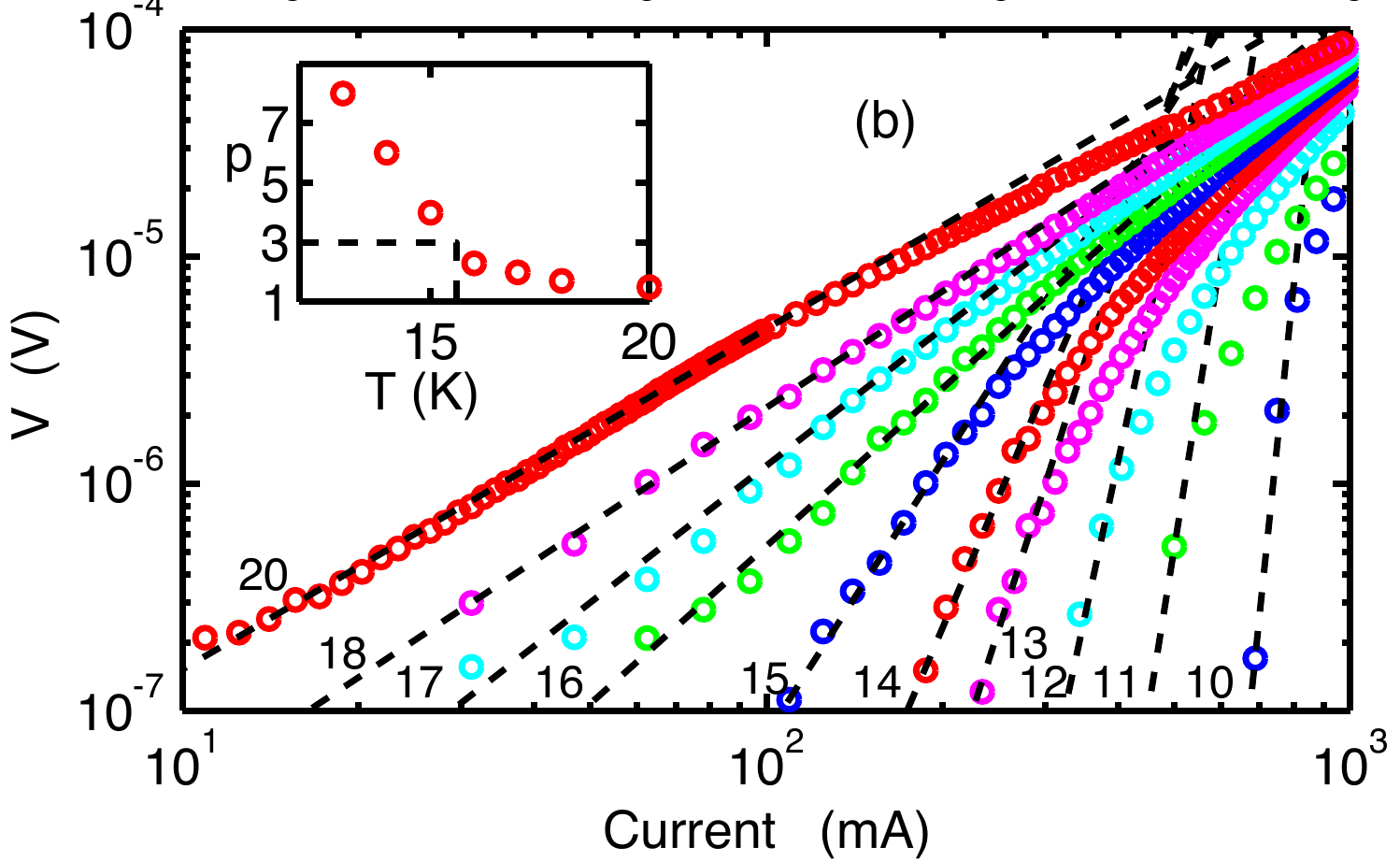}
  \caption{Log-log plot of in-plane voltage, $V$, vs.\ current, $I$, at temperatures from 20 to 10 K. Each curve is labeled by $T$ in K. Dashed lines are approximate fits to the slopes at low current; slope $=p$, with $V\sim I^p$. Inset: plot of $p$ vs $T$. Dashed line indicates that $p$ crosses 3 at $T = 15.6$~K. For an isolated 2D system, a KT transition is characterized by a jump of $p$ from 1 to 3, with a non-universal evolution of $p$ below the transition.  From Li {\it et al.}  \cite{li07}. }
  \label{fg:KT}
 \end{figure}

One can see from Fig.~\ref{fg:rho_chi} that $\rho_{\rm ab}$ heads towards zero as $T$ approaches 16~K, while $\rho_c$ is still quite large.  This suggests a transition to 2D superconductivity within the CuO$_2$ layers.  In general, one would expect 2D superconducting order to develop through a Kosterlitz-Thouless (KT) transition, across which the transport goes from linear to nonlinear.  Evidence for such a transition is presented in Fig.~\ref{fg:KT}.

While the superconductivity in the cuprates is determined by pairing interactions within the CuO$_2$ layers, it is extremely unusual to observe a 2D superconducting transition.   Josephson coupling between the layers causes the appearance of 3D superconducting order as soon as the superconducting correlations within the layers start to grow large.  The absence of 3D order implies a frustration of Josephson coupling, but what could cause this?

A sharp drop in Josephson coupling due to a dopant-induced change in structure from LTO to LTT was first observed in a study of $c$-axis optical conductivity by Tajima {\it et al.} \cite{taji01}.   With the assumption that this change is associated with the onset of stripe order, Himeda {\it et al.} \cite{hime02} proposed a type of superconducting stripe order that, when combined with the LTT structure, could explain the frustrated Josephson coupling.  This pair-density-wave (PDW) superconducting order was rediscovered by Berg {\it et al.} \cite{berg07,berg09b} in light of the LBCO results.  

\begin{figure}[t]
  {\includegraphics[width=0.4\textwidth]{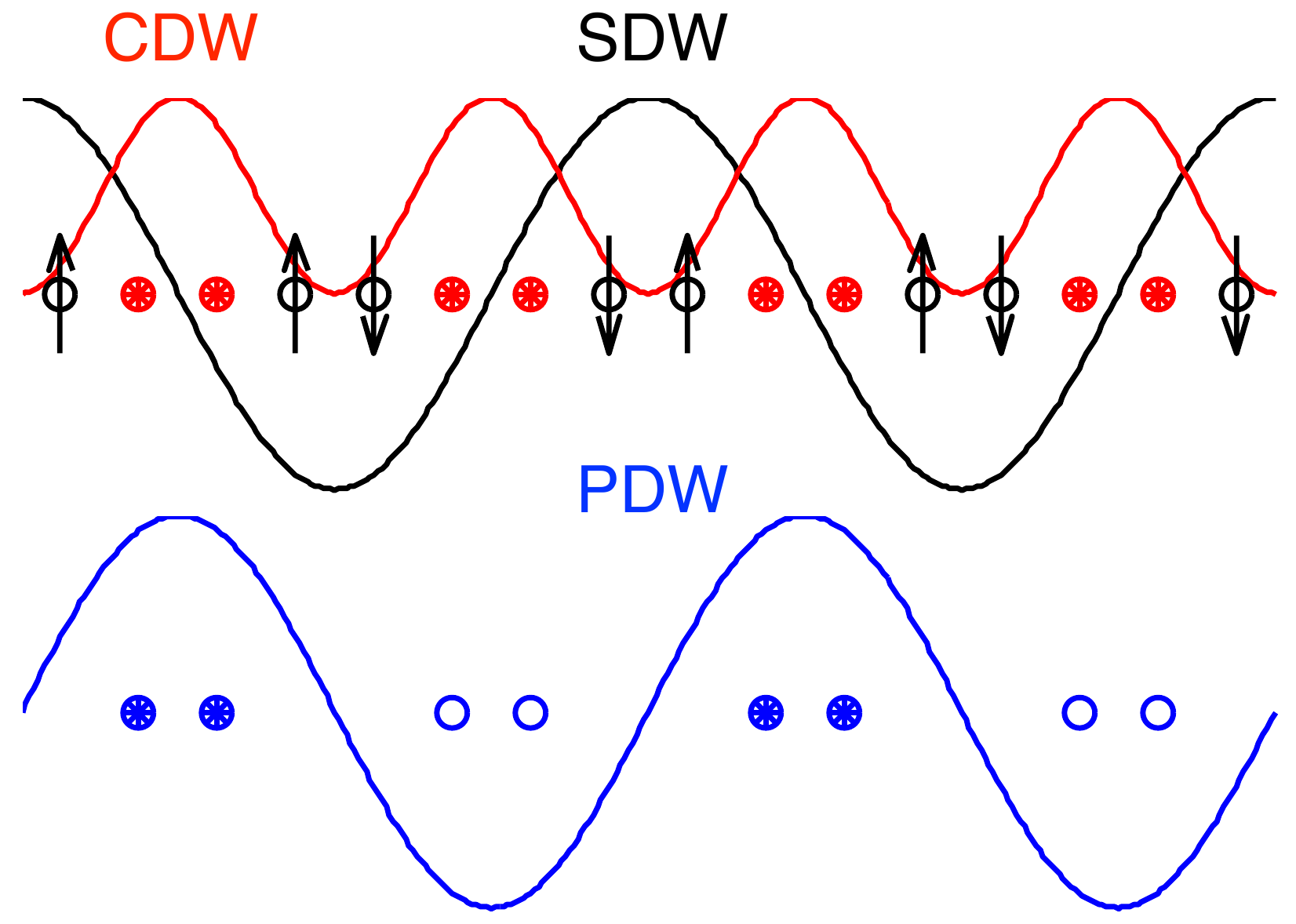}\hskip60pt
  \includegraphics[width=0.3\textwidth]{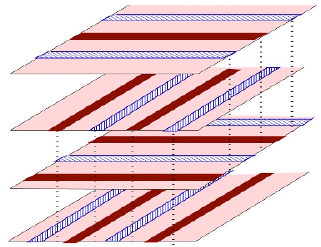}}
  \caption{(Left)  Schematic diagram of CDW, SDW, and PDW orders, indicating the relationships among the phases of the modulations \cite{fuji12a}.  (Right) Stacking of stripe planes, with alternating superconducting phase in the charge stripes indicated.  From Berg {\it et al.} \cite{berg07}.}
  \label{fg:pdw}
\end{figure}

The concept is illustrated schematically in Fig.~\ref{fg:pdw}.   The pair wave function is large in each charge stripe, but it changes sign between neighboring stripes, going through zero in the antiferromagnetic stripes.  The zero in the AF regime may be necessary to allow coexistence with the ordered spins.  The period of the PDW is the same as that of the spin order.  Because of the rotation of the stripe direction from layer to layer, the oscillations of the pair wave function cause the net Josephson coupling to be zero.

\subsection{ARPES studies}

\begin{figure}[t]
 \includegraphics[width=0.5\textwidth]{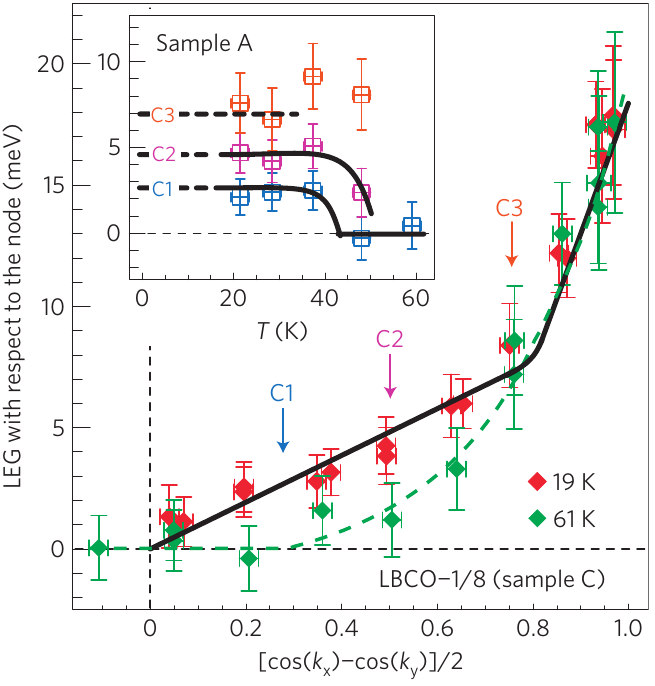}
  \caption{Energy gap (measured in terms of the leading edge of the spectral function) plotted vs.\ electron wave vector in a form such that the data points should follow a straight line in the case of a $d$-wave gap.  From ARPES measurements on LBCO with $x={\scriptstyle\frac18}$ by He {\it et al.}  \cite{he09}.  Reprinted by permission from Macmillan Publishers Ltd.  Solid line is a guide to the eye through the data at 19 K; dashed line characterizes 61~K.  Inset shows the temperature dependence of the gap at wave vectors indicated by arrows in the main panel. }
  \label{fg:arpes}
 \end{figure}

The first ARPES study of a stripe-ordered sample was performed by Valla {\it et al.} \cite{vall06} on LBCO with $x={\scriptstyle\frac18}$.  The low-temperature measurements were compatible with a $d$-wave-like gap near the Fermi level, similar to that seen in other cuprates.  Evidence that the gap is symmetric in energy about the Fermi level was provided in the same paper by scanning tunneling microscopy (STM) measurements.  The temperature dependence of the gap has been studied by He {\it et al.} \cite{he09}, as indicated in Fig.~\ref{fg:arpes}.  The $d$-wave gap begins to close near the node on warming above $\sim40$~K, the temperature where the strong 2D superconducting fluctuations disappear.  

The gap function expected for a PDW state has been calculated by Baruch and Orgad \cite{baru08}.  They find that the gap should be large in the antinodal regions, but that there should be a gapless nodal arc.  A gap of this sort is typically seen in the pseudogap phase, for example in Bi$_2$Sr$_2$CaCu$_2$O$_{8+\delta}$ \cite{kani06}, and it has also been observed in a study of La$_{1.48}$Nd$_{0.4}$Sr$_{0.12}$CuO$_4$ \cite{chan08b}.    Nevertheless, a puzzle is left by the ARPES results on LBCO \cite{he09}.  The observed $d$-wave-like gap below 40~K, where frustrated Josephson coupling is indicated \cite{berg07}, is inconsistent with the prediction of a gapless nodal arc.  This puzzle has not yet been resolved.

It is intriguing to note that the ARPES results on LBCO look rather similar to measurements on other superconducting cuprate compounds.   There is nothing in these measurements that provides a special signature of stripe order.  It follows that ARPES measurements do not seem to be especially sensitive to the presence of stripe order.

\subsection{Field-induced stripe order}

As first shown by Lake {\it et al.} \cite{lake02}, a magnetic field $H_\bot$ applied perpendicular to the CuO$_2$ planes can induce static, incommensurate spin order in underdoped LSCO \cite{chan08}.  In Bi2212, modulations of the low-energy electronic density of states have been observed in the vicinity of magnetic vortices by STM \cite{hoff02}.  It is tempting to view these measurements as evidence of field-induced stripe order.

\begin{figure}[t]
 \includegraphics[width=0.7\textwidth]{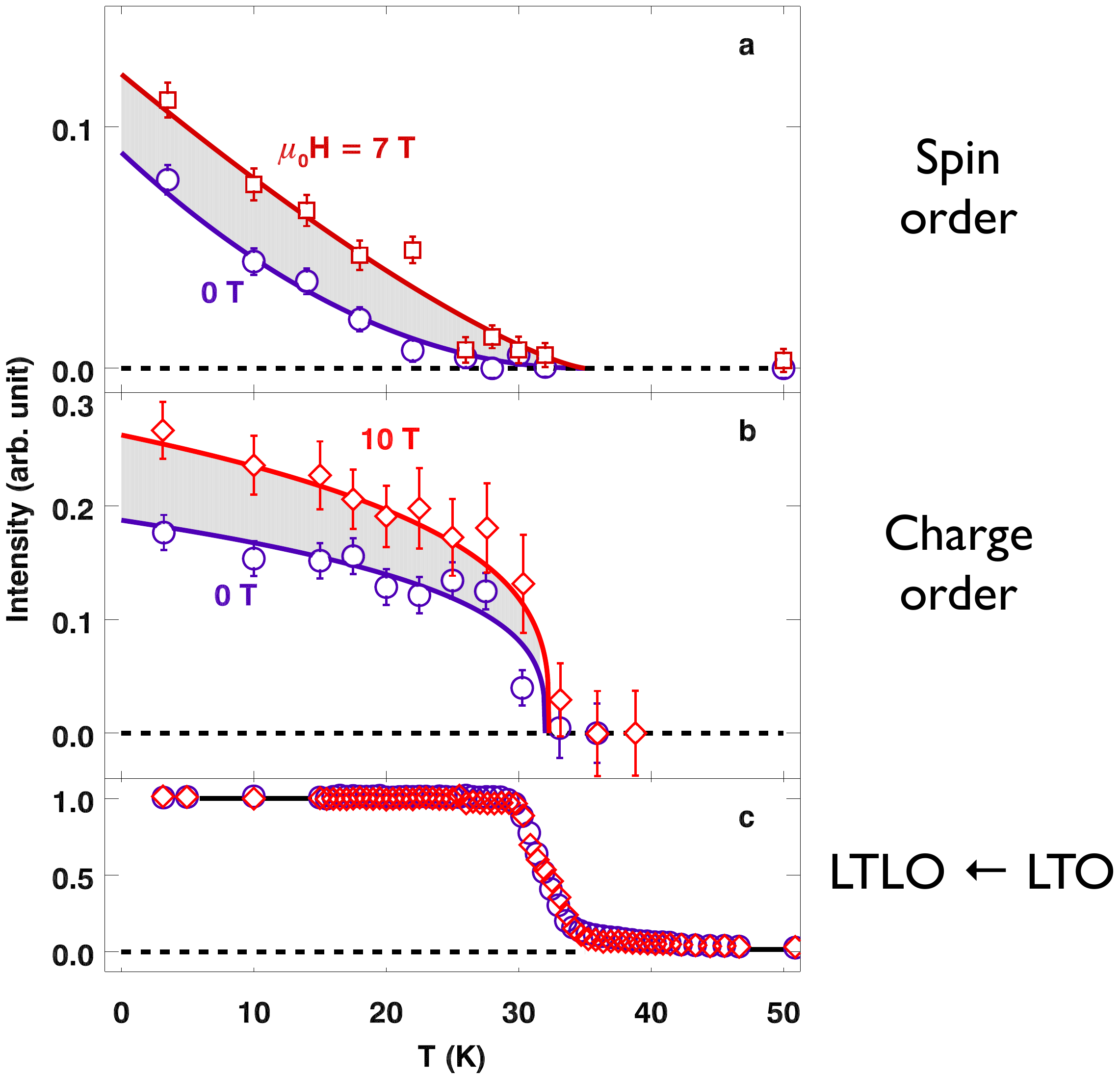}
  \caption{Measurements of superlattice intensities in LBCO with $x=0.095$ in zero field (circles) and in a magnetic field applied perpendicular to the planes for (a) spin order, (b) charge order, and (c) lattice structure. From Wen {\it et al.}  \cite{wen12}. }
  \label{fg:field_ind}
 \end{figure}

Can charge and spin stripes be enhanced by application of $H_\bot$?  A test has been performed on LBCO with $x=0.095$, a composition for which the stripe order in zero field is very weak.  Figure~\ref{fg:field_ind} shows that, indeed, application of $H_\bot$ enhances both charge and spin order \cite{wen12}.

\begin{figure}[t]
 \includegraphics[width=0.6\textwidth]{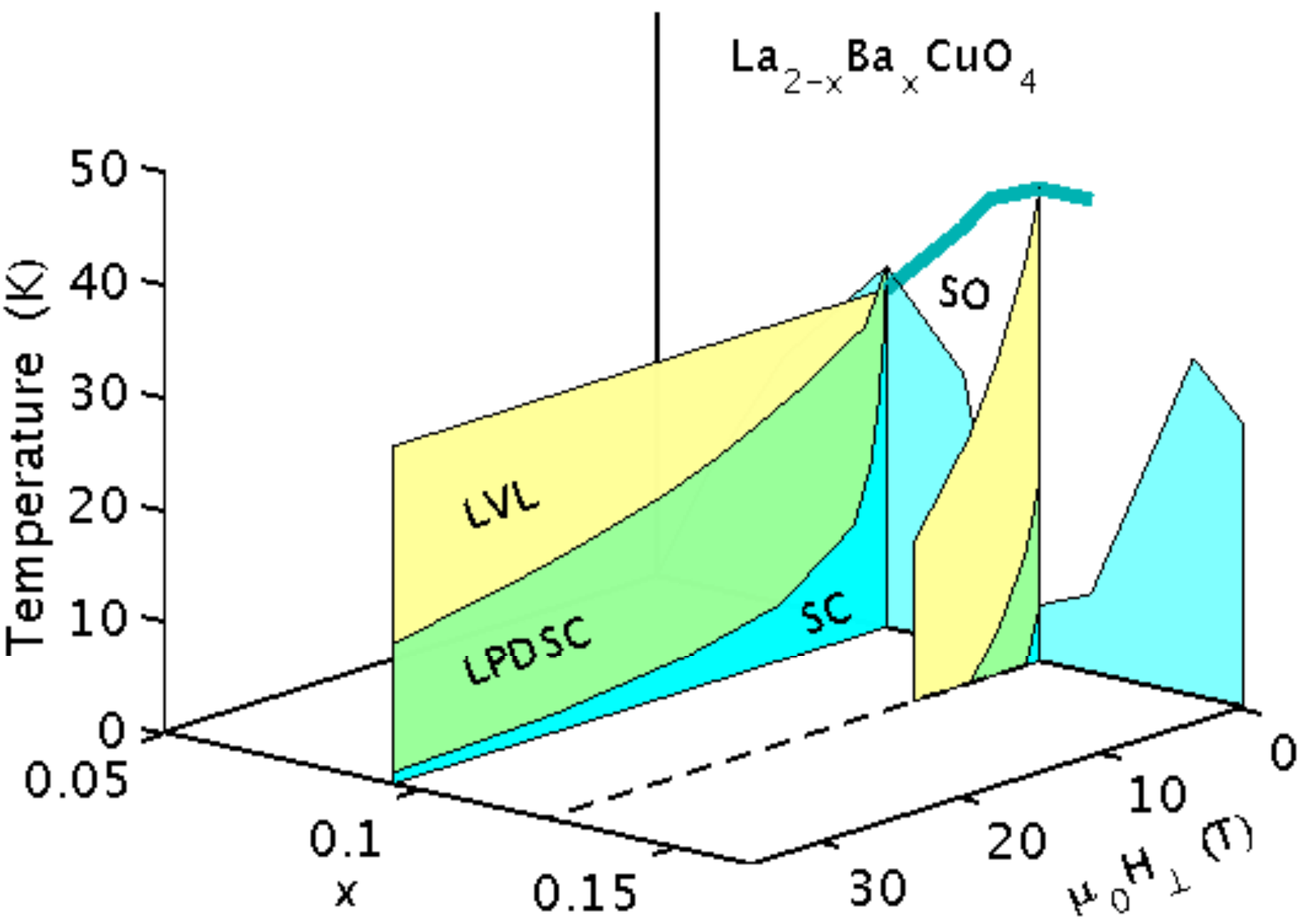}
  \caption{Phase diagram for LBCO as a function of $T$ , $H_\bot$, and $x$ comparing results for various samples. For the $T$-$x$ plane at $H_\bot = 0$, spin order (SO) sets in below the thick line, and superconductivity (SC) occurs in the shaded region below the thin
line. The phase observed in finite $H_\bot$ for $x=0.095$ include the layered vortex liquid (LVL) and the layered, phase-decoupled superconductor (LPDSC). From Stegen {\it et al.}  \cite{steg13}. }
  \label{fg:htx}
 \end{figure}

Besides enhancing stripe order, application of $H_\bot$ to LBCO with $x=0.095$ has also been found to induce a decoupling of the superconducting layers \cite{wen12,steg13}, similar to the decoupling observed for $x={\scriptstyle\frac18}$ at zero field.  Figure~\ref{fg:htx} shows an effective phase diagram.  Consider LBCO $x=0.095$ at a large, fixed $H_\bot$ as a function of temperature.  On cooling, one initially observes an onset of strong superconducting correlations parallel to the planes, but with no sign of coupling between them.  This phase has been labelled a layered vortex liquid (LVL).  With further cooling, the layers appear to develop superconducting order, but still with no coherence between them.  This state is labelled a layered, phase-decoupled superconductor (LPDSC).  With further cooling, interlayer coherence eventually develops and 3D superconducting order is established.

\section{General relevance of stripes to cuprates}

We have seen that stripe order can be detected in a few special cuprate families, and, while stripe order can coexist with some form of 2D superconductivity, it frustrates interlayer coherence.  While there is some exotic behavior associated with stripes, do they have any general relevance to the superconductivity in cuprates?

There is presently no consensus on this question, but it is possible to put together a positive argument for the relevance of stripes.  As a starting point, we have seen that there is a common hourglass shape to the magnetic spectrum of superconducting cuprates.  This is exactly the same spectrum observed in stripe-ordered LBCO, where the low-energy excitations emerge from elastic superlattice peaks.  Charge-stripe correlations provide a natural way to explain both the survival of superexchange-coupled spin correlations and the universal doping dependence of the low-energy spin incommensurability.\footnote{There are certainly other perspectives on the low-energy spin excitations.  For example, see \cite{esch06}.}

It is clear that static stripes are not common among the cuprates.  The connection is likely to involve dynamic stripes.  By analogy with the liquid-crystal phases formed from large molecules, Kivelson, Fradkin, and Emery proposed a model of electronic liquid crystals to describe the phases of a doped Mott insulator \cite{kive98}.   Static stripes correspond to a smectic phase, while slowly fluctuating stripes, which break rotation but not translation symmetry, are associated with a nematic phase.  Further discussion of stripe fluctuations is provided by Kivelson {\it et al.} \cite{kive03}.

Are stripes directly relevant to pairing?  This question is difficult to answer experimentally, but there are theoretical proposals in this direction.  For example, Emery, Kivelson, and Zachar \cite{emer97} argued that the 1D electron gas in each charge stripe should be very susceptible to pairing, especially in the presence of the neighboring AF spin correlations.  The development of 2D superconductivity would be limited by the Josephson coupling between neighboring stripes.  Further elaboration on this topic is provided by Carlson {\it et al.} \cite{carl03} and by Kivelson and Fradkin \cite{kive07}.


\begin{theacknowledgments}
I have had the privilege to work with and learn from a great many excellent collaborators and colleagues, and their names can be seen on many of the references cited here.
My neutron scattering work at Brookhaven National Laboratory is supported by the U.S. Department of Energy, Office of Basic Energy Sciences, Division of Materials Sciences and Engineering under Contract No.\ DE-AC02-98CH10886, while collaborative work on transport properties is supported through the Center for Emergent Superconductivity, an Energy Frontier Research Center.
\end{theacknowledgments}



\bibliographystyle{aipproc}   

\bibliography{LNO,theory}

\end{document}